\documentclass[11pt]{article}
\pdfoutput=1
\usepackage{jheppub}
\usepackage[T1]{fontenc}
\usepackage{cancel,tensor,enumitem,fix-cm}
\usepackage{epsfig}
\usepackage{graphicx}
\usepackage[english]{babel}
\usepackage{amsmath}
\usepackage{textcomp}
\usepackage{multirow}
\usepackage{booktabs}
\usepackage{tikz}
\usepackage{overpic}

\newcommand{\bea}{\begin{eqnarray}}
\newcommand{\eea}{\end{eqnarray}}
\newcommand{\be}{\begin{equation}}
\newcommand{\ee}{\end{equation}}
\newcommand{\beq}{\begin{equation}}
\newcommand{\eeq}{\end{equation}}

\def\({\left(}
\def\){\right)}
\def\[{\left[}
\def\]{\right]}

\newcommand{\bra}[1]{\langle #1|}
\newcommand{\ket}[1]{|#1\rangle}

\def\g{\gamma}
\def\half{\frac{1}{2}}

\def\bx{\bar{x}_0}

\def\tz0{\tilde{z}_0}

\def\zm{z_{m}}
\def\tzx{\tilde{z}_{*}}

\def\tA{\tilde{A}}

\definecolor{palatinatepurple}{rgb}{0.41, 0.16, 0.38}

\definecolor{uglybrown}{rgb}{0.8,  0.7,  0.5}

\def\ee{\epsilon}

\def\hn{\hat{n}}
\def\htt{\hat{\tau}}
\def\l{\lambda}

\subheader{\begin{flushright}YITP-18-37
\end{flushright}}

\title{Geometric Aspects of Holographic Bit Threads}

\author[a]{Cesar A. Ag\'on,}
\author[b]{Jan de Boer}
\author[b]{and Juan F. Pedraza}
\affiliation[a]{C.~N. Yang Institute for Theoretical Physics, State University of New York\\
Stony Brook, NY 11794, USA}
\affiliation[b]{Institute for Theoretical Physics, University of Amsterdam\\
Amsterdam, 1090 GL, The Netherlands}
\emailAdd{cesar.agon@stonybrook.edu}
\emailAdd{j.deboer@uva.nl}
\emailAdd{jpedraza@uva.nl}

\abstract{We revisit the recent reformulation of the holographic prescription to compute entanglement entropy in terms of a convex optimization problem, introduced by Freedman and Headrick. According to it, the holographic entanglement entropy associated to a boundary region is given by the maximum flux of a bounded, divergenceless vector field, through the corresponding region. Our work leads to two main results: (i) We present a general algorithm that allows the construction of explicit thread configurations in cases where the minimal surface is known. We illustrate the method with simple examples: spheres and strips in vacuum AdS, and strips in a black brane geometry. Studying more generic bulk metrics, we uncover a sufficient set of conditions on the geometry and matter fields that must hold to be able to use our prescription. (ii) Based on the nesting property of holographic entanglement entropy, we develop a method to construct bit threads that maximize the flux through a given bulk region. As a byproduct, we are able to construct more general thread configurations by combining (i) and (ii) in multiple patches. We apply our methods to study bit threads which simultaneously compute the entanglement entropy and the entanglement of purification of mixed states and comment on their interpretation in terms of entanglement distillation. We also consider the case of disjoint regions for which we can explicitly construct the so-called multi-commodity flows and show that the monogamy property of mutual information can be easily illustrated from our constructions.}

\begin{document}
\maketitle
\flushbottom

\section{Introduction}

An important lesson from the past decade's research program in holography is that quantum information theory provides a powerful tool to structure our thinking about quantum gravity and quantum field theories. By now, significant progress has been made in understanding how quantum entanglement is encoded in the bulk \cite{Ryu:2006bv,Ryu:2006ef,Hubeny:2007xt,Lewkowycz:2013nqa,Dong:2016hjy} and, in particular, how Einstein's equations emerge from its dynamics \cite{Lashkari:2013koa,Faulkner:2013ica,Faulkner:2017tkh}. See \cite{Nishioka:2009un,Takayanagi:2012kg,Headrick:2013zda,Rangamani:2016dms} for useful reviews. According to the RT prescription \cite{Ryu:2006bv}, entanglement entropy of boundary regions is encoded by the area of certain codimension two surfaces in the bulk, in a manner reminiscent of the well-known Bekenstein-Hawking formula of black hole entropy. Due to its elegance and simplicity, entanglement entropy has become a powerful tool to investigate some of the most fundamental aspects of the holographic correspondence, from bulk reconstruction to the emergence of spacetime. Accordingly, an enormous amount of work has been devoted to finding both analytic and numerical solutions of minimal surfaces in different gravitational settings.

Very recently, Freedman and Headrick suggested that the computation of entanglement entropy can be reinterpreted in terms of a specific convex optimization problem \cite{Freedman:2016zud}. More specifically, they showed that the problem can be cast in terms of finding a bulk vector field $V$ with maximum flux through the corresponding boundary region, divergenceless $\nabla\cdot V=0$, and with an upper bound set by the Planck scale $|V|\leq 1/4G_{N}$.\footnote{These divergenceless flows are Hodge dual to closed forms called calibrations \cite{Harvey:1982xk}. In fact, entanglement entropy has been recently recast in this language in \cite{Bakhmatov:2017ihw}.} Besides the enormous conceptual advantage that this new picture brings into the bulk interpretation of different quantum information properties, it also presents an alternative method with potential utility for the computation of holographic entanglement entropy of boundary regions.

From a more practical standpoint, one can use the tools borrowed from Riemannian geometry and apply them to the construction of the auxiliary objects ---referred to as flows or bit threads--- of the max-flow min-cut proposal. Convex optimization techniques have been shown to be useful in the proof of statements about such objects \cite{Headrick:2017ucz}. However, explicit examples of such vector fields do not make a significant appearance in the entanglement entropy literature up to now. An obvious reason for this is the fact that such vector field configurations are not unique and therefore have no physical meaning on their own. Nevertheless, it is illustrative to work out examples of such objects (representatives of the homology class of vector fields that satisfy the previously described conditions) to gain intuition about their behavior and hopefully be able to make more general claims that are independent of the specific class of vector fields used in the constructions.

In this work we will start filling this gap by constructing explicit examples of such vector fields configurations, starting from situations in which the minimal surface is known. Even though this is a crucial restriction, the main goal of this work is to initiate a broader program that aims at the understanding of more general bit threads configurations. Indeed, in all of our constructions, we are able to see a large amount of non-uniqueness as it is expected from the previous arguments. By focusing on particular examples and studying the freedom allowed in the constructions, we then are able to uncover a large class of new possibilities for consistent bit thread configurations.

In summary, the kind of constructions that we study throughout this paper fall in one of the following three categories:
\begin{itemize}
  \item Symmetric or geodesic flows: these are obtained from general geodesic foliations of a suitable (possibly unphysical) bulk geometry. The geodesics here are identified as the integral lines of the corresponding vector field, while its norm is chosen such that the divergenceless condition is satisfied locally.
  \item Maximally packed flows: these are obtained based on the entanglement wedge nesting property of holographic entanglement entropy. The integral lines of the vector field are taken to be orthogonal to the geodesics (or minimal surfaces) associated to a set of nested subregions. By construction, the norm of the vector saturates the bound $|V|=1$.
  \item Mixed flows: these are obtained by consistently cutting and pasting symmetric and/or maximally packed threads in different patches of the bulk geometry.
\end{itemize}
The last category is the most generic case that we consider, however, we emphasize that there are infinitely many more configurations that are not covered by our constructions. In the future, it would be interesting to study more in detail this degree of ambiguity, e.g, by determining the most general kind of transformations that can be implemented to a particular set of threads and determine how would they modify our specific vector field configurations.\footnote{The most general local ambiguity is of the following form: since $V$ is divergenceless it obeys $d\ast V=0$ and we can always change $V\rightarrow V+\ast d X$; this will e.g. be allowed for sufficiently small generic $|X|$ in regions where $|V|<1$. It is, however, difficult to classify these ambiguities in full generality due to the constraint $|V|\leq 1$.}

This paper is organized as follows. In section \ref{section2} we propose a general algorithm to construct flows in cases where the minimal surface is known. Specializing to the case where the integral curves are taken as geodesics (or minimal surfaces) of the underlying geometry, we then present several explicit examples of the symmetric flows advertised above, including spheres and strips in vacuum AdS, and strips in a black brane geometry. In section \ref{sec:foli} we study general constraints on curvature and matter fields that must be fulfilled in a generic gravitational setting such that a similar construction based on geodesic foliations leads to a valid flow. Based on this study, we revisit the explicit examples constructed in section \ref{section2} and explain why they work based solely on the geometric properties of the underlying geometry.
In section \ref{NPMPF} we describe the construction of maximally packed flows in a general setting and explain the reasoning behind these constructions. Finally, in section \ref{sec:5} we present our two main applications, which are based on mixed flows: specific flow configurations that illustrate the computation of entanglement of purification and the monogamy property of mutual information. We close in section \ref{sec:con} with a summary of our results and a discussion about potential topics for future research.

\section{Symmetric flows: a general algorithm\label{section2}}
In \cite{Freedman:2016zud} it was proposed that the entanglement entropy of a holographic theory associated to a spatial boundary region $A$ is given by the maximum flux of a bounded, divergenceless, bulk vector field through the region in question. As noted there, there are infinite many vector fields that maximize that flux and respect the defining properties. Indeed such non-uniqueness play an important role in its tempting interpretation as a holographic representation of the pattern of spatial entanglement of the field theory degrees of freedom.

In this section, we would like to consider a particular construction of a representative of the homology class of vector fields that solve the above problem. We will start by  proposing a general method before specializing to the AdS case. For this construction we assume that the following data is given: a space-time metric $g_{ab}$ with asymptotic boundary at $z\to 0$, i.e. $g_{ab}\to z^{-2} \eta_{ab}$, a connected boundary region $A$ and its corresponding minimal area bulk surface $m(A)$ (such that $m(A)|_{z\to0}\sim \partial A$). The method has two parts. First, we propose a family of integral curves (also referred to as flow lines or threads) with certain properties (discussed in detail below), from which we extract their tangent vector $\hat{\tau}$. And second, we find the magnitude $|V|$ by imposing the divergenceless condition. With these two ingredients, we can then construct the desired vector field $V=|V|\hat{\tau}$.

Let us explain in more detail the method. The first step is to propose a family of integral curves that satisfy the following properties:
\begin{enumerate}
  \item They  must be continuous and not self-intersecting so that the tangent vector $\hat{\tau}$ is unique and well defined everywhere. Notice that we do not need to impose that the integral curves foliate the full manifold. Points that are not reached by any of the curves will have by definition a vanishing tangent vector $\hat{\tau}=0$.
  \item At the minimal surface $m(A)$, the tangent vector $\hat{\tau}$ must be equal to the unit normal of the minimal surface $\hat{n}_{m}$. This is required since the bound on the magnitude of the vector field must be saturated there, a key fact in the equivalence between the minimal surface and the maximum flux prescriptions for the holographic entanglement entropy.
  \item And finally, since the vector field must be divergenceless, the integral curves should start and end at the boundary.
\end{enumerate}
These three conditions are completely generic, however, since the choice of integral curves is highly non-unique we will impose a further constraint to reduce the arbitrariness in the possible vector fields, without eliminating it completely:
\begin{enumerate}
\setcounter{enumi}{3}
\item We will consider curves that leave from region $A$ and end in region $\bar{A}$. Physically, this condition is equivalent to restrict our attention to microstates without entanglement between degrees of freedom contained in region $A$ or $\bar{A}$.
\end{enumerate}

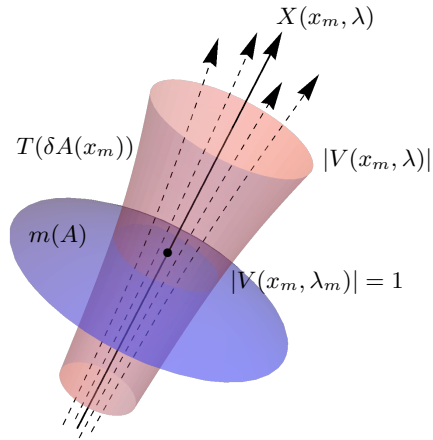
\begin{figure}
\centering
 \begin{overpic}[width=1.7in
 ]{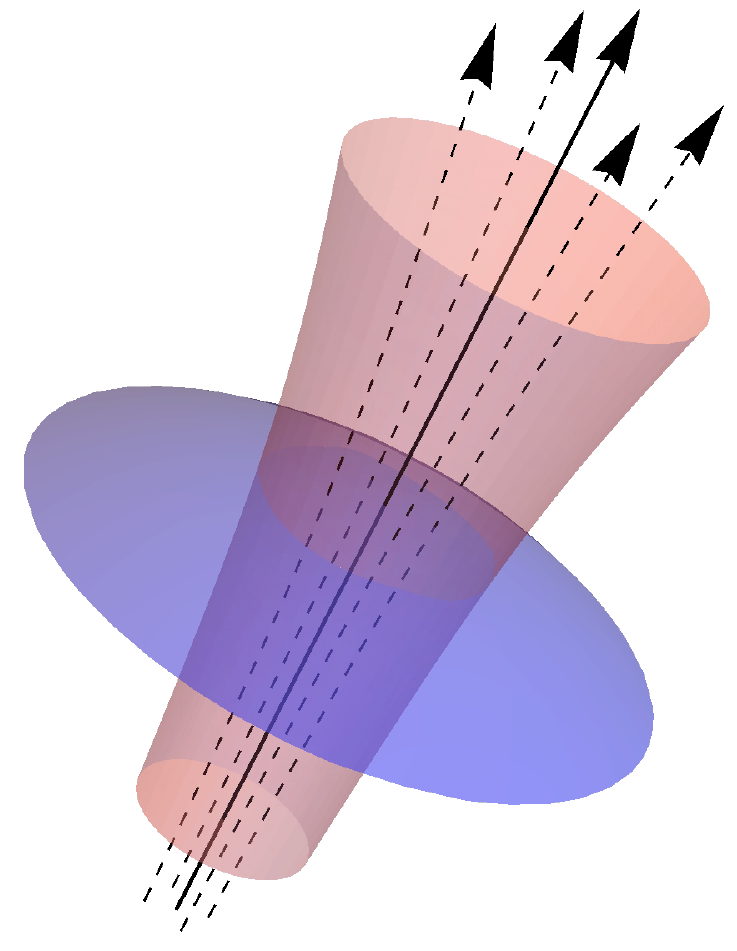}
 \put (7,49) {\footnotesize{$m(A)$}}
 \put (55,38) {\footnotesize{$|V(x_m,\lambda_m)|=1$}}
  \put (77,67) {\footnotesize{$|V(x_m,\lambda)|$}}
  \put (66,101) {\footnotesize{$X(x_m,\lambda)$}}
  \put (4,70) {\footnotesize{$T(\delta A(x_m))$}}
  \put (38.75,45) {\scriptsize{$\bullet$}}
 \end{overpic}
 \caption{Schematic representation of the conservation of flux, for an infinitesimal tube made of integral curves around a point $x_m$. The construction is local, so the $|V|$ obtained with this method is valid for all points away from the minimal surface. The final non-trivial check is to verify that the bound $|V|\leq1$ is respected everywhere.\label{consflux}}
\end{figure}

Once a family of integral curves with the above conditions has been found, the second step is to find the appropriate norm $|V|$. A natural way to do it is by enforcing that the flux through an infinitesimal area element is constant throughout the threads. This construction automatically ensures that the divergenceless condition is satisfied. In practice, we proceed in the following way: first, we parametrize the curves $X(x_m,\lambda)$ by the point at which they intersect the minimal surface, $x_{m}$, and a parameter $\lambda$ that runs along each curve. 
Then, we follow the set of integral curves that leave from an infinitesimal region $\delta A(x_{m})$ around the point $x_{m}$ at $m(A)$.
This area will propagate along the threads and define a co-dimension one bulk region with the topology of a cylinder, which we denote by $T(\delta A (x_m))$.  Since the surface of $T(\delta A (x_{m}))$ is made out of integral curves, this means that there is no flux leaving or entering that surface. Therefore, the divergenceless condition can be imposed by choosing $|V|$ such that the flux through any transverse section of $T(\delta A (x_{m}))$ is constant ---see figure \ref{consflux} for a schematic representation. Mathematically, this condition implies
\bea\label{divergentless}
\int_{\delta A(x_{m})} |V| \sqrt{h |_{\lambda}} d^{d-1}x_{m}={\rm constant}\,,
\eea
where $h|_{\lambda}$ is the induced metric on the plane orthogonal to the flow line at the point $X(x_{m},\lambda)$.
Using the fact that at the minimal surface $|V(x_m,\lambda_m)|=1$, and our knowledge of the exact location of that surface, one can write an explicit expression for $|V|$ along the integral curves, this is
\bea\label{magnitudeV}
 |V(x_{m},\lambda)|  =\frac{\sqrt{h(x_{m},\lambda_{m}) }}{\sqrt{h(x_{m},\lambda)}}\,,
\eea
where $\lambda_{m}$ is the parameter at which the flow line intersects the minimal surface $m(A)$.

The final non-trivial check is to verify that the magnitude of the vector field respects the upper bound away from the minimal surface, $|V|\leq1$. If this is not the case then, the choice of integral curves should be modified and the process should be repeated until the bound is achieved. In that sense, the critical/essential step of our construction relies on the choice of what we call ``good'' integral curves.

\subsection{Explicit constructions in pure AdS}

For concreteness, we would like to start by illustrating the method for the case of pure AdS$_{d+2}$ in Poincar\'e coordinates and for highly symmetric cases: spherical regions (\ref{spheres}) and strips (\ref{strips}). In cases like these the natural choice is to consider curves that respect the symmetry, so the problem simplifies and becomes effectively lower dimensional.

\subsubsection{The sphere \label{spheres}}

The sphere is the most symmetric region on can think of. The minimal surface as well as the explicit bulk and boundary modular Hamiltonians are known explicitly and have been widely exploited leading to many deep results connecting properties of entanglement entropy in field theory with the dynamics of the emergent bulk spacetime.

\vspace{5mm}
\textbf{Minimal surface:}
Consider the minimal surface associated to a spherical region $A$ of radius $R$ located at the boundary of AdS$_{d+2}$. In Poincare coordinates (a constant-$t$ slice of) the metric is given by\footnote{We will set the AdS radius to unity throughout this paper, $L=1$, but it can be easily restored at any point by dimensional analysis.}
\bea
ds^2=\frac{1}{z^2}\(d\vec{x}^2+dz^2\)=\frac{1}{z^2}\( dr^2+r^2d\Omega^2_{d-1}+dz^2\)\,.
\eea
The minimal surface is given by a hemisphere of radius $R$ that extends into the extra dimension $z\in\mathbf{R}^{+}$. The minimal surface can be defined implicitly by the collection of points $(r_m,z_m)$ that satisfy
\bea\label{minsphere}
r_m^2+z_m^2=R^2\,,
\eea
where $r_m=\sqrt{\sum_i (x^i_m)^2}$ and $x^i_m$ are the standard cartesian coordinates. Notice that, without loss of generality, we have chosen to locate our minimal surface centered at the origin of the Poincar\'e coordinates.

For later convenience, we give here the outward-pointing unit normal vector $\hn_m$ at a point $(r_m,z_m)$, which in the above coordinate system is given by
\bea\label{nasphere}
\hn_m^{a}=\frac{z_m}{R}\(r_m,z_m\)\,.
\eea
For simplicity we have chosen to suppress all angular coordinates, so the index $a$ runs over the coordinates $(r,z)$.

\vspace{5mm}
\textbf{Integral curves:}
Since the problem has spherical symmetry, we restrict our attention to one plane, i.e., we set $x_1=r$ and $x_i=0$ $\,\forall\,i>1$. The angular dependence can be easily restored at the end by implementing a rotation on the resulting integral curves.\footnote{In addition, notice that since there is a reflection symmetry in this plane, $x_1\to-x_1$, we can further restrict our attention to $x_1=r\in\mathbf{R}^{+}$.}

Now, let us consider the space of geodesics that lie on the $(r,z)$ plane and find the ones that intersect $m(A)$ at $(r_m,z_m)$ and whose tangent vector $\hat{\tau}$ is parallel to the normal $\hn_m$ at that point. We will argue below that such geodesics are a natural candidate for the integral curves of our vector field $V$.

The two-dimensional effective metric is the following:
\bea
ds^2=\frac{1}{z^2}\(dr^2+dz^2\)\,,
\eea
and the set of geodesics that lie in this plane is given by the two-parameter family of circumferences defined implicitly by
\bea\label{geodesic}
(r-r_s)^2+z^2=R_s^2
\eea
where $r_s$ is the center of the circle on the $r$-axis and $R_s$ is its radius. The tangent vector with unit norm at an arbitrary point is given by
\bea\label{tau}
\hat{\tau}^a=\frac{z}{R_s}\(z , r_s-r\)\,.
\eea
Enforcing that $\hat{\tau}=\hn_m$ at a point $(r_m,z_m)$ on the minimal surface leads to
\bea\label{80}
R_s=\frac{z_m}{r_m} R=\frac{\sqrt{R^2-r_m^2}}{r_m} R \qquad \textrm{and} \qquad r_s=r_m+\frac{z_m^2}{r_m}=\frac{R^2}{r_m}\,.
\eea
Finally, plugging (\ref{80}) into  (\ref{geodesic}) we obtain an implicit expression for the family of geodesics orthogonal to $m(A)$, parametrized by the point $r_m$ on the minimal surface.

Before proceeding further, we need to check if the proposed flow lines satisfy our criteria. Based on our construction, the first condition (namely, that the curves do not intersect with each other) is the only non-trivial requirement. In order to check this, we parametrize the curves with respect to the point $r_a$ at which they intersect $A$,
\bea\label{r0}
r_a=r_s-R_s=\frac{R}{r_m}\(R-\sqrt{R^2-r_m^2} \)\,.
\eea
Similarly, we can find the dual point $r_{\bar{a}}$ at which the curves intersect $\bar{A}$,
\bea\label{barr0}
r_{\bar{a}}=r_s+R_s=\frac{R}{r_m}\(R+\sqrt{R^2-r_m^2} \)\,.
\eea
Self-intersection is avoided if and only if the curves parametrized by $r_m$ are nested. Since the proposed curves are geodesics, this condition is guaranteed provided that $dr_a/dr_m>0$ and $dr_{\bar{a}}/dr_m<0$. Indeed, a quick calculation leads to
\begin{align}
\frac{dr_a}{dr_m}&=\frac{R^2}{r_m^2} \frac{(R-\sqrt{R^2-r_m^2})}{\sqrt{R^2-r_m^2}}>0\,,\\
\frac{dr_{\bar{a}}}{dr_m}&=-\frac{R^2}{r_m^2} \frac{(R+\sqrt{R^2-r_m^2})}{\sqrt{R^2-r_m^2}}<0\,,
\end{align}
which satisfy the above conditions. Therefore our choice of integral curves is validated.

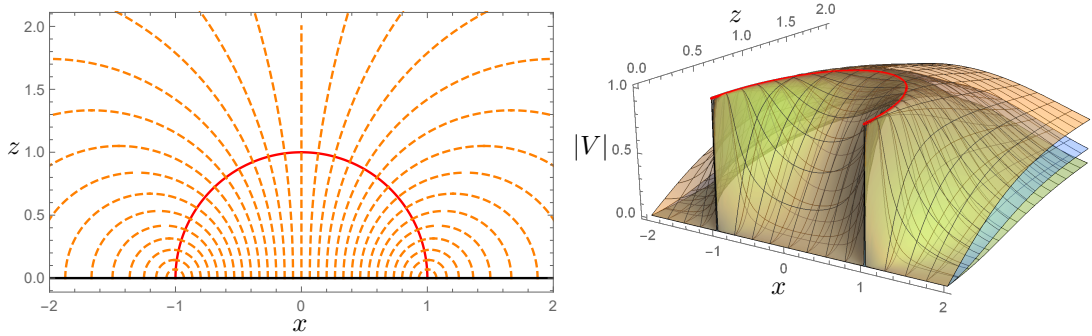
\begin{figure}
\centering
 \includegraphics[width=2.8in
 ]{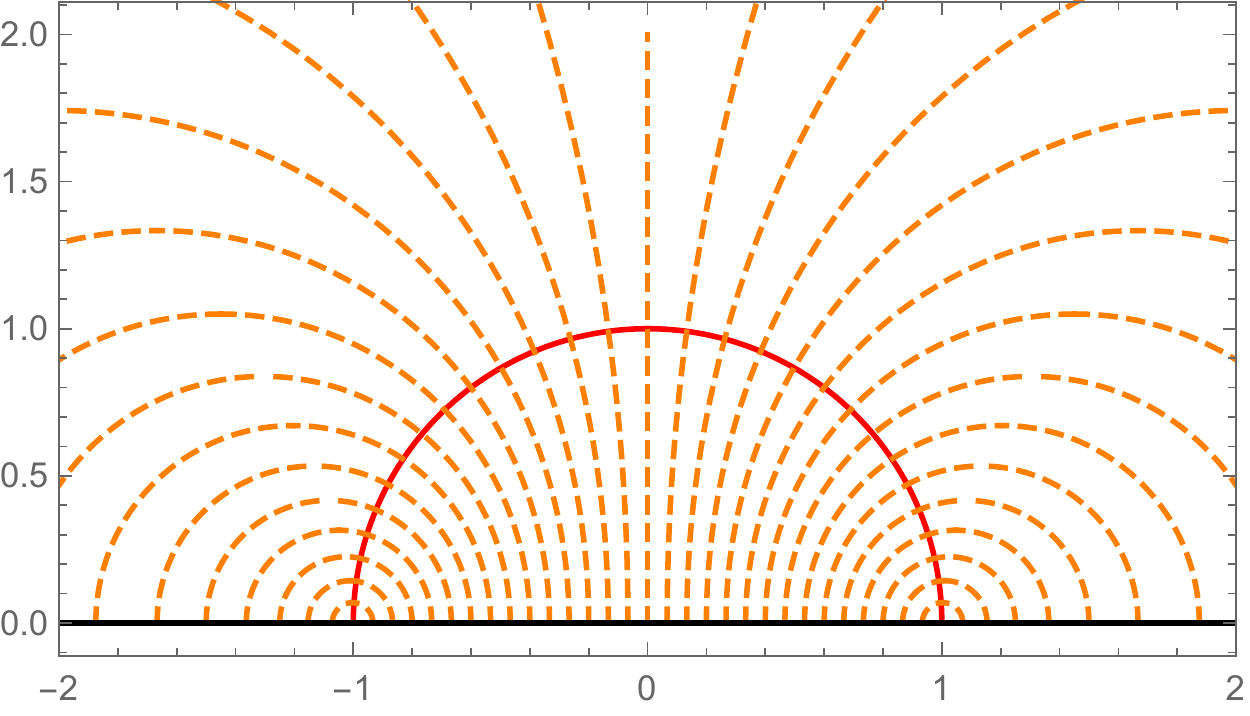}
  \includegraphics[width=3in
 ]{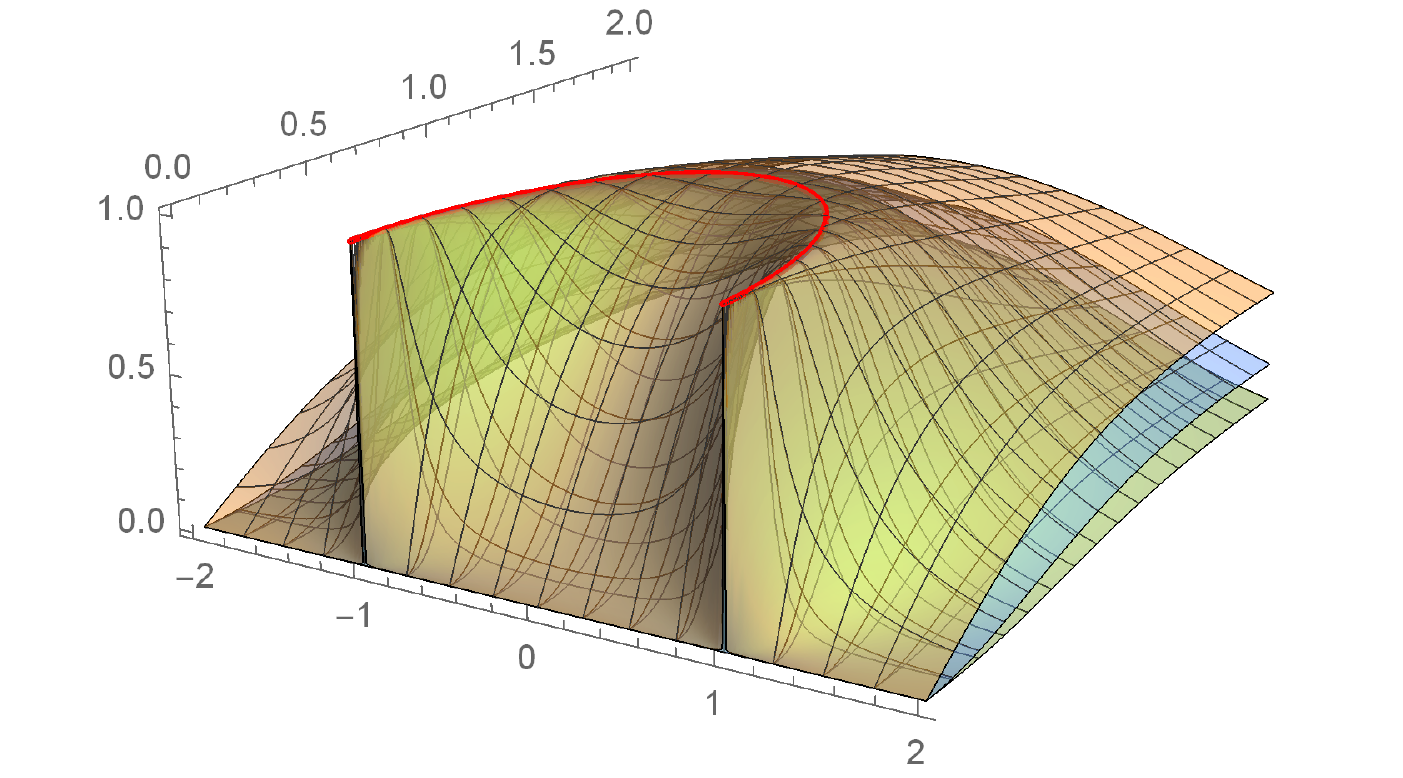}
 \begin{picture}(0,0)
\put(-160,112){{\small $z$}}
\put(-219,62){{\small $|V|$}}
\put(-144,9){{\small $x$}}
\put(-433,62){{\small $z$}}
\put(-325,-5){{\small $x$}}
\end{picture}
 \caption{Vector lines and magnitude $|V|$ for a sphere in $d=1$ (orange), $d=2$ (blue) and $d=3$ (green) spatial dimensions, respectively. The vector field $V$ exhibits spherical symmetry so, for simplicity, we have plotted only one of the spatial axis in all case. The solid red line corresponds to the minimal surface, $m(A)$. This curve also signals the location where the magnitude of the vector field attains its maximal value, $|V|=1$.}\label{Vspheres}
\end{figure}

\vspace{5mm}
\textbf{Vector field:} We can now proceed to find the appropriate norm of the vector field $|V|$. In order to compute the right hand side of (\ref{magnitudeV}) we consider the family of integral curves parametrized by the point $r_m$ at which the curves cross the minimal surface $m(A)$. These curves are given implicitly by equation (\ref{geodesic}), with $R_s(r_m)$ and $r_s(r_m)$ given in (\ref{80}). Then, we compute the orthogonal metric at different points along the curve,
\bea\label{orthospheres}
h_{ab}(r_m;r,z)=g_{ab}-\hat{\tau}_a \hat{\tau}_b\,,
\eea
where $\hat{\tau}$ is the unit tangent vector, given in (\ref{tau}). A brief calculation leads to
\bea\label{dsperp1}
ds_\perp^2\equiv h_{ab}dx^a dx^b&=&\frac{1}{z^2}\left[\(1-\frac{z^2}{R_s^2}\)dr^2+\(1-\frac{(r-r_s)^2}{R_s^2}\)dz^2+\frac{2z(r-r_s)}{R_s^2}dz dr\right]\nonumber\\
&=&\frac{1}{z^2}\frac{1}{R_s^2}\left[(r-r_s)dr+z dz \right]^2\,.
\eea
Now, we can use the geodesic equation (\ref{geodesic}) to $(i)$ eliminate one of the variables, and $(ii)$ express (\ref{dsperp1}) as a differential along the transverse coordinate, namely $dr_m$. In practice, it is convenient to eliminate $z$ (in order to avoid multi-valuedness), and choose $r=\lambda$ as the affine parameter. We therefore solve for $z(r)$ in (\ref{geodesic}), which can be done analytically:
\bea\label{geodesic2}
z(r)=\sqrt{R_s^2-(r-r_s)^2}=\sqrt{\frac{2 R^2 r}{r_m}-R^2-r^2}\,.
\eea
On the other hand, from implicit differentiation of (\ref{geodesic}) we obtain
\bea
(r-r_s)dr+z dz=R_s dR_s +(r-r_s)dr_s=-\frac{r R^2}{r_m^2}dr_m\,,
\eea
where in the last equality we have used the explicit expressions for $R_s(r_m)$ and $r_s(r_m)$ given in (\ref{80}).
Putting all together, and restoring the angular dependence, we find that
\bea
ds_\perp^2=\frac{1}{z(r)^2}\(\frac{r^2R^2}{r_m^2(R^2-r_m^2)}dr_m^2+r^2d\Omega_{d-1}^2\) \,.
\eea
The magnitude of the vector field follows from equation (\ref{magnitudeV}), which leads to
\bea\label{magVsph}
|V|=\frac{\sqrt{h(r_m;r_m,z_m)}}{\sqrt{h(r_m;r,z)}}=\(\frac{r_m z}{z_m r}\)^{d}\,,
\eea
where $z_m=z(r_m)$. Finally, we would like to express our vector field as a function of $(r,z)$ without reference to the minimal surface. This can be achieved by solving for $r_m=r_m(r,z)$ from the geodesic (\ref{geodesic2}) and plugging it back into (\ref{magVsph}). A short calculation leads to
\bea
|V|=\(\frac{2Rz}{\sqrt{(R^2+r^2+z^2)^2-4R^2r^2}}\)^{d}\,.
\eea
It is easy to check that $|V|\leq1$ everywhere, an is only saturated at the position of the minimal surface $m(A)$. This confirms that our construction was successful.
Finally, we can plug the function $r_m(r,z)$ into the expression for $\hat{\tau}$ (\ref{tau}), to obtain
\bea
\htt^a =\frac{2Rz}{\sqrt{(R^2+r^2+z^2)^2-4R^2r^2}}\( \frac{r z}{R}\, ,\, \frac{R^2-r^2+z^2}{2R}\)
\eea
The full vector field $V=|V|\htt$ is then given by
\bea\label{Vd2}
V^a=\(\frac{2Rz}{\sqrt{(R^2+r^2+z^2)^2-4R^2r^2}}\)^{d+1}\(\frac{r z}{R}\, , \frac{R^2-r^2+z^2}{2R} \)\,.
\eea
In Figure \ref{Vspheres} we show plots of the vector lines as well as the magnitude of $V$ for $d=1$, $d=2$ and $d=3$ spatial dimensions.

\subsubsection{The strip \label{strips}}

Given the success achieved in the spherical case, it is natural to conjecture that geodesics would always be good candidates for the integral curves of an arbitrary entangling region. However, 
it is easy to check that geodesics do not always satisfy the minimal requirements outlined above. Moving beyond spherical symmetry, the simplest class of subsystems that can be studied analytically are strips. Indeed, it is possible to show that for strip geometries it is not possible to build a family of non-intersecting geodesics which are perpendicular to the minimal surface for dimensions other than $d=1$ and $d=2$. A proof by explicit construction is given in appendix \ref{A}. The goal of this section is to propose a family of integral curves that satisfy all the condition for strips in arbitrary dimensions.

\vspace{5mm}
\textbf{Minimal surface:}
The strip is defined by the collection of points with $-\ell/2\leq x_1\leq\ell/2$ and $-\ell_\perp/2\leq x_i\leq\ell_\perp/2$ ($\forall\,i>1$), with $\ell_\perp\to\infty$. The corresponding minimal surface can be obtained analytically by exploiting the translational symmetry of the problem, see e.g. \cite{Hubeny:2012ry}, and it is given by
\bea\label{minstrip}
\pm x_m(z_m)=\frac{\sqrt{\pi}}{ 2d} \frac{\Gamma(\frac 12 +\frac{1}{2d})}{\Gamma(1+\frac{1}{2d})}z_*-\frac{z_m}{d+1}\(\frac{z_m^d}{z_*^d}\)\,_2F_1\(\frac 12,\frac12+\frac1{2d},\frac 32+\frac1{2d},\frac{z_m^{2d}}{z^{2d}_*}\),
\eea
where $z_{*}$ is given in terms of the strip length by
\bea\label{zstar}
z_*=\frac{ d}{\sqrt{\pi}} \frac{\Gamma(1+\frac{1}{2d})}{\Gamma(\frac 12 +\frac{1}{2d})}\ell\,.
\eea
The two signs $\pm$ in (\ref{minstrip}) correspond to two different branches of $x_m(z_m)$; the `$+$' corresponds to the $x>0$ part of the surface, while the `$-$' corresponds to $x<0$. However, since there is a reflection symmetry around the origin, $x\to-x$, we will restrict our attention to the right branch, with `$+$' sign and $x>0$. We note that the minimal surface for the strip (\ref{minstrip}) is more complicated than the minimal surface for the sphere (\ref{minsphere}). In particular for general $d$ we cannot invert (\ref{minstrip}) to obtain $z_m=z_m(x_m)$. We will see below that this represents an obstacle for finding a closed expression for vector field $V(x,z)$.
Finally, we give the outward-pointing unit normal vector $\hn_m$ at a point $(x_m,z_m)$ on $m(A)$,
\bea\label{nastrip}
\hn_m^{a}=\frac{z_m}{z_*^d}\(\sqrt{z_*^{2 d}-z_m^{2 d}},z_m^d\)\,.
\eea
As a check, notice that if we set $d=1$ and $z_*=R$ we recover (\ref{nasphere}).

\vspace{5mm}
\textbf{Integral curves:}
The first step in our construction is to propose a family of integral curves that satisfy the consistency requirements outlined at the begging of section \ref{section2}. As mentioned above the geodesics are not a good choice for strips in general dimensions. A second natural proposal is to consider curves in the $(x,z)$ plane based on minimal surfaces for strips. This new choice is equally intuitive since these surfaces satisfy all the symmetry requirements and can be defined covariantly from the given bulk geometry. More specifically, our proposed family of integral curves will be given by the curves of the form
\bea\label{geodesicstrip}
x(z)=x_s \pm \alpha \,x_m\(\frac{z}{\alpha}\)\,,
\eea
where $x_m(z)$ is a minimal surface of the form (\ref{minstrip}), centered at $x=0$ and with a maximum depth $z_*$. It is clear $x_{s}$ corresponds to a shift along the $x$-direction, while $\alpha$ rescales the depth,
\bea
z_*\to \tilde{z}_*\equiv\alpha z_{*}\,.
\eea
The choice of sign $\pm$ corresponds to the two different branches of $x_m(z)$; the `$+$' corresponds to the $x>x_s$ part of the surface, while the `$-$' corresponds to $x<x_s$. It is important to keep both signs in this case in order to have a full coverage of the region $x>0$. It is easy to check that the above curves are geodesics in the following two-dimensional effective metric:\footnote{This two-dimensional effective metric has the form of a hyperscaling violating metric. Indeed, hyperscaling violation is known to arise generically from the dimensional reduction of higher dimensional AdS spaces \cite{Gouteraux:2011ce}.}
\bea\label{effmetricstrip}
ds^2=\frac{V_\perp^2}{z^{2d}}\(dx^2+dz^2\)\,,
\eea
where the extra $d$ in the denominator and the constant $V_\perp\equiv\ell_\perp^{d-1}$ arise from integrating out the transverse coordinates.

In section \ref{strips2} we will justify the above choice of integral curves in more detail. In particular, we will show that a foliation with minimal surfaces generally leads to a good family of integral curves for a strip entangling region in \emph{any} translational invariant bulk geometry. For the time being, we will proceed with the explicit construction of the vector $V$ in the particular case in consideration, i.e., empty AdS.

The next step is to impose the orthogonality of the integral curves with respect to $m(A)$. In order to do so we first compute the unit tangent vector along the curves,
\bea\label{taustrip}
\hat{\tau}^a=\frac{z}{\tilde{z}_*^d}\(z^d,\mp\sqrt{\tilde{z}_*^{2 d}-z^{2 d}}\)\,.
\eea
Notice that both (\ref{taustrip}) and (\ref{nastrip}) are normalized with respect to the original AdS metric.
The `$-$' here correspond to the $x>x_s$ branch, while the `$+$' corresponds to $x<x_s$. It is clear that we need the latter sign to enforce orthogonality, since we are focusing on the $x>0$ portion of $m(A)$ and the outward-pointing normal vector $\hn_m$ for this branch (\ref{nastrip}) has this sign. Enforcing that $\hat{\tau}=\hn_m$ at a point $(x_m,z_m)$ on the minimal surface leads to
\bea\label{alpha}
\tilde{z}_*=\frac{z_*z_m}{(z_*^{2d}-z_m^{2d})^{\frac{1}{2 d}}}\,,\qquad x_s=x_m(z_m)+\frac{z_m}{(z_*^{2d}-z_m^{2d})^{\frac{1}{2 d}}} x_m\left((z_*^{2d}-z_m^{2d})^{\frac{1}{2 d}}\right)\,.
\eea
Notice that if we set $d=1$, $z_*=R$, $\tilde{z}_*=R_s$ and $x_s=r_s$ we recover the expressions in (\ref{80}).
Finally, plugging (\ref{alpha}) into  (\ref{geodesicstrip}) we can obtain an explicit expression for the family of geodesics orthogonal to $m(A)$, parametrized by the point $z_m$ on the minimal surface:
\bea\label{xzstrip}
x(z_m;z)=x_m(z_m)+\frac{z_m}{(z_*^{2d}-z_m^{2d})^{\frac{1}{2 d}}}\[ x_m\left((z_*^{2d}-z_m^{2d})^{\frac{1}{2 d}}\right)\pm  x_m\left(\frac{(z_*^{2d}-z_m^{2d})^{\frac{1}{2 d}}z}{z_m}\right)\]\,,
\eea
where $x_m(z)$ is the function given in (\ref{minstrip}) (with `$+$' sign) and $z \in(0,\tzx)$. Again, we have two branches depending on the sign of the last term in (\ref{xzstrip}). The `$-$' branch intersects $A$ at
\bea\label{nested-1}
x_{a}=x_m(z_m)+\frac{z_m}{(z_*^{2d}-z_m^{2d})^{\frac{1}{2 d}}}\[ x_m\left((z_*^{2d}-z_m^{2d})^{\frac{1}{2 d}}\right)-\frac{\ell}{2}\]\,,
\eea
while the `$+$' branch intersects $\bar{A}$ at
\bea\label{nested-2}
x_{\bar{a}}=x_m(z_m)+\frac{z_m}{(z_*^{2d}-z_m^{2d})^{\frac{1}{2 d}}}\[ x_m\left((z_*^{2d}-z_m^{2d})^{\frac{1}{2 d}}\right)+\frac{\ell}{2}\]\,.
\eea
With these expressions we can now check if the curves parametrized by $z_m$ are properly nested.
Since the curves are minimal surfaces, this condition is guaranteed provided that $dx_a/dz_m < 0$ and
$dx_{\bar{a}}/dz_m>0$. Indeed, after some algebra we find that
\bea
\frac{dx_{a}}{d\zm}=-\frac{z_*^{2d}}{(z_*^{2d}-z_m^{2d})^{1+\frac{1}{2d}}}\left[ \frac{\ell}{2}-x_m\((z_*^{2d}-z_m^{2d})^{\frac{1}{2 d}}\)\right]<0\,,\eea
and
\bea
\frac{dx_{\bar{a}}}{d\zm}=\frac{z_*^{2d}}{(z_*^{2d}-z_m^{2d})^{1+\frac{1}{2d}}}\left[ \frac{\ell}{2}+x_m\((z_*^{2d}-z_m^{2d})^{\frac{1}{2 d}}\)\right]>0\,,
\eea
which justify the choice of minimal surfaces as good integral curves for the case of the strip.

\begin{figure}
\centering
 \includegraphics[width=2.8in
 ]{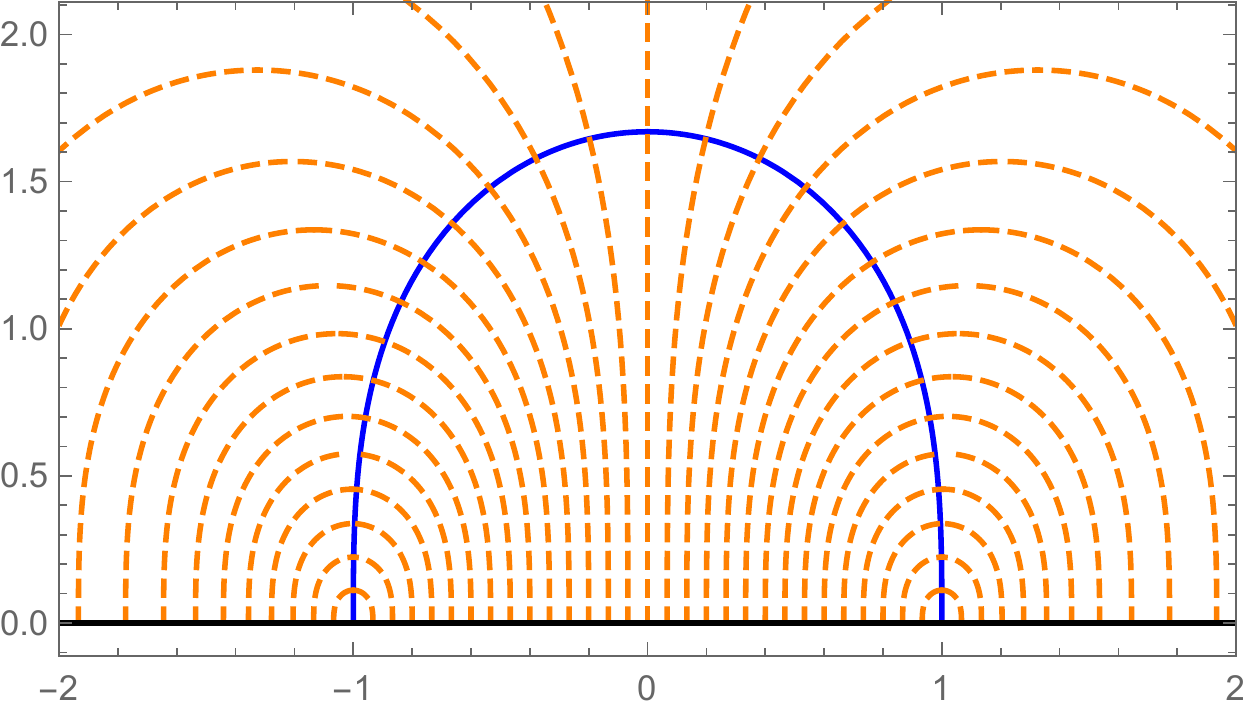}
  \hspace{0.2cm}\includegraphics[width=3in
 ]{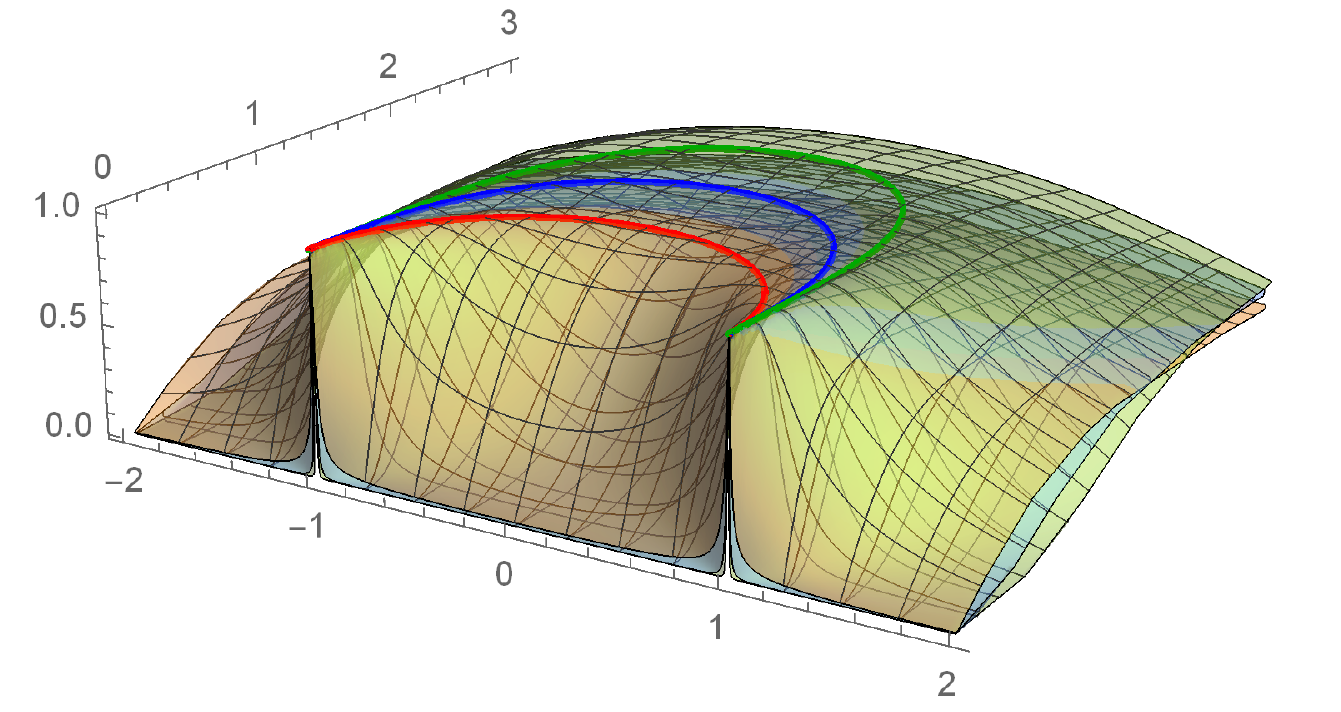}
  \begin{picture}(0,0)
\put(-173,105){{\small $z$}}
\put(-228,62){{\small $|V|$}}
\put(-143,12){{\small $x$}}
\put(-438,62){{\small $z$}}
\put(-330,-5){{\small $x$}}
\end{picture}
 \caption{Vector lines for a strip in $d=2$ spatial dimensions (to be compared with the $d=1$ case, shown in Figure \ref{Vspheres}) and magnitude $|V|$ for $d=1$ (orange), $d=2$ (blue) and $d=3$ (green), respectively. The vector fields $V$ exhibit translational invariance along the transverse directions, which have been omitted for simplicity. The solid line(s) corresponds to the minimal surface(s), $m(A)$. These curves also signal the location where the magnitude of the vector fields attains its maximal value, $|V|=1$. In general, we observe that the integral curves get elongated and reach deeper into the bulk as the number of dimensions is increased. This property is inherited from their corresponding minimal surfaces.}\label{Vstripes}
\end{figure}
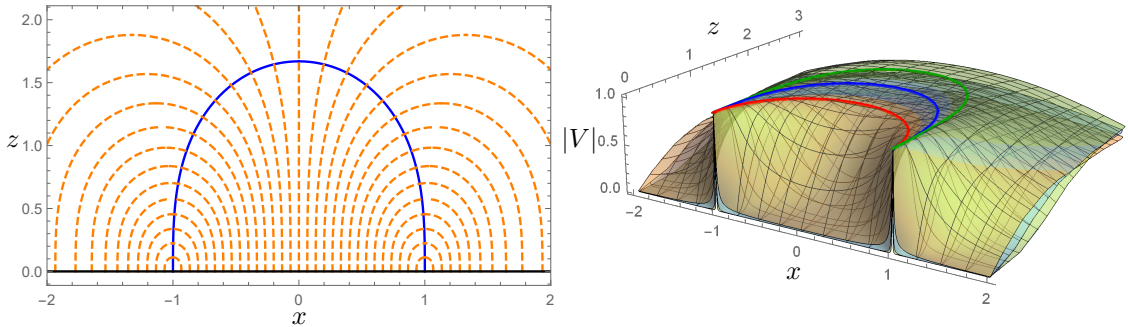

\vspace{5mm}
\textbf{Vector field:}
We can now proceed to find the appropriate norm of the vector field $|V|$, through equation (\ref{magnitudeV}). In order to do so,
we first compute the orthogonal metric at different points along the curve parametrized by $z_m$,
\bea\label{habstrip}
h_{ab}(z_m;x,z)=g_{ab}-\hat{\tau}_a\hat{\tau}_b\,,
\eea
where $\hat{\tau}$ is the unit tangent vector, given in (\ref{taustrip}).
After some algebra we can put the orthogonal metric in the following form:
\bea
ds_\perp^2\equiv h_{ab}dx^a dx^b=\frac{1}{z^2}\frac{1}{\tzx^{2d}}\(\sqrt{\tzx^{2d}-z^{2d} }dx \pm z^{d}dz \)^2\,,
\eea
where $\tzx$ is given in (\ref{alpha}).  This line element can be rewritten in terms of the transverse element $dz_m$ by implicitly differentiating (\ref{xzstrip}), which yields
\bea
\frac{1}{\tzx^{2d}}\(\sqrt{\tzx^{2d}-z^{2d} }dx\pm z^{d}dz\)^2
=\(\frac{\sqrt{\tzx^{2d}-z^{2d} }}{\tzx^{d}} \partial_{\zm} x(z_m;z)\)^2d \zm^2\,.
\eea
Putting all together, and restoring the transverse coordinates, we find that
\bea
ds_\perp^2=\frac{1}{z^2}\left[\(\frac{\sqrt{\tzx^{2d}-z^{2d} }}{\tzx^{d}} \partial_{\zm} x(z_m;z)\)^2d\zm^2 +d\vec{x}_{\perp}^2\right]\,.
\eea
The magnitude of the vector field follows from equation (\ref{magnitudeV}), which leads to\footnote{We note that the same result can be obtained by performing the same analysis starting directly with the effective metric (\ref{effmetricstrip}). This includes: normalizing the vectors $\hat{n}$ and $\hat{\tau}$ with respect to this metric and using it as a background metric in (\ref{habstrip}). We will come back to this point in  section \ref{strips2}.}
\bea\label{Vstrip}
|V|=\left|\(\frac{z}{\zm}\)^d \sqrt{\frac{\tzx^{2d}-\zm^{2d}}{\tzx^{2d}-z^{2d}} }\frac{\( \partial_{\zm}x_{-}(z_m;z) \)|_{z=\zm}}{\!\!\!\!\!\!\!\!\!\!\!\!\!\!\( \partial_{\zm}x_{\pm}(z_m;z) \)}\right|\,.
\eea
The subscript in $x(z_m;z)$ refers to the choice of sign in (\ref{xzstrip}). In particular, it is important that $x_{-}(z_m;z)$ appears in the numerator of (\ref{Vstrip}) since this is the branch that intersects the minimal surface $m(A)$. The two signs of $x_{\pm}(z_m;z)$ in the denominator of (\ref{Vstrip}) cover the regions $x>x_s$ and $x<x_s$, respectively, and together they span the whole axis $x\in\mathbf{R}^{+}$. The explicit expressions for $|V|$ as a function of $z$ and $z_m$ are straightforward to obtain but are lengthy and not particularly illuminating, so we will
refrain from writing out these results here. The final step would be to solve explicitly for $z_m(x,z)$ from the geodesic equation (\ref{xzstrip}) and replace the result in equation (\ref{Vstrip}) to obtain $|V|$ purely in terms of $x$ and $z$. As mentioned above, this is not possible to do analytically for $d>1$. However, we can perform a parametric plot by varying $z$ and $z_m$. The final result is plotted in figure \ref{Vstripes} for strips in various dimensions. From these plots we can easily check that the magnitude of the vector is indeed bounded, $|V|\leq1$, and is saturated precisely at the location of the minimal surface.

\subsection{Explicit constructions in Schwarzschild-AdS\label{sec2btz}}

\subsubsection{The strip \label{strips}}

In terms of minimal area surfaces, going beyond pure AdS is challenging and often require the use of numerical techniques. Of particular physical relevance
is the case of minimal area surfaces in an AdS black hole geometry, from which one can compute entanglement entropy in an excited state at finite temperature. Perhaps the only non-trivial example that can be treated analytically is the case of strips in a black brane geometry which was recently explored in \cite{Erdmenger:2017pfh}. In $d=1$ the problem simplifies drastically. On one hand, the codimension-two surfaces required to compute entanglement entropies are space-like geodesics. On the other hand, Schwarzschild-AdS$_3$ (the BTZ black hole) is diffeomorphic to AdS$_3$ so the geodesics can be obtained in a compact form. In the following we will specialize for simplicity to the $d=1$ case. However, the generalization to higher dimensional cases is straightforward, since the minimal surfaces are known analytically for any $d$ \cite{Erdmenger:2017pfh}. We expect that the general method and the qualitative results will apply also to these cases.

\vspace{5mm}
\textbf{Minimal surface:}
Consider the minimal surface associated to an interval of length $\ell$ in (a time slice of) a BTZ black hole,
\bea\label{BTZmetric}
ds^2=\frac{1}{z^2}\(dx^2+\frac{dz^2}{f(z)}\)\,,\qquad\qquad f(z)=1-\frac{z^2}{z_h^2}\qquad\text{with}\qquad z_h=\frac{\beta}{2\pi}\,.
\eea
The metric is invariant under translations in $x$, so without loss of generality we focus on an interval centered at the origin. The minimal surface is given by the collection of points $(x_m,z_m)$ that satisfy
\bea\label{geoBTZ}
z_m=\sqrt{\frac{z_*^2+z_h^2}{2}-\frac{z_h^2-z_*^2}{2} \cosh\left(\frac{2x_m}{z_h}\right)}\,,
\eea
or, equivalently,
\bea\label{geoBTZ2}
x_m=\pm z_h \log\left(\frac{\sqrt{z_h^2-z_m^2}+\sqrt{z_*^2-z_m^2}}{\sqrt{z_h^2-z_*^2}}\right)\,.
\eea
In either of these expressions, $z_*$ indicates the maximum depth of the geodesic. This parameter is related to the length of the interval $\ell$ through any of the following equivalent relations:
\bea\label{ellBTZ}
\ell=z_h\,\text{arccosh}\left(\frac{z_h^2+z_*^2}{z_h^2-z_*^2}\right)=z_h \log\left(\frac{z_h+z_*}{z_h-z_*}\right) =2z_h\, \text{arctanh}\left(\frac{z_*}{z_h}\right)\,.
\eea
The  outward-pointing normal unit vector $\hn_m$ at the point $(x_m,z_m)$ is given by
\bea\label{normalBTZ}
\hn_m=\frac{\left\{z_h z_m \left(z_h^2-z_*^2\right)\sinh\left(\frac{2 x_m}{z_h}\right),2 z_m^2 \left(z_h^2-z_m^2\right)\right\}}{z_h \sqrt{4 z_m^2 (z_h^2 - z_m^2) + (z_h^2 - z_*^2)^2\sinh^2\left(\frac{2 x_m}{z_h}\right)}}.
\eea
For future reference we also give an explicit expression for the slope $s_m=n_m^z/n_m^x$ of the normal vector, at a given point of the minimal surface
\bea\label{slopeBTZ}
s_m=\frac{2 z_m \left(z_h^2-z_m^2\right)\text{csch}\,\left(\frac{2 x_m}{z_h}\right)}{z_h \left(z_h^2-z_*^2\right)}=\frac{z_m}{z_h} \coth\left(\frac{x_m}{z_h}\right)=\pm\frac{z_m}{z_h}\sqrt{\frac{z_h^2-z_m^2}{z_*^2-z_m^2}}\,.
\eea

\vspace{5mm}
\textbf{Integral curves:}
Now lets consider the space of geodesics that lie on the $(x,z)$ plane that intersect $m(A)$ at $(x_m,z_m)$ and whose tangent vector is parallel to the normal $\hn_m$ at that point. These geodesic will represent the integral curves of our vector field $V$.

As we will see below, for the case of the BTZ black hole, some integral curves will \emph{necessarily} end up at the horizon, while others will go back to the boundary. For this reason it will be convenient to express the geodesics in two different parametrizations, as $z(x)$ and as $x(z)$. The former parametrization will be useful to describe threads that start and end at the boundary, while the latter will be more convenient for threads with one end at the horizon.
\begin{itemize}
  \item \underline{Parametrization I}:
  A general geodesic that goes through the point $(x_0,z_0)$ with a slope $dz/dx=s_0$ can be written as follows:
    \bea\label{fluxparaI}
  z(x)=\sqrt{\mathcal{C}_1+\mathcal{C}_2 \cosh\left[\frac{2 (x-x_0)}{z_h}\right]+\mathcal{C}_3\sinh\left[\frac{2 (x-x_0)}{z_h}\right]}
  \eea
  where the constants $\mathcal{C}_i$ are given by
  \bea\nonumber
  \mathcal{C}_1=\frac{(z_h^2+z_0^2)(z_h^2-z_0^2)+s_0^2 z_0^2 z_h^2}{2(z_h^2-z_0^2)}\,,\qquad \mathcal{C}_2=-\frac{(z_h^2-z_0^2)^2+s_0^2 z_0^2 z_h^2}{2(z_h^2-z_0^2)}\,,\qquad \mathcal{C}_3=s_0 z_0 z_h\,.
  \eea
  As a simple check, notice that if we set $x_0=0$, $z_0=z_*$ and $s_0=0$, we recover the minimal surface (\ref{geoBTZ}). Enforcing that the tangent to the geodesic  is equal to the normal of the minimal surface (\ref{normalBTZ}) at the point $(x_m,z_m)$ we arrive to the corresponding family of integral curves. These can be obtained simply by setting $x_0=x_m$, $z_0=z_m$ and $s_0=s_m(x_m,z_m)$ in (\ref{fluxparaI}), with $s_m(x_m,z_m)$ given in (\ref{slopeBTZ}). Furthermore, using the equation for minimal surface one can eliminate either $x_m$ or $z_m$ and write the final result as $z(x_m;x)$ or $z(z_m;x)$.
  \item \underline{Parametrization II}:
  In the second parametrization, a general geodesic that goes through the point $(x_0,z_0)$ with a slope $dx/dz=1/s_0$ can be written as follows:
  \bea\label{fluxparaII}
  x(z)=x_0+\sigma z_h\log\left[\frac{\sqrt{z_h^2-z^2}+\sqrt{\mathcal{D}^2+z_0^2-z^2}}{\sqrt{z_h^2-z_0^2}-\sigma \mathcal{D}}\right],
  \eea
  where $\mathcal{D}$ and $\sigma$ are the following constants:
  \bea
  \mathcal{D}=\frac{s_0 z_0 z_h}{\sqrt{z_h^2-z_0^2}}\,,\qquad \sigma=\pm1\,.
  \eea
  The solution has two branches, depending on the sign of $\sigma$. For $s_0>0$ the correct solution has $\sigma=-1$, while for $s_0<0$ the right choice is $\sigma=1$. As a check, notice that if we set $x_0=0$, $z_0=z_*$ and $s_0=0$, we recover the minimal surface (\ref{geoBTZ2}). Enforcing that the tangent to the geodesic is equal to the normal of the minimal surface (\ref{normalBTZ}) at the point $(x_m,z_m)$ we arrive to the corresponding family of integral curves. These can be obtained simply by setting $x_0=x_m$, $z_0=z_m$ and $s_0=s_m(x_m,z_m)$ in (\ref{fluxparaI}), with $s_m(x_m,z_m)$ given in (\ref{slopeBTZ}). Finally, using the equation for minimal surface one can eliminate either $x_m$ or $z_m$ and write the final result as $x(x_m;z)$ or $x(z_m;z)$.
\end{itemize}
A comment about the parametrizations is in order. As mentioned above, the parametrization II is more convenient for geodesics that reach the horizon. However, if the geodesic has the two endpoints at the boundary one can still use the two branches in equation (\ref{fluxparaII}) to fully describe them. Notice that the fact that we need two solutions is to be expected because $x(z)$ is double valued for this type of geodesics. The extra branch in each case describes the analytic continuation of the geodesic after the maximum depth is reached which, by symmetry, is located at the center of the interval. Moreover, at this point one has that $dz/dx=0$ (or, equivalently, $dx/dz\to\infty$). Similarly, for geodesics that reach the horizon one can alternatively use equation (\ref{fluxparaI}) from the parametrization I to describe them, but in that case one would need to truncate the solution at $z=z_h$. Beyond this point the solution would be unphysical.

Before proceeding further, let us investigate the exact conditions that must be satisfied in order to have threads connecting the boundary and the horizon. Assuming that $s_0\neq0$, from our solution (\ref{fluxparaII}) we obtain that the point at which $dx/dz\to\infty$ is given by:
\bea
z_{\text{max}}=z_0\sqrt{\frac{z_h^2(1+s_0^2)-z_0^2}{z_h^2-z_0^2}}\,.
\eea
For $z_{\text{max}}<z_h$ the curve has its two endpoints at the boundary, but for $z_{\text{max}}\geq z_h$ one of its endpoints inevitably reaches the horizon. The transition takes place exactly when $z_{\text{max}}=z_h$, which leads to a simple relation between the maximum slope and the radial depth $z_0$,
\bea
s_{\text{max}}=\frac{z_h^2-z_0^2}{z_h z_0}\,.
\eea
For $|s_0|<s_{\text{max}}$ the curve has its two endpoints at the boundary, however, for steep enough slopes $|s_0|>s_{\text{max}}$ the curve will necessarily have one of its endpoints at the horizon.

\begin{figure}[t!]
\centering
 \includegraphics[width=2.8in
 ]{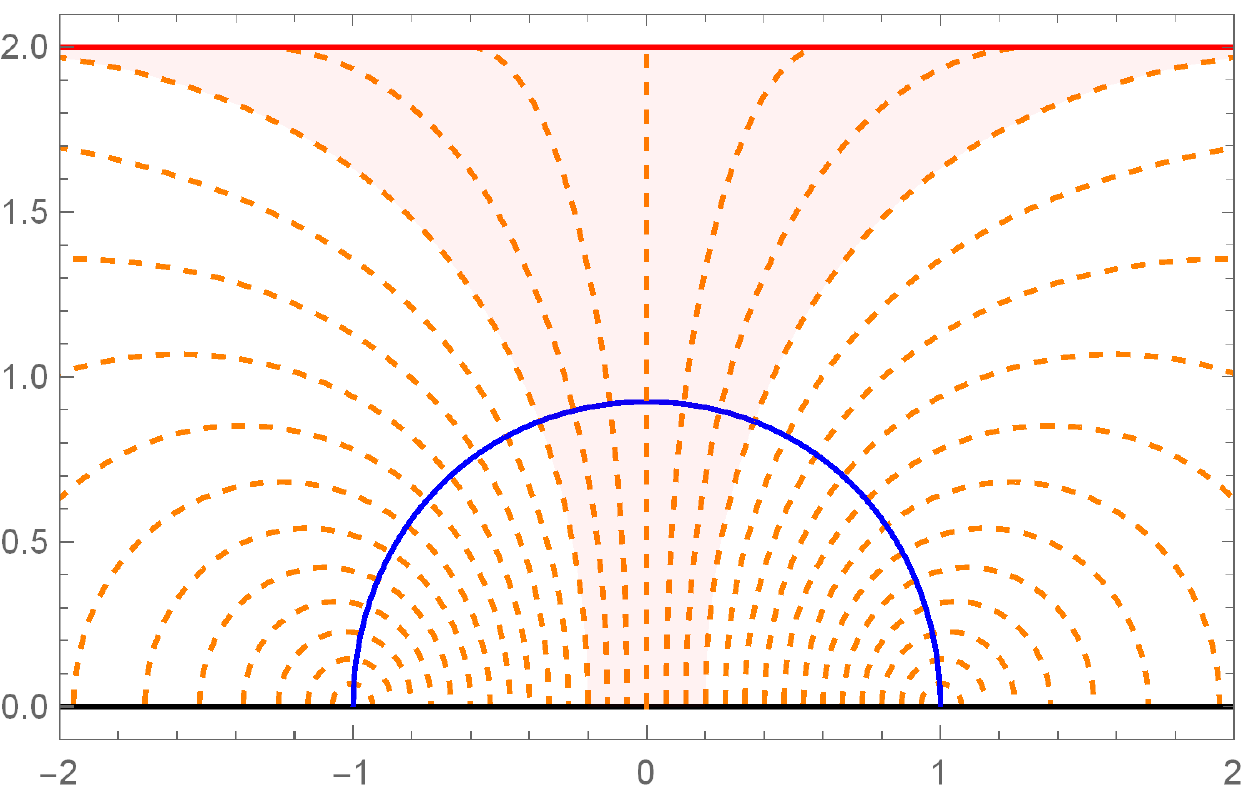}
  \hspace{0.2cm}\includegraphics[width=3in
 ]{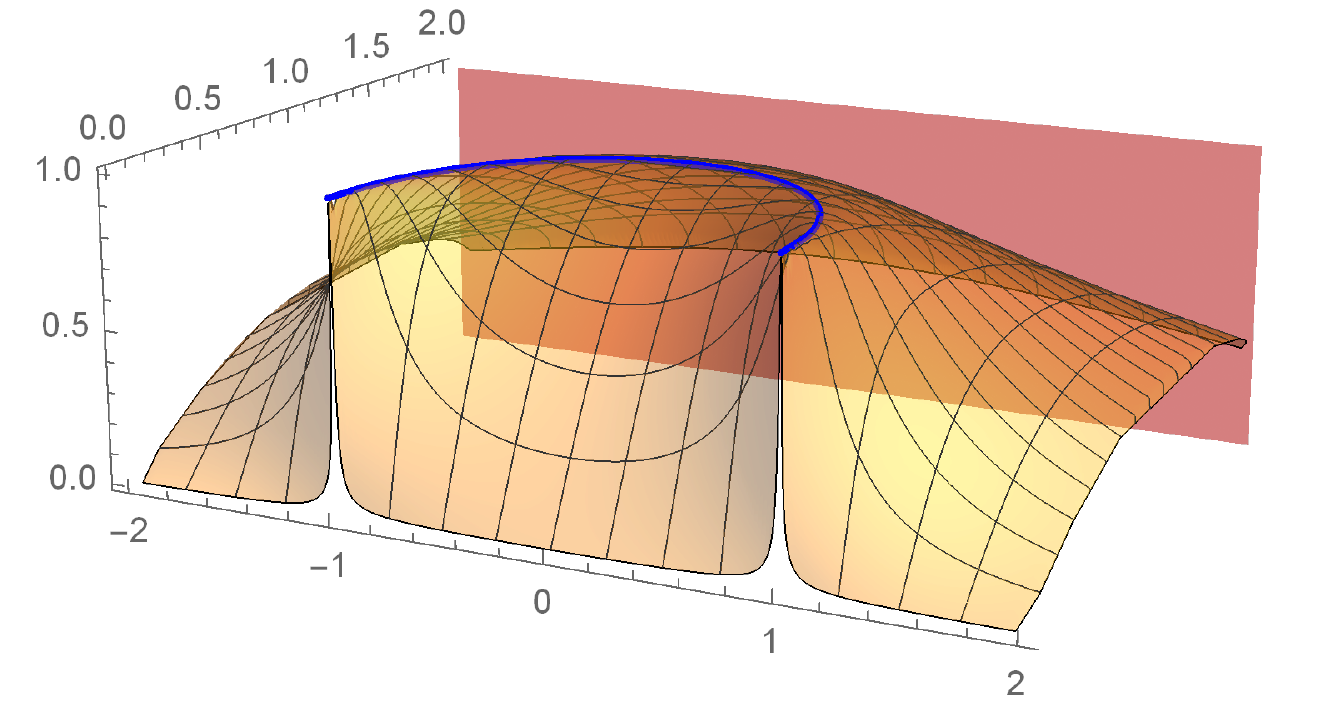}
  \begin{picture}(0,0)
\put(-177,111){{\small $z$}}
\put(-228,60){{\small $|V|$}}
\put(-135,8){{\small $x$}}
\put(-439,66){{\small $z$}}
\put(-331,-5){{\small $x$}}
\end{picture}
 \caption{Typical integral curves and magnitude $|V|$ in a BTZ black hole geometry. The solid lines in blue correspond to the minimal surface, $m(A)$, and precisely at this location the magnitude of the vector field attains its maximal value, $|V|=1$. The vector lines in the shaded region correspond to threads that end at the horizon, while the the ones in the white area correspond to threads that go back to the boundary.\label{BTZlines}}
\end{figure}
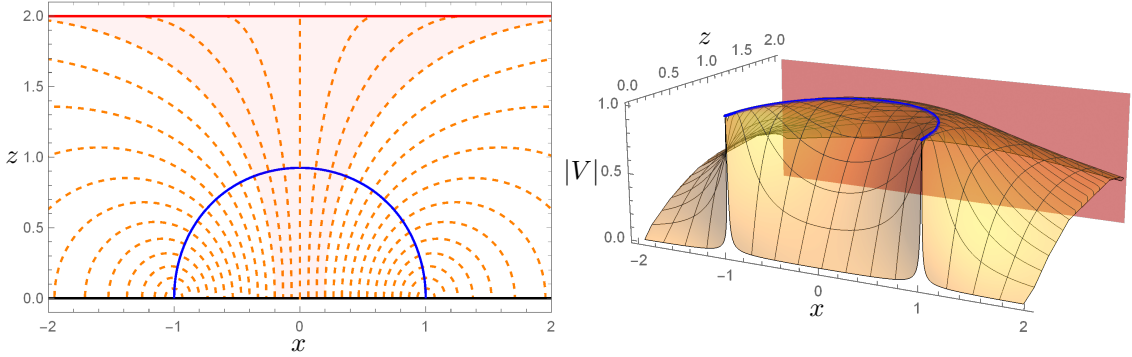

The above bounds hold for \emph{any} constant-$t$ geodesic in the BTZ background (\ref{BTZmetric}). On the other hand, using explicitly the information about the minimal surface (\ref{geoBTZ}) and the equation for the normal slope (\ref{slopeBTZ}) we can arrive to the following bounds for threads of our vector field $V$:
\bea\label{zmax}
z_\text{max}=\frac{z_h z_*}{\sqrt{z_h^2+z_*^2}}\,,
\eea
which translates into
\bea\label{xmax}
x_\text{max}=z_h\, \text{arcsech}\left(\sqrt{1-z_*^4/z_h^4}\right)=z_h\, \text{arccoth}\left(\frac{z_h^2}{z_*^2}\right)\,.
\eea
For points of the minimal surface $m(A)$ close to the center $|x|<x_\text{max}$ ($z>z_\text{max}$) the normal vector is steep enough so that $|s_m|>s_{\text{max}}$, hence the corresponding threads reach the horizon. Conversely, points of $m(A)$ near the edges, with $\frac{\ell}{2}>|x|>x_\text{max}$ ($z<z_\text{max}$), have threads that connect the region $A$ with its complement $\bar{A}$. This phenomenon is illustrated in figure \ref{BTZlines}.

Finally, in order to see if these geodesics are indeed good candidates for integral curves we need to check that they are properly nested. There is a subtlety, however, since some of the threads end at the horizon. Without loss of generality, we focus on the region $x_m>0$ and split the analysis in two, $0<z_m<z_{\text{max}}$ and $z_{\text{max}}<z_m<z_*$:
\begin{itemize}
  \item \underline{$0<z_m<z_{\text{max}}$}: In this range the threads connect a point of region $A$,
  \bea\label{xabtz}
  x_a(z_m)=x_{-}(z_m;z=0)\,,
  \eea
  with a point in the complementary region $\bar{A}$,
  \bea
  x_{\bar{a}}(z_m)=x_{+}(z_m;z=0)\,.
  \eea
  The subscript in $x(z_m;z)$ in these expressions refers to the choice of $\sigma$ or, equivalently, the branch of the geodesic. By the property of entanglement wedge nesting, we know that these integral curves are properly nested if and only if $dx_a(z_m)/dz_m<0$ and $dx_{\bar{a}}(z_m)/dz_m>0$. Indeed, a brief calculation leads to:
  \bea\label{dxadzm}
  \frac{dx_a(z_m)}{dz_m}= -\frac{z_h^2 z_*^2}{\left(z_* \sqrt{z_h^2-z_m^2}+z_h z_m\right) \sqrt{\left(z_h^2-z_m^2\right) \left(z_*^2-z_m^2\right)}}<0\,,
  \eea
  and
  \bea
  \!\!\frac{dx_{\bar{a}}(z_m)}{dz_m}&=&\frac{z_h z_m z_*^2 \text{csch}\left(\frac{x_m}{z_h}\right)}{\sqrt{\left(z_h^2-z_m^2\right) \left(z_*^2-z_m^2\right)}}\times\nonumber\\
  &&\left[\frac{\text{csch}\left(\frac{x_m}{z_h}\right)}{z_h^2-z_*^2 \coth \left(\frac{x_m}{z_h}\right)}+\frac{z_* \coth \left(\frac{x_m}{z_h}\right)}{z_m^2 z_* \text{csch}\left(\frac{x_m}{z_h}\right)+z_h z_m \sqrt{z_h^2-z_*^2}}\right]>0\,.
  \eea
  The only term that is not explicitly positive definite is the denominator of the first term in the square brackets, however, it is easy to check from (\ref{xmax}) that if $x_m>x_{\text{max}}$ then $z_h^2>z_*^2 \coth(x_m/z_h)$. The above two inequalities then show that the corresponding integral curves are nested.
  \item \underline{$z_{\text{max}}<z_m<z_*$}: In this range the threads connect a point of region $A$, $x_a(z_m)$ given in (\ref{xabtz}), with a point at the horizon,
  \bea
  x_{h}(z_m)=x_{-}(z_m;z=z_h)\,.
  \eea
  Besides the inequality (\ref{dxadzm}) it is possible to show that
  \bea\label{dxhdzm}
  \frac{dx_h(z_m)}{dz_m}= -\frac{z_h^3 z_*^2 z_m}{\sqrt{\left(z_h^2-z_m^2\right) \left(z_*^2-z_m^2\right)} \left[z_m^2 \left(z_h^2+z_*^2\right)-z_h^2 z_*^2\right]}<0\,.
  \eea
  All terms in this expression are strictly positive, except for the piece of the denominator in the square brackets. However from (\ref{zmax}) one can check that if $z_m>z_{\text{max}}$ then $z_m^2 (z_h^2+z_*^2)>z_h^2 z_*^2$. In order to prove the nesting of the integral curves we point out that the excised geometry outside the bulk horizon has been conjectured to be dual to a pure state with extra degrees of freedom living at the (stretched) horizon \cite{Takayanagi:2017knl,Nguyen:2017yqw}.\footnote{This equivalence was established in \cite{Takayanagi:2017knl,Nguyen:2017yqw} using intuition from tensor networks.} In this setup, geodesics that end at the horizon can indeed be interpreted as entanglement entropies in the purified state \cite{Espindola:2018ozt}, so from the property of entanglement wedge nesting, we can conclude that (\ref{dxadzm}) and (\ref{dxhdzm}) are indeed sufficient to prove the nesting of the corresponding integral curves.
\end{itemize}

\vspace{5mm}
\textbf{Vector field:}
We can now proceed to find the appropriate norm of the vector field $|V|$, through equation (\ref{magnitudeV}). To do so,
we find it convenient to work with the second parametrization, so we label the geodesics as $x(z_m;z)$. The orthogonal metric at different points along one of these geodesics is
\bea\label{habbtz}
h_{ab}(z_m;x,z)=g_{ab}-\hat{\tau}_a\hat{\tau}_b\,,
\eea
where $\hat{\tau}$ is the corresponding unit tangent vector. In this parametrization,
this is given by
\bea\label{taubtz}
\hat{\tau}^a=\frac{z}{\tilde{z}_*}\left(z,\mp\sqrt{f(z)(\tilde{z}_*^2-z^2)}\right)\,,\qquad \tilde{z}_*\equiv\frac{z_* z_m}{\sqrt{z_*^2-z_m^2}}\,.
\eea
Since $\hat{\tau}^{z}=0$ precisely at $z=\tilde{z}_*$, this means that $\tilde{z}_*$ gives the maximum depth of the geodesic labeled by $z_m$. Plugging (\ref{taubtz}) into (\ref{habbtz}) we arrive to
\bea
ds_\perp^2\equiv h_{ab}dx^a dx^b=\frac{1}{z^2}\frac{1}{\tzx^{2}f(z)}\(\sqrt{f(z)(\tzx^{2}-z^{2}) }dx \pm z dz \)^2\,.
\eea
This line element can be rewritten in terms of the transverse element $dz_m$ by implicitly differentiating the geodesic $x(z_m;z)$, which yields
\bea
ds_\perp^2=\frac{1}{z^2}\(\frac{\sqrt{\tzx^{2}-z^{2} }}{\tzx^{}} \partial_{\zm} x(z_m;z)\)^2d \zm^2\,.
\eea
The magnitude of the vector field can be read from equation (\ref{magnitudeV}), which leads to
\bea\label{Vbtz}
|V|=\left|\frac{z}{\zm}\sqrt{\frac{\tzx^{2}-\zm^{2}}{\tzx^{2}-z^{2}} }\frac{\( \partial_{\zm}x_{-}(z_m;z) \)|_{z=\zm}}{\!\!\!\!\!\!\!\!\!\!\!\!\!\!\( \partial_{\zm}x_{\pm}(z_m;z) \)}\right|\,.
\eea
Again, the subscript in $x(z_m;z)$ refers to the choice of $\sigma$ or the branch of the geodesic. In particular, it is important that $x_{-}(z_m;z)$ appears in the numerator of (\ref{Vbtz}) since this is the branch that intersects the minimal surface $m(A)$. The two signs of $x_{\pm}(z_m;z)$ in the denominator of (\ref{Vbtz}) cover the full axis $x\in\mathbf{R}^{+}$. The explicit expression for $|V|$ as a function of $z_m$ and $z$ is straightforward to obtain but is lengthy and not particularly illuminating, so we will not write the final result here. The final step would be to invert the geodesic equation $x(z_m;z)$ to obtain $z_m(x,z)$, and replace the result in equation (\ref{Vbtz}) to obtain $|V|$ purely in terms of $x$ and $z$. However, this is not possible to do for the case in consideration. Here we proceed as we did for the strip geometry in higher dimensions, i.e., by varying $z$ and $z_m$ and performing a parametric plot. The final result is shown in figure \ref{BTZlines}. From these plots we can easily check that the magnitude of the vector is indeed bounded, $|V|\leq1$, and is saturated precisely at the location of the minimal surface.

\section{Constraints on geometry and matter\label{sec:foli}}
In the previous section, we provided a general algorithm to construct bounded, divergenceless vector fields, with maximum flux through a given boundary region, for a given bulk geometry. With this algorithm studied specific examples of vector fields $V$ representing holographic bit threads, i.e., satisfying $\nabla_\mu V^\mu=0$ and  $|V|\leq 1$, for some particular cases of interest: spheres and strips on empty AdS in arbitrary dimensions and strips on an AdS black brane geometry. The purpose of this section is to relax some of the conditions used there and investigate the possibility of implementing a similar construction in more general bulk geometries.

Let us first summarize the main ingredients of our constructions:
 \begin{itemize}
   \item  Perhaps the most essential feature of our examples was the high degree of symmetry in both, the geometry of the problem (including the bulk geometry and the shape of the entangling region $A$) and the vector field $V$. Indeed, assuming that the symmetries of the problem are inherited by the vector field $V$, our analysis became effectively two dimensional. For spheres, we studied vector field configurations by focusing on a given plane and assuming that $V$ had no angular components nor angular dependence. Similarly, for strips we assumed that $V$ had no components along the tangent directions so we were also able to reduce our analysis to a fixed plane.
   \item Another important ingredient was our canonical choice of integral curves. In all cases, the integral curves
were taken to be geodesics of a suitable auxiliary geometry. For spheres such curves were given by the geodesics of the global metric, while for the case of strips they were given by geodesics of a 2-dimensional effective metric. Once the choice of integral curves was made, the final step was to obtain the norm $|V|$ from the continuity equation (\ref{magnitudeV}). From this equation it was clear that $|V|\leq 1$, if and only if the area transverse satisfies $ \sqrt{h(x,\lambda)} \geq \sqrt{h(x,\lambda_m)}$ everywhere away from the minimal surface. We emphasize that this condition was verified only \emph{a posteriori}, but was not explicitly imposed in our constructions.
 \end{itemize}

In this section we will investigate more generally when the bound $|V|\leq 1$ is satisfied for this kind of constructions. More specifically, we will consider the same class of foliations in a generic bulk geometry and find constraints on the curvature in order to satisfy the aforementioned bound. We do so by studying the transverse area of the hypersurfaces orthogonal to the given foliations and finding the requirements that must be satisfied such that the transverse area element increases as one moves away from the minimal surface. In addition, we will use the full Einstein equations to translate these constraints in terms of the stress-energy tensor of the bulk matter fields. On general grounds, we expect that such constraints must hold for (semi-)classical gravity in the large $N$ limit ($G_N\to0$), but should be modified once quantum corrections in the bulk are taken into account.

\subsection{General geodesic foliations}
Lets consider a manifold $\Sigma$ with boundary $\partial \Sigma$ endowed with a Riemannian metric $g_{ab}$, having the property that given a minimal co-dimension one bulk surface $m(A)$ anchored at the boundary $\partial \Sigma$, this is $m(A)\Big|_{\partial \Sigma}=\partial A$ such that $A \subset \partial \Sigma$ is a connected boundary region, there exists a global or partial foliation of the manifold by geodesics that intersect orthogonally the surface $m(A)$.\footnote{The foliation does not necessarily have to cover the full manifold, only a neighboring region near the minimal surface $m(A)$. For portions of the manifold that are not covered by the foliation we can either impose that $V=0$ or appropriately glue a portion of another vector field. Examples of geometries that do not admit a global covering include spacetimes with cosmological horizons \cite{Engelhardt:2013jda,Fischler:2013fba} and more general surface barriers \cite{Engelhardt:2013tra}.}  If such a foliation exists, then, it is possible to prove a theorem for the area of the family of hypersurfaces that are orthogonal to the foliation. 

For the class of foliations that we consider, the most general form of the metric is the following:
\bea\label{metricl}
ds^2=d\lambda^2+\g_{ij}(x,\lambda)dx^i dx^j\,,
\eea
where $x^i$ are coordinates labeling points along the minimal surface $m(A)$ and $\lambda$ is the affine parameter that runs along the geodesics, transverse to the surface $m(A)$. Without loss of generality, we take $\lambda=0$ to be the location of the minimal surface, so that $|\lambda|$ measures the proper length from $m(A)$ to an arbitrary slice of the foliation. We further assume that $\lambda>0$. This means that we generically require two different patches to cover the regions inside and outside of $m(A)$.

Let us write our vector field $V$ as $V=|V|\xi$, and assume that $\xi$ is affinely parametrized.
Thus, in the above coordinate system $\xi$ takes a simple form:
\bea
\xi=\partial_\lambda\,.
\eea
We would like to find a coordinate invariant criterion for the transverse area element induced by $\xi$. Namely, we would like to find an equation for
\bea
\partial_\lambda \ln\sqrt{\g}=\frac{1}{2}\g^{ij}\partial_\l \g_{ij}\,,
\eea
where $\g=\det{\g_{ij}}$. Indeed, since we know that at the minimal surface $\partial_\lambda \ln \sqrt{\g}=0$, what we really want is to argue that $\partial_\lambda \ln\sqrt{\g}>0$ away from the minimal surface so that $|V|$ decreases. We note that this condition may be too strong in certain situations. Namely, there can be cases where $\partial_\lambda \ln\sqrt{\g}<0$ locally, but $|V|<1$ globally nevertheless. For simplicity, we will assume that $|V|$ decreases \emph{monotonically} along our geodesic foliations.

We notice that our problem can be recast as an electrostatics problem with a \emph{constant surface charge density} at the location of the minimal surface. In flat space, the magnitude of the electric field indeed decreases \emph{monotonically} away from the sources, given that i) the transverse area increases with the distance and ii) by Gauss's Law, the magnitude of the electric field is inversely proportional to the transverse area. Here, we want to find an equivalent statement for the transverse area in a general Riemannian metric $g_{ab}$. In order to do so, we will impose the condition that $\partial_\l^2\ln\sqrt{\g}>0$ everywhere so that the magnitude $|V|$ decreases monotonically.\footnote{This condition is sufficient and fully covariant, but might be too strong in the most general case.}

We begin by studying how the curvature of the ambient space $g_{ab}$ varies along the geodesic flow. This is, we compute the Ricci tensor $R_{ab}$ and study its $\l \l$ component,
\bea
R_{ab}\xi^a\xi^b=R_{\l \l}=\partial_a\Gamma^{a}_{\l \l}-\partial_\l \Gamma^a_{a \l}+\Gamma^a_{\l \l} \Gamma^b_{a b}-\Gamma^a_{b \l}\Gamma^b_{a\l}\,.
\eea
Using the explicit expressions for $\Gamma^a_{bc}$ it is easy to check that:
\bea
\Gamma^a_{\l\l}=0, \quad \textrm{and} \quad \Gamma^i_{\l j}=\frac 12 \g^{ik}\partial_\l \g_{kj}\,,
\eea
leading to the following relations:
\bea
R_{\l \l}&=&-\partial_\l  \Gamma^i_{\l i}-\Gamma^i_{ \l j}\Gamma^j_{\l i}\,, \\
R_{\l \l}&=&-\partial_\l \(\frac 12 \g^{ij}\partial_\l \g_{ij}\)-\frac 14 \(\g^{ik}\partial_\l \g_{jk}\)\(\g^{jl}\partial_\l \g_{il}\)\,, \\
R_{\l \l}&=&-\partial_\l \(\frac 12 \g^{ij}\partial_\l \g_{ij}\)+\frac 14\partial_\l \g^{ij}\partial_\l \g_{ij}\,,\label{Rll3}
\eea
where in the last line we have used that $\g^{ik}\partial_\l \gamma_{il}=-\gamma_{il}\partial_\l \g^{ik}$. Equation (\ref{Rll3}) can be rewritten in the following way,
\bea \label{arealaw}
\partial_\l^2 \ln \sqrt{\g}=-R_{\l \l}+\frac 14\partial_\l \g^{ij}\partial_\l \g_{ij}\,,
\eea
which is the Riemannian version of the Raychauduri equation. Alternatively, this equation can be put in a coordinate invariant way by noticing that the extrinsic curvature tensor and its trace are given by $K_{ij}=\half \partial_\lambda \g_{ij}$ and $K=\g^{ij} K_{ij}=\partial_\lambda \ln \sqrt{\g}$, respectively,
which lead to the more familiar form,
\bea\label{Ray1}
\partial_\l K=-R_{ab} \xi^a \xi^b -K^{ab}K_{ab}\,.
\eea

We can now study the geometry of the constant-$\l$ slices. The induced metric on these hypersurfaces is the following:
\bea\label{indmetric}
d\tilde{s}^2=\gamma_{ij}(x,\l)dx^i dx^j\,.
\eea
It can be shown that Ricci tensor $\tilde{R}_{ij}$ associated to this metric satisfies the following relation:
\bea\label{indRicci}
\gamma^{ij} \tilde{R}_{ij}=\gamma^{ij}R_{ij}-R_{\l\l}+K^2-K_{ij}K^{ij} \,
\eea
which relates its components with those of the global Ricci tensor and the extrinsic curvature $K_{ij}$. To arrive at this equation one can start from the definition of $R_{ab}$ in the coordinates $(\l, x^i)$ and then separate the components involving only $\g_{ij}, \partial_k\g_{ij}$ which will form the $\tilde{R}_{ij}$. The remaining terms will involve functions of the extrinsic curvature $K_{ij}$. With this relation at hand we can rewrite the Raychauduri equation (\ref{Ray1}) as
\bea\label{Ray2}
\partial_\l K=-\gamma^{ij}\(R_{ij}-\tilde{R}_{ij}\)-K^2\,.
\eea
Finally, adding (\ref{Ray1}) and (\ref{Ray2}) we obtain
\bea\label{Ray3}
\partial_\l K=-\half(R-\tilde{R}) -\half K^{ab}K_{ab} -\half K^2\,.
\eea
where $\tilde{R}=\gamma^{ij}\tilde{R}_{ij}$ is the Ricci scalar associated to the induced metric (\ref{indmetric}). This equation has the advantage that involves only geometric quantities.

In the following, we will use equations (\ref{Ray1}), (\ref{Ray2}) and (\ref{Ray3}) to argue for the monotonicity of the transverse area in various cases of interest.

\subsection{Strips in a general translationally invariant background\label{strips2}}

For strip entangling surfaces in a general translationally invariant background the problem can be formulated in terms of an effective two-dimensional bulk geometry, which can be obtained by dimensional reduction of the original metric.  We assume that this effective geometry is foliated by geodesics so that
\bea
ds^2=d\l^2+\g(\l,x) dx^2\,,
\eea
where $\g$ is the determinant of the induced geometry for the constant-$\lambda$ slices. A geometry of this type can be obtained by integrating out the transverse coordinates, and then foliating the effective geometry with geodesics, which are nothing but codimension-one minimal surfaces in the ambient space. Now, since the constant-$\lambda$ hypersurfaces are one dimensional in this case, a number of simplifications take place: i) the extrinsic curvature has only one component $K_{ab}=K_{xx}$ and ii) the induced Ricci tensor vanishes identically $\tilde{R}_{ij}=0$\footnote{In fact, the Riemann tensor in a one dimensional space is identically zero since the covariant derivatives will always commute.}. Therefore, we can use the general formula  (\ref{Ray3}) in our analysis, which in this case reduces to
\bea
\partial_\l K=-\half R-K^2\,.
\eea
Rewriting this equation purely in terms of $\g$ we obtain,
\bea
\frac1\g \partial^2_\l \g -\frac1{2\g^2} \(\partial_\l\g\)^2 &=&-R\,,   \nonumber \\
\frac{2}{\sqrt{\g}}\partial^2_\l \sqrt{\g}
&=&-R\,.\label{curvaturecond2d}
\eea
Therefore, if $R<0$ we obtain that $\partial_\l^2 \sqrt{\g}>0$. Since $\partial_\l\sqrt{\g}=0$ at the minimal surface this implies that $\partial_\l \sqrt{\g}>0$ everywhere so
that $\sqrt{\g}$ is a monotonically increasing function of $\lambda$, i.e. it increases as one moves away from the minimal surface.

In the last part of this section, we will revisit the vacuum solutions studied in section \ref{section2} and then study more general backgrounds supported by matter fields. In the latter case, we will be able to translate the curvature condition found above in terms of an upper bound for the energy density of the fields.

\vspace{5mm}
\textbf{Vacuum solutions:}
Let us start by revisiting the known cases, strips in pure AdS and in a black brane background. As it was shown in the previous section, the two dimensional effective geometry for the former case is the following:
\bea\label{twometads}
ds^2=\frac{V_\perp^2}{z^{2d}}(dx^2+dz^2)\,.
\eea
The factor of $d$ in the denominator and the constant $V_\perp\equiv\ell_\perp^{d-1}$
arise from integrating out the transverse coordinates. One can easily repeat the same exercise for a strip in an AdS black brane, which yields
the following effective metric:
\bea\label{twometbh}
ds^2=\frac{V_\perp^2}{z^{2d}}\(dx^2+\frac{dz^2}{f(z)}\)\,,\qquad f(z)=1-\(\frac{z}{z_h}\)^{d+1}\,.
\eea
It is easy to see that (\ref{twometads}) is a special case of (\ref{twometbh}), where one sends $z_h\to\infty$. So without loss of generality we can focus only on the second case. A short calculation shows that the Ricci scalar associated to the metric (\ref{twometbh}) is:
\bea
R=-\frac{2d\, z^{2(d-1)}}{V_\perp^2}\[1+\frac{d-1}{2}\(\frac{z}{z_h}\)^{d+1}\]<0\,.
\eea
Since it is negative definite, then it follows that $|V|$ decreases monotonically away from the minimal surface. Therefore $|V|\leq1$ everywhere for strips in pure AdS or in an AdS black brane in arbitrary dimensions.

\vspace{5mm}
\textbf{Non-vacuum solutions:}
To begin with, consider a generic static gravitational system in $3$-dimensions. Let us assume that the full bulk metric
\bea\label{metric}
ds^2\equiv G_{\mu\nu}dx^{\mu}dx^{\nu}=-\psi(\l,x) dt^2+d\l^2+\g(\l,x) dx^2\,,
\eea
is a solution of Einstein's equations with an appropriate stress-tensor:\footnote{We have set $8\pi G_N=1$ for simplicity.}
\bea\label{EEQ-CC}
\mathcal{R}_{\mu \nu}-\half \mathcal{R} G_{\mu \nu}+\Lambda G_{\mu \nu}=T_{\mu\nu}\,.
\eea
A quick calculation shows that the induced Ricci on a constant-$t$ slice is:
\bea
R=\frac{2}{\psi(\l,x)}(T_{00}(\lambda,x)+\Lambda \psi(\l,x))\,,
\eea
hence, for negative cosmological constant $\Lambda<0$, we have that $R<0$ if and only if the local energy density is bounded from above:
\bea\label{boundE}
\varepsilon(\lambda,x)\equiv -T^0_{\;\;0}(\lambda,x)<-\Lambda\,.
\eea
Notice that the right hand side is parametrically large in the $G_N\to0$ limit. Indeed, in this limit the negative curvature is fully supported by the cosmological constant term, so a flow based on geodesic foliations would be allowed for arbitrary matter content. On the other hand, if one takes $G_N$ to be small but finite, equation (\ref{boundE}) would indeed provide a sharp upper bound for the energy density. Once this bound is violated, geodesics start to focus and the transverse area start to decrease instead of increase.

It is possible to generalize this inequality to higher dimensions. To show this, we dimensionally reduce
a translationally invariant metric in $3+k$ dimensions down  to $3$ dimensions. We write the full metric as follows:
\bea\label{fullmetric}
ds^2=e^{-2k\tau(\lambda,x)}ds^2_{3}+e^{2\tau(\lambda,x)}dy^2_k\,,
\eea
where $ds^2_{3}$ is of the form (\ref{metric}). The Einstein-Hilbert term in the action transforms as follows:
\bea
\frac{1}{2\kappa^2}\int d^{3+k}x\sqrt{-g}\mathcal{R}=\frac{1}{2\kappa^2_{3}}\int d^{3}x\sqrt{-g_{3}}\left(\mathcal{R}_{3}-k(k+1)\partial_m\tau\partial^m\tau\right)\,,
\eea
where $\kappa_3^2=\kappa^2/V_\perp$.
Assuming that the $3+k$ dimensional metric satisfies Einstein's equations (\ref{EEQ-CC}), after some algebra we arrive to the following relation:
\bea
R=\frac{2}{\psi(\l,x)}T_{00}(\lambda,x)+\frac{2\Lambda}{e^{2k\tau}}+k(k+1)\left[(\partial_\lambda\tau)^2+\frac{(\partial_x\tau)^2}{\gamma(\l,x)}\right]\,,
\eea
where $R$ is the induced Ricci on a constant-$t$ slice of the effective 3-dimensional geometry. Hence, we have that $R<0$ if and only if the local energy density is bounded from above:
\bea
\varepsilon(\lambda,x)<-\Lambda-\frac{k(k+1)}{2}\partial_\mu\tau\partial^\mu\tau\,.
\eea
The last term in this equation is contracted using the full metric (\ref{fullmetric}) so the criterium is fully covariant. This term is strictly negative, so in higher dimensions there is an interplay between the cosmological constant term and the kinetic term of the scalar field that arises from the dimensional reduction. In particular, we notice that at finite $G_N$ the bound becomes more rigid as one increases $k$.

\subsection{Spheres in a general rotationally invariant background}

For spherical entangling surfaces one can try to repeat the same steps as for the case of strips. However, since the construction here is directly based on geodesics of the full geometry (instead of minimal surfaces), it is easy to see that it is not possible to recast the problem solely in terms of a two dimensional effective geometry.
Nevertheless, we will see below that the spherical symmetry is powerful enough to allow some simplification.

To begin the analysis, we start from a generic metric of the form:
\bea\label{metricsph}
ds^2=d\lambda^2+\gamma(\lambda,r)dr^2+e^{2\tau(\l,r)}d\Omega_k^2\,,\qquad ds_2^2\equiv d\lambda^2+\gamma(\lambda,r)dr^2\,.
\eea
We emphasize that the two-dimensional metric $ds_2^2$ \emph{cannot} be obtained by a dimensional reduction of the full geometry, as was done in the case of strips, but instead should be thought of as an auxiliary object that is part of the full metric (\ref{metricsph}). With the above ansatz, one can compute the quantities of interest, namely the curvature tensors $R_{ij}$, $\tilde{R}_{ij}$ and the extrinsic curvature tensor $K_{ij}$.
After some algebra, we can rewrite (\ref{Ray3}) as follows:
\bea\label{critesph}
2(\partial_\lambda K+K^2)=\frac{2}{\sqrt{\det \gamma_{ij}}}\partial_\lambda^2\sqrt{\det\gamma_{ij}}=-R_2+2k\left[\partial^2_\lambda\tau+\partial_\lambda\tau(k+\partial_\lambda\log\gamma)\right]\,.
\eea
In this equation $\gamma_{ij}$ involves the metric functions of the $r$-part of the metric as well as the angular components; moreover, $R_2$ refers to the induced metric of the auxiliary 2-dimensional metric defined in (\ref{metricsph}). We must require the sum of these terms to be positive, in order to have a transverse area that grows away from the minimal surface. This implies that for spheres, the condition in terms of curvature generalizes to:
\bea\label{R2cond}
R_2<2k\left[\partial^2_\lambda\tau+\partial_\lambda\tau(k+\partial_\lambda\log\gamma)\right]\,.
\eea
As expected, for $k=0$ we recover the 2-dimensional condition (\ref{curvaturecond2d}) which implies negative curvature. One way to realize the above condition for general $k$ is to assume that $R_2<0$ and simultaneously have all terms in the right hand side of (\ref{R2cond}) to be positive. As we will see below, this is indeed the case for spheres in empty AdS. We leave a more general analysis of (\ref{R2cond}) and its geometrical and physical interpretation to future work.


In the last part of this section we will revisit the case of spheres in empty AdS and offer some comments on more generic cases.

\vspace{5mm}
\textbf{Vacuum solutions:}
Let us revisit the examples studied in section \ref{section2}, namely, spheres in empty AdS for an arbitrary number of dimensions.
In this case, it can be shown that the spatial part of the metric can be put into the more symmetric form
\bea\label{metsph}
ds^2=d\lambda^2+\gamma(\l,r)\( dr^2+r^2 d\Omega^2_{k}\)\,,
\eea
so that
\bea
e^{2\tau(\l,r)}=r^2\gamma(\l,r)\,.
\eea
To prove this we start with the spatial part of the metric of an hyperbolic AdS black hole:
\bea\label{hyperbmet}
ds^2=\frac{1}{\sin^2\chi}\(d\chi^2+du^2+\sinh^2u d\Omega_k^2\)\,.
\eea
Indeed, this metric covers the inside of the entanglement wedge associated to a spherical region in empty AdS. Since the 2-dimensional metric (at constant angles) is invariant under translations in $u$, it is clear that the vertical geodesics with $u=$ constant correspond to the integral curves of $V$ in this coordinate system. Finally, the coordinate transformation
\bea
\frac{d\chi^2}{\sin^2\chi}=d\lambda^2\,,\qquad \frac{du^2}{\sinh^2u}=\frac{dr^2}{r^2}\,,
\eea
brings the metric (\ref{hyperbmet}) into the form (\ref{metsph}). Since
\bea
R_2=-2<0\,,
\eea
and
\bea
\partial_\lambda\tau=\partial_\lambda\log\sqrt{\gamma}>0\,,\qquad\partial^2_\lambda\tau=\partial^2_\lambda\log\sqrt{\gamma}>0\,,
\eea
then, it follows immediately that the right hand of (\ref{critesph}) is positive. Hence, the transverse area grows monotonically away from the minimal surface.

The other example in the vacuum that one can consider is the case of an AdS black brane. However, we do not know the minimal surface associated to a spherical region in this case, nor the explicit spacelike geodesics. It is also not clear that the foliation admits a simplification of the form (\ref{metsph}).
It would be interesting to attempt such a construction numerically.

\vspace{5mm}
\textbf{Non-vacuum solutions:} Adding time to the metric (\ref{metricsph}) we get
\bea
ds^2=-\psi(\l,r)dt^2+d\lambda^2+\gamma(\lambda,r)dr^2+e^{2\tau(\l,r)}d\Omega_k^2\,.
\eea
One can assume that this metric satisfies Einstein equations (\ref{EEQ-CC}), and proceed in the same way as we did for the case of strips. By doing so, one obtains an upper bound on the energy density $\varepsilon(\l,r)$ in terms of the cosmological constant and derivatives of the metric functions, $\gamma(\l,r)$ and $\tau(\l,r)$. However, the final result is not fully covariant and does not seem to be particularly illuminating. Hence, we will refrain from writing out these results here.

\section{Nesting property and maximally packed flows \label{NPMPF}}

So far we have considered only one kind of thread configurations, which are based on specific foliations of the spacetime by geodesics or minimal surfaces. This particular class of constructions ensures that the flux across the minimal surface is maximal, but they generically lead to a flow that decreases in magnitude as one moves away from the minimal surface. In certain cases, the entanglement wedge itself might contain one or multiple \emph{bottle necks} that might admit more flux than the one supplied by the geodesic flow. The idea here is to propose a strategy to \emph{simultaneously} maximize the flux across the minimal surface and another minimal surface of interest. In order to do so, we will use a general property of holographic entanglement entropy, which is known as entanglement wedge nesting.

Let us begin for simplicity with an arbitrary two dimensional Riemannian geometry with a boundary; this can be, for instance, a constant-$t$ slice of pure $AdS_3$ or a BTZ black brane. Consider a family of intervals of lengths $l_n$, such that their left boundary it located at $x_L=0$ while the right boundary is at $x_R=l_n$, such that $l_{n+1} >l_n$ with $n \in \{0,1,\cdots, N\}$. This family will have minimal surfaces $m(A_n)$ such that their entanglement wedges $r(A_n)$ are nested, i.e., $r(A_{n+1}) \supset r(A_n)$.
Now, the nesting property for bit threads implies that one can find a flow that maximizes the flux through $A_N$, while simultaneously maximizing the flux through all $A_n$ with $n <N$. If one takes $l_n$ to vary continuously then the above construction would then generate a thread configuration with magnitude $|V|=1$ everywhere in $r(A_N)\setminus r(A_0)$, which smoothly interpolates between the unit normal on $m(A_0)$, $\hat{n}_0$, and the one on $m(A_N)$, $\hat{n}_N$. Since the norm of vector saturates the inequality $|V|\leq1$ in this region, we call this class of constructions \emph{maximally packed flows}.\footnote{This term was originally used in \cite{Headrick:2017ucz}, where a particular maximally packed flow was constructed.  Recently similar constructions involving this kind of flows were used in \cite{Hubeny:2018bri} to provide a constructive proof of the monogamy property of mutual information.}

Let us make the above example more precise. Assuming that the geometry is given by a time slice of AdS$_{3}$ and taking the intervals to have endpoints at:
\bea
x_{L}=0\,,\qquad x_R=l_n=l_0+\frac{n}{N}(l_N-l_0)\,,\qquad l_N>l_0\,,
\eea
then, the corresponding family of minimal surfaces will be given by:
\bea
z=\sqrt{R_n^2+\(x-x_n\)^2}\,,\qquad\text{with}\qquad R_n\equiv l_n\,,\qquad x_n\equiv l_n\,,
\eea
i.e., a family of semicircles with both, radii and centers given by $l_n$. The continuum limit of this family can be easily obtained by taking the limit $N\to\infty$ and defining a parameter $\lambda\equiv n/N\in[0,1]$, such that:
\bea\label{llambda}
x_{L}(\lambda)=0\,,\qquad x_R(\lambda)=l(\lambda)=l_0+\lambda(l_N-l_0)\,.
\eea
Taking this limit leads to the continuous foliation by semicircles:
\bea\label{foliationcircle}
z=\sqrt{R(\lambda)^2+\(x-x(\lambda)\)^2}\,,\qquad\text{with}\qquad R(\lambda)=l(\lambda)\,,\qquad x(\lambda)=l(\lambda)\,.
\eea
For this foliation, the outward-pointing unit normal vector $\hat{n}(\lambda)$ at a point $(x,z)$ of a constant-$\lambda$ slice is given by:
\bea\label{normalfoliation}
\hat{n}^a(\lambda)=\frac{z}{R(\lambda)}\(x-x(\lambda),z\)=\frac{z}{l(\lambda)}\(x-l(\lambda),z\)\,.
\eea
Since $V$ is equal to the normal vector $\hat{n}(\lambda)$ in the region $r(A_N)\setminus r(A_0)$, it would be desirable to eliminate $\lambda$ from from this equation
and obtain an expression that does not refer to a specific slice of the foliation. We can do so by explicitly inverting $l(\lambda)$ from equation (\ref{foliationcircle}). A brief calculation yields
\bea\label{llexplicit}
l(\lambda)=\frac{x^2+z^2}{2 x}\,.
\eea
Plugging (\ref{llexplicit}) into (\ref{normalfoliation}) we then arrive to the following formula for the vector field,
\bea\label{vnesting}
V^a\equiv\hat{n}^a=\frac{z}{x^2+z^2}\(x^2-z^2,2 x z\)\,.
\eea
Notice that since $V$ has been constructed from the unit normal vector $\hat{n}_n$ associated to a set of minimal surfaces, then, the divergenceless condition of $V$ follows trivially. The argument is the following: at every minimal surface $m(A_n)$ one can define its extrinsic curvature $K_n=\nabla_a \hat{n}_n^a$. However, the minimality condition implies that the extrinsic curvature vanishes identically, therefore $\nabla\cdot V=0$. Since $V$ is smooth in the interpolating region, divergenceless and satisfies the bound $|V|\leq1$, then it represents a valid thread configuration.
\begin{figure}[t!]
\centering
 \includegraphics[width=5in]{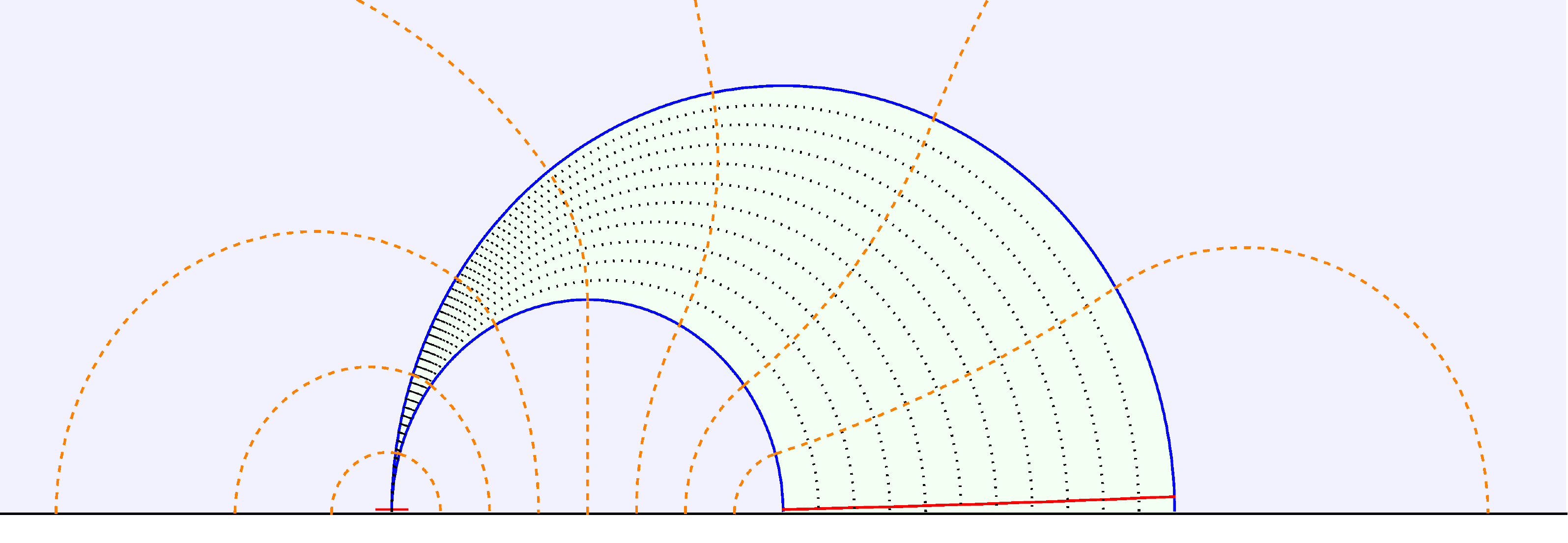}
  \begin{picture}(0,0)
\put(-287,-3){{\small $x=0$}}
\put(-187,-3){{\small $l_0$}}
\put(-96,-3){{\small $l_N$}}
\put(-93,11){{\small {\color{red}$\epsilon(x)$}}}
\put(-281,11){{\small {\color{red}$\epsilon$}}}
\end{picture}
 \caption{\label{nesting} An explicit construction of a maximally packed flow. In this example we have considered a family of intervals $A_n$ with left boundary located at $x_L=0$ and right boundary at $x_R=l_n$, with $l_{n+1}>l_n$. The two limiting minimal surfaces $m(A_0)$ and $m(A_N)$ bound the portion of the bulk that is shaded in green. In this region, the vector $V$ has maximal norm, i.e., $|V|=1$ and is orthogonal to the intermediate minimal surfaces $m(A_n)$. The UV cutoff that leaves the flux across the different surfaces constant is shown in red, but other choices are also allowed. The regions inside of $m(A_0)$ and outside of $m(A_N)$, which are shaded in blue, are continued with the geodesic flows constructed with the algorithm of section \ref{section2}. }
\end{figure}
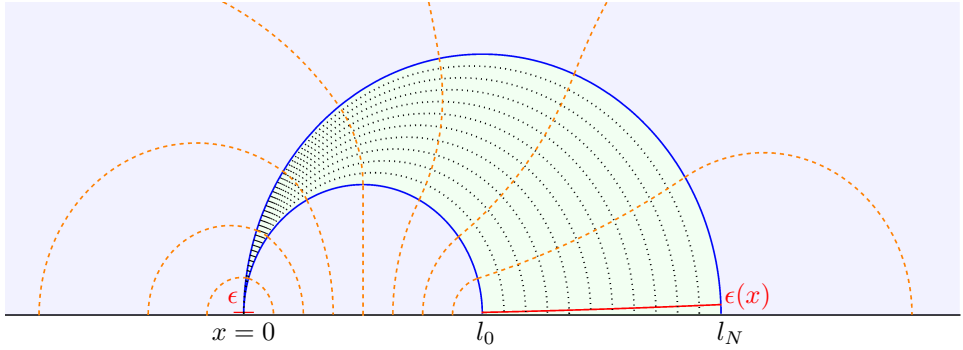

\begin{figure}[t!]
\centering
 \includegraphics[width=5in]{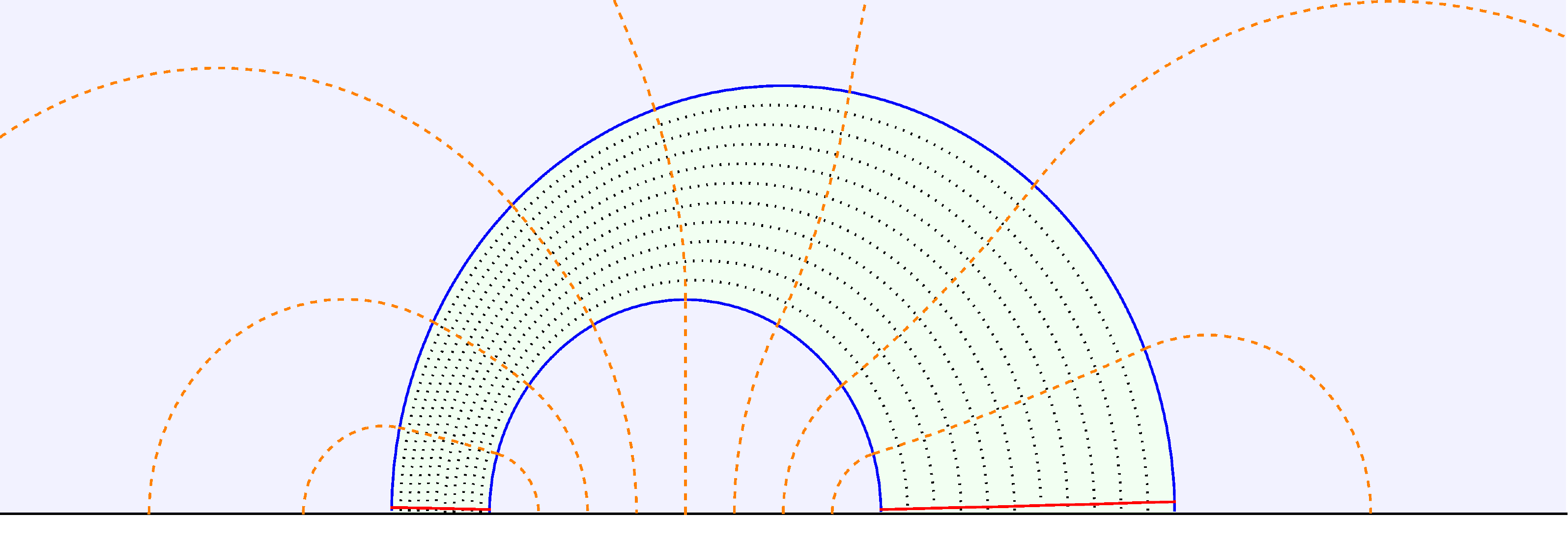}
   \begin{picture}(0,0)
\put(-276,-3){{\small $l^L_N$}}
\put(-254,-3){{\small $l^L_0$}}
\put(-164,-3){{\small $l^R_0$}}
\put(-96,-3){{\small $l^R_N$}}
\put(-93,11){{\small {\color{red}$\epsilon(x)$}}}
\put(-293,11){{\small {\color{red}$\epsilon(x)$}}}
\end{picture}
 \caption{\label{nesting2} A second example of a maximally packed flow. In this example, we have considered a family of intervals $A_n$ with variable left and right boundaries. The two limiting minimal surfaces $m(A_0)$ and $m(A_N)$ bound the portion of the bulk that is shaded in green. In this region, the vector $V$ has maximal norm, i.e., $|V|=1$ and is orthogonal to the intermediate minimal surfaces $m(A_n)$. The UV cutoff that leaves the flux across the different surfaces constant is shown in red, but other choices are also allowed. The regions inside of $m(A_0)$ and outside of $m(A_N)$, which are shaded in blue, are continued with the geodesic flows constructed with the algorithm of section \ref{section2}. }
\end{figure}
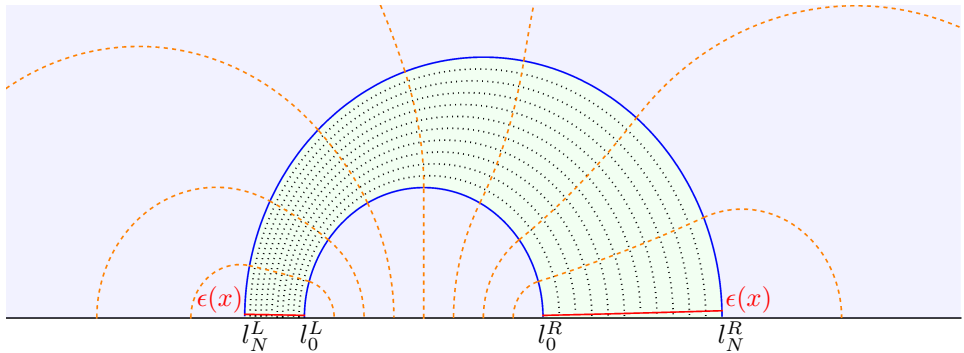

By construction, then,  we have that the portion of the vector field $V$ that goes from $m(A_0)$ to $m(A_N)$ is equal to the unit normal vector $\hat{n}_n$ at the intermediate minimal surfaces. We illustrate this example graphically in figure \ref{nesting}. A few important observations are in order:
\begin{itemize}
  \item The integral curves can be easily obtained from the vector field itself (\ref{vnesting}). For the above example they satisfy the following equation:
  \bea\label{intceq}
  \frac{dz}{dx}=\frac{2 x z}{x^2-z^2}\,,
  \eea
  which has solutions of the form:
  \bea\label{solintcurves}
  x^2 + z^2 = 2\,c\, z\,,\qquad c=\text{constant}\,.
  \eea
        These curves describe circles centered at $x_c=0$, $z_c=c$ with radius $c$.
  \item Since the volume of space is infinite near the boundary, the flux across any of the regions $A_n$ is also infinite. Therefore, we have to be careful with the regularization. A particular choice of cutoff can be obtained if one imposes that the flux across all surfaces are the same. In this case the cutoff surface should follow one of the integral curves defined by (\ref{solintcurves}), i.e., it must be orthogonal to all slices of the foliation. If we assume that the cutoff at the $A_0$ slice is $\epsilon$, then a brief calculation shows that
      \bea
      \epsilon(x)=\frac{x^2 \epsilon}{4 l_0^2}+\mathcal{O}(\epsilon ^2)\,.
      \eea
      In other words, since the area of nested intervals grows monotonically with the length, it means that the cutoff should also increase so that the flux is preserved. However, other choices of cutoff can be made by extending this construction, but will generally lead to different fluxes across the surfaces. For example, for the standard choice $\epsilon(x)=$ constant, the flux increases as the region increases.
  \item Notice that so far we did not say anything about what $V$ should be in the region inside $m(A_0)$ or the region outside $m(A_N)$. One simple option is to glue part of a geodesic flow constructed in the previous sections. We do this explicitly in figure \ref{nesting}. Since the magnitude and tangent vectors across the minimal surfaces are continuous, the vector field $V$ is continuous and once differentiable at the gluing surfaces. 
 \item We can generalize the above construction to arbitrarily nested intervals. For a two dimensional geometry, the most general case has endpoints at $x_L=l^{L}_n$ and $x_R=l^R_n$ such that $l^L_{n+1}<l^L_{n}$ and $l^R_{n+1}>l^L_{n}$. These two conditions guarantee that the corresponding intervals are properly nested. Notice that this simple generalization already lead to infinite many new possibilities, which are inequivalent in the continuum limit. Fixing one endpoint uniquely determines the flow, because \emph{all} choices of $l(\lambda)$ would be equivalent to each other by considering reparametrizations of $\lambda$. Conversely, having two arbitrary functions $l^L(\lambda)$ and $l^R(\lambda)$ make the choices inequivalent, up to a function that continuously maps points in the range of $l^L(\lambda)$ to points in the range of $l^R(\lambda)$. As an example, consider a time slice of AdS$_3$. Repeating the steps of the previous example leads to a vector field of the form:
     \bea\label{Vgeneralfoli}
     V^a=\frac{z}{R(\lambda)}\(x-x(\lambda),z\)\,,\qquad R(\lambda)\equiv \frac{l^{R}(\lambda)-l^{L}(\lambda)}{2}\,,\quad x(\lambda)\equiv\frac{l^{R}(\lambda)+l^{L}(\lambda)}{2}\,.
     \eea
     Having two functions $l^L(\lambda)$ and $l^R(\lambda)$ related through only one equation, namely, the minimal surface equation,
     \bea\label{geoeqfoli}
     z=\sqrt{R(\lambda)^2-(x-x(\lambda))^2}\,,
     \eea
     it is clear that generically it is not possible to explicitly solve for $V(x,z)$. This is clear from the previous argument, since we generically expect that different choices of functions would lead to different vector fields $V$. For particular choices of $l^L(\lambda)$ and $l^R(\lambda)$, however, it is still possible to invert (\ref{geoeqfoli}) to obtain $\lambda(x,z)$. Having done that, then one can simply replace the result in (\ref{Vgeneralfoli}) to obtain an explicit equation for $V(x,z)$.
     Figure \ref{nesting2} illustrates one of these generalized maximally packed flows.
 \end{itemize}

Needless to say, the above constructions can be easily generalized to arbitrary 2-dimensional geometries, and to higher dimensional cases for any family of nested subregions. In the following section, we will use both, maximally packed flows and geodesic flows, to study more general thread configurations for particular cases of interest.

\section{Applications of flow constructions\label{sec:5}}

\subsection{Entanglement of Purification\label{sec:puri}}
An interesting byproduct of the bit threads picture of holographic entanglement entropy was the discovery of the holographic dual to the quantum information quantity known as entanglement of purification \cite{purification}. Already in \cite{Freedman:2016zud}, the authors studied the minimal cross section for disjoint regions and interpreted it as the maximum flux among the thread configurations that live completely inside the corresponding entanglement wedge, connecting the two disconnected regions. Using intuition from tensor networks, this minimal cross section was later identified with the concept of entanglement of purification in \cite{Takayanagi:2017knl,Nguyen:2017yqw}.
Interestingly, the minimal cross-section of the entanglement wedge is an example of a \emph{bottle neck}, discussed in the previous section. Hence, we expect that a particular construction with the so-called maximally packed flows can be designed to specifically compute this quantity. We will devote this subsection to construct and study this kind of flows.

The definition of entanglement of purification is the following. Consider a quantum system $Q$ bipartitioned into sets of degrees of freedom $A$ and $B$, in a state described by a density matrix $\rho_{AB}$. If the state is mixed, the `entanglement entropies' associated to $A$ and $B$ differ from each other, $S_{A}\neq S_{B}$. More importantly, these entropies quantify both quantum \emph{and} classical correlations between $A$ and $B$. In order to quantify the amount of correlations that are purely quantum, one option is to purify the system, i.e., to consider a set $Q'$ of additional degrees of freedom, and a choice of pure state $\ket{\psi}$ for the overall system $QQ'$, such that $\text{Tr}_{\,Q'}\ket{\psi}\bra{\psi}=\rho_{AB}$. If we further partition the auxiliary system $Q'$ into $A'$ and $B'$, we can compute the entanglement entropies $S_{AA'}=S_{BB'}$, which should indeed arise from purely quantum correlations. By optimizing among \emph{all} possible purifications and \emph{all} possible partitions $A'$ and $B'$, the entanglement of purification between $A$ and $B$ is then defined as
\bea\label{eopdef}
\mathcal{E}(A:B)\equiv\min_{\scriptstyle \ket{\psi},A'} S_{AA'}~.
\eea
According to the proposal of \cite{Takayanagi:2017knl,Nguyen:2017yqw}, the holographic dual of this quantity is given by the minimal cross section $\Sigma$ of the entanglement wedge of $AB$, namely
\bea\label{eopholo}
\mathcal{E}(A:B)=\min\frac{\text{Area}(\Sigma_{AB})}{4G}\,.
\eea

\begin{figure}
\centering
 \includegraphics[width=2.8in
 ]{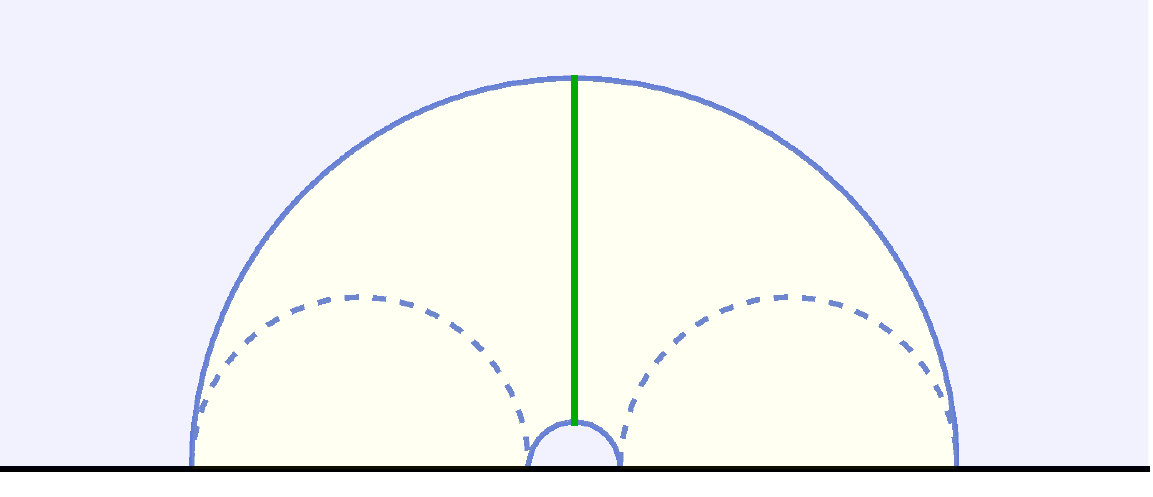}$\quad$
  \includegraphics[width=2.8in
 ]{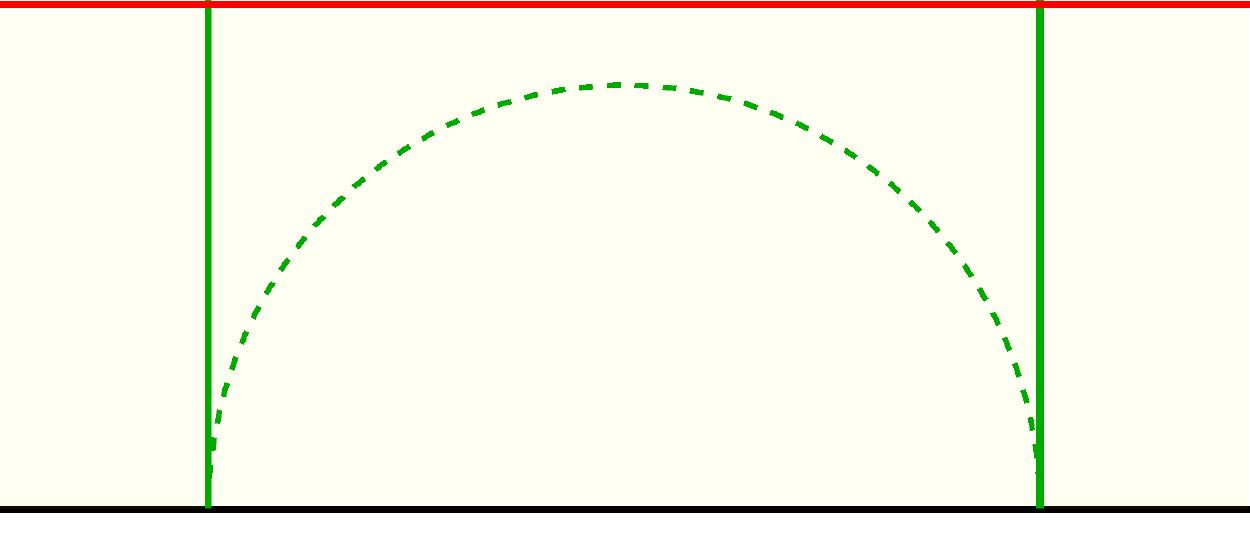}
 \begin{picture}(0,0)
\put(-321,50){{\small $\Sigma_{AB}$}}
\put(-36,50){{\small $\Sigma_{AB}^{(1)}$}}
\put(-193,50){{\small $\Sigma_{AB}^{(1)}$}}
\put(-114,60){{\small $\Sigma_{AB}^{(2)}$}}
\put(-365,-5){{\small $A$}}
\put(-288,-5){{\small $B$}}
\put(-109,-5){{\small $A$}}
\put(-25,-5){{\small $B$}}
\put(-193,-5){{\small $B$}}
\end{picture}
 \caption{Holographic entanglement of purification for two particular cases of interest: two disconnected intervals in pure AdS and an interval and its complement in a black brane background. The corresponding entanglement wedges are shaded in yellow, while the candidates for minimal cross-section $\Sigma_{AB}$ are shown in green. In the first case, there is only one bottle neck in $r(AB)$, for the case connected case and hence one candidate for $\Sigma_{AB}$. For large enough separations, the entanglement wedge undergoes a first order transition, and becomes disconnected. In this case the entanglement of purification vanishes. In the second case there are two bottle necks in $r(AB)$ and hence two candidates for the minimal cross-section $\Sigma_{AB}$. The entanglement of purification is this case is the minimum of the two, which depends on the ratio between the length of the interval and inverse temperature.}\label{Fig:Purif}
\end{figure}
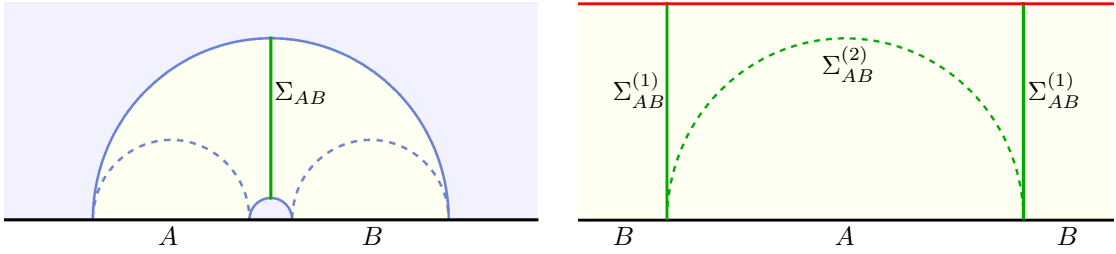

In figure \ref{Fig:Purif} we show two examples that illustrate the calculation of the minimal cross section. The first example consists of two disconnected intervals in empty AdS$_3$, $A=[a_1,a_2]$ and $B=[b_1,b_2]$. For simplicity, we have shown the case where the two intervals have equal lengths. In this case, one starts with a pure state and traces over the complement of $AB$. Hence, the resulting density matrix $\rho_{AB}$ describes a mixed state. It is well known that, the entanglement wedge of the system $AB$ undergoes a first order transition. When $A$ and $B$ are close enough $r(AB)$ is connected, however, when $A$ and $B$ are sufficiently far $r(AB)$ is disconnected.  More specifically, the transition depends on the cross ratio
\bea
z\equiv\frac{(a_2-a_1)(b_2-b_1)}{(b_1-a_2)(b_2-a_1)}\,,
\eea
and it is only for $z\geq1$ that the corresponding entanglement wedge is connected. An example of a connected $r(AB)$ is shown in figure \ref{Fig:Purif}. Since the intervals are symmetric, the minimal cross-section $\Sigma_{AB}$ in this example is simply given by the green vertical line centered at the origin. For the disconnected case, the minimal cross-section would be identically zero.

The second example shown in figure \ref{Fig:Purif} consists of a single interval in a black brane background $A=[a_1,a_2]$ and its complement $B=\bar{A}$. In this case, one starts directly with a mixed state and, in particular, $\rho_{AB}$ is given by a thermal density matrix. There are two bottle necks in this case, the two vertical lines at the edges of $A$, $\Sigma_{AB}^{(1)}$, and the minimal surface that computes entanglement entropy of region $A$, $\Sigma_{AB}^{(2)}$. The entanglement of purification is then the minimum of these two,
\bea
\mathcal{E}(A:B)=\frac{1}{4G}\min \left\{\text{Area}\(\Sigma_{AB}^{(1)}\),\text{Area}\(\Sigma_{AB}^{(2)}\)\right\}\,.
\eea
It can be shown that the two above areas exchange dominance at a particular value of the interval length $\ell$, in units of the inverse temperature $\beta$, so that for large enough $\ell/\beta$ the disconnected solution is favored. In that case the thread configuration that computes the entanglement of purification has an interesting interpretation, when compared with the one that computes the entanglement entropy of region $A$. If one recalls the interpretation of the threads as quantum bits of holographic entanglement, then it means that the threads that go through the surfaces $\Sigma_{AB}^{(1)}$ compute the \emph{maximum} amount of entanglement entropy present in $S(A)$ whose source is purely quantum mechanical. In other words those threads could be related to the maximum number of Bell pairs which can be distilled from $\hat{\rho}_{AB}$ using only local operations and classical communication (LOCC). The threads that go into the horizon, on the other hand, could be interpreted as the \emph{minimum} amount of correlations present in $S(A)$ that are thermal or classical. The above separation makes sense since, as we will see below, one can explicitly construct flows which simultaneously compute the entanglement entropy $S(A)$ (as the flux through the boundary region $A$) and the entanglement of purification ${\mathcal{E}}(A:B)$ (as the part of the flux that goes through $\Sigma_{AB}^{(1)}$).

We will now proceed to construct the vector fields that computes the entanglement of purification in these two examples. We will start with one interval in a BTZ black hole, since most of the formulas needed for this case were already worked out in section \ref{sec2btz}. Then, we will consider the case of two disjoint intervals in AdS. The constructions of this example will in turn set the grounds for the topic that we will discuss next, namely, the monogamy property of mutual information.

\subsubsection{One interval in a BTZ black brane\label{QCE}}

Consider an interval of length $\ell$ centered at the origin, and its complement, in a constant-$t$ slice of the BTZ geometry (\ref{BTZmetric}). This example was studied in detail in section \ref{sec2btz} so we will borrow some of the formulas presented in that section. There are two bottle necks for this configuration, as shown in figure \ref{Fig:Purif}. The first one consists of two straight surfaces located at $x=\ell/2$ and $x=-\ell/2$, respectively. They have area:
\bea
\text{Area}\(\Sigma_{AB}^{(1)}\)=2\log\( \frac{\beta}{\pi \epsilon}\)\,,
\eea
where $\epsilon$ is a UV regulator and $\beta$ is the inverse temperature $\beta=2\pi z_h$.
The second bottle neck corresponds to the minimal area surface that computes entanglement entropy for region $A$. This minimal surface can be written either as (\ref{geoBTZ}) or (\ref{geoBTZ2}), depending on the parametrization, and is characterized by a single constant $z_*$ that indicates the maximum depth of the surface. This constant is related to the length of the interval $\ell$ through any of the equivalent expressions presented in equation (\ref{ellBTZ}). The area of this surface can be obtained by direct integration and can be written in terms of the interval length $\ell$ as follows
\bea
\text{Area}\(\Sigma_{AB}^{(2)}\)=2\log\[ \frac{\beta}{\pi \epsilon} \sinh\(\frac{\pi \ell}{\beta}\) \]\,.
\eea
Comparing the two areas, we arrive to the following expression:
\bea
{\mathcal{E}}(A:B)=\left\{\begin{array}{ll}
\frac{c}{3}\log\left\{\frac{\beta}{\pi \epsilon}  \right \}\,, &  \quad \ell> \ell_c\,,\\
\frac{c}{3}\log\left\{  \frac{\beta} {\pi \epsilon} \sinh\(\frac{\pi \ell}{\beta}\) \right \}\,, & \quad \ell< \ell_c\,,
\end{array}\right.
\eea
where $c=3/2G$ is the central charge of the 2-dimensional CFT and $\ell_c \equiv \frac{\beta}{\pi} \log(\sqrt{2}+1)\approx 0.28\beta$ (this yields a critical radial depth $z^c_*=z_h/\sqrt{2}\approx0.71z_h$). Thus, for large enough regions,
the surface $\Sigma_{AB}^{(1)}$ gives the minimal cross-section and hence the entanglement of purification,\footnote{Notice that this surface is \emph{not} homologous to $A$ so it does not compute entanglement entropy of region $A$.} while for small regions $\Sigma_{AB}^{(2)}$ computes both entanglement entropy of region $A$ and its entanglement of purification. In this latter case, then, there should be particular microstates where $S(A)$ can be interpreted \emph{entirely} as quantum correlations between Bell pairs in region $A$ and its complement $B$. In the former case, on the other hand, one finds that there is a \emph{maximum} number of Bell pairs that can be distilled from the mixed state, so at least part of $S(A)$ must be thermal in nature.

In the following, we will construct explicitly flows that maximize the fluxes through $A$ and $\Sigma_{AB}^{(1)}$ simultaneously, for $\ell > \ell_c$. In other words, we will find  thread configurations that most efficiently avoid the horizon (since those would necessarily cross $\Sigma_{AB}^{(1)}$) when maximizing the flux on $A$ and whose magnitude saturates the bound $|V|=1$ at $\Sigma_{AB}^{(1)}$. Interestingly, one can see that the maximally packed flows studied in the previous section provide a clean construction of such configurations.

\vspace{3mm}
\textbf{Quantum and classical entanglement in $S(A)$:}
We would like to construct a vector field that computes ${\mathcal{E}}(A:B)$ for an interval of length $\ell> \ell_c$. Without loss of generality, we will discuss the construction of the $x>0$ portion of the vector field, given that the configuration has reflection symmetry around the origin.

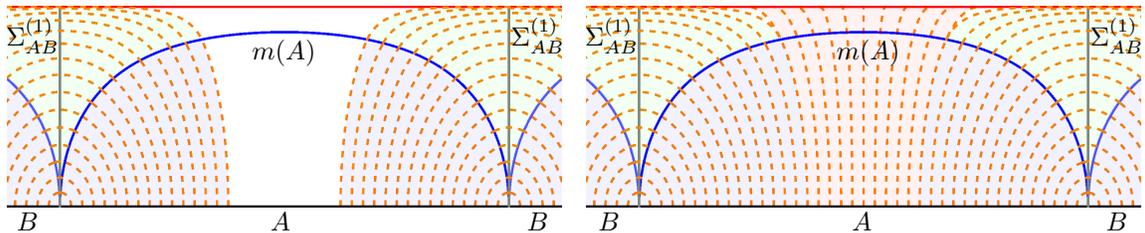
\begin{figure}[t!]
\centering
\centering
 \includegraphics[width=2.9in
 ]{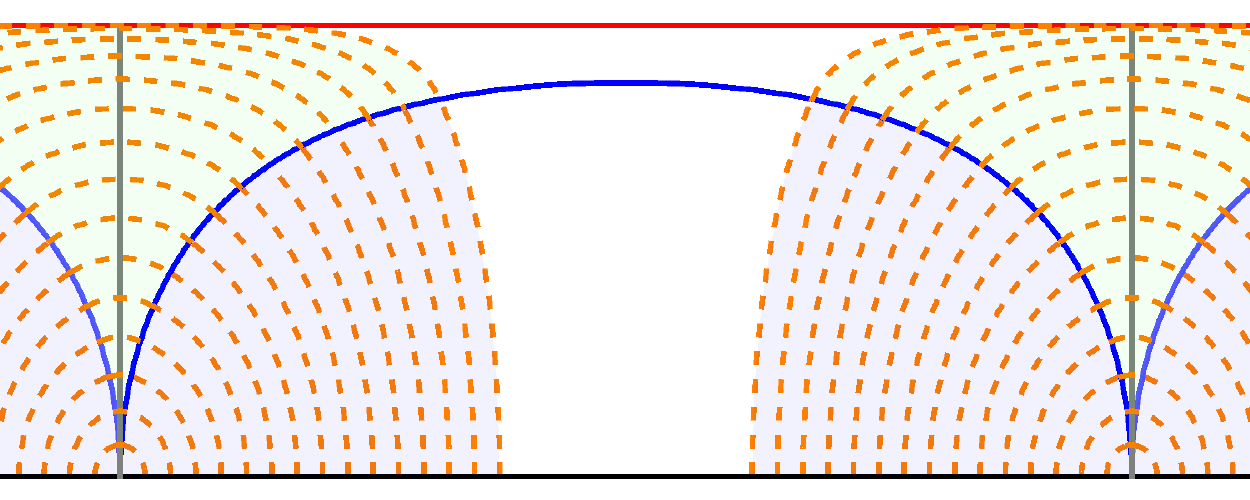}
  \hspace{0.2cm}\includegraphics[width=2.9in
 ]{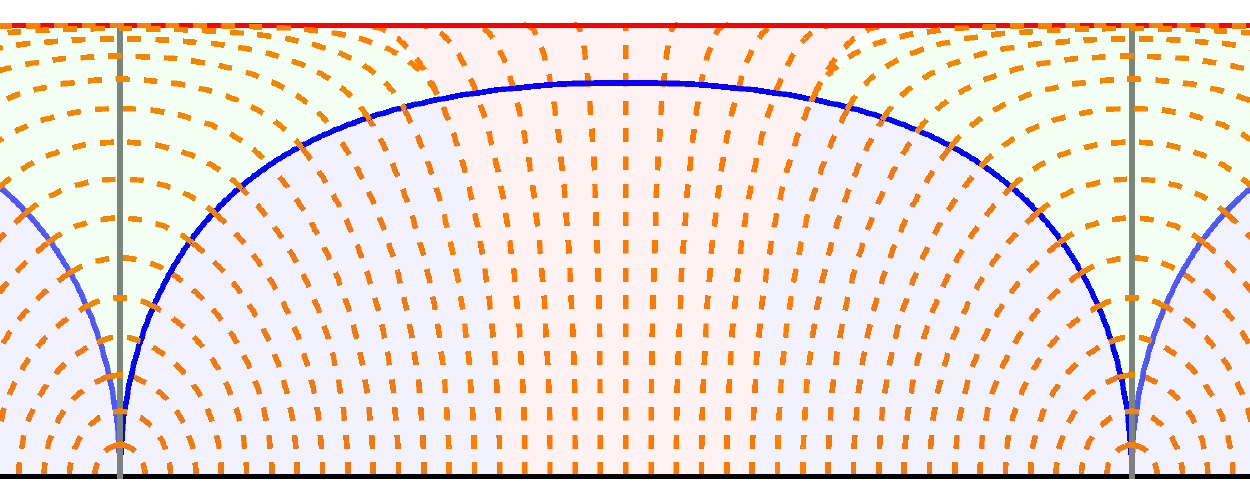}
\put(-109, -5){{\small ${A}$}}
\put(-329, -5){{\small ${A}$}}
\put(-13,-5){{\small $B$}}
\put(-205, -5){{\small ${B}$}}
\put(-232, -5){{\small ${B}$}}
\put(-425,-5){{\small $B$}}
\put(-115, 60){{\small ${m(A)}$}}
\put(-336, 60){{\small ${m(A)}$}}
\put(-19,65){{\small $\Sigma_{AB}^{(1)}$}}
\put(-210,65){{\small $\Sigma_{AB}^{(1)}$}}
\put(-238,65){{\small $\Sigma_{AB}^{(1)}$}}
\put(-429,65){{\small $\Sigma_{AB}^{(1)}$}}
 \caption{\label{EoP-BTZ} Integral curves (dashed orange) of the vector field configurations whose flux through $A$ computes the entanglement of purification ${\mathcal{E}}(A:B)$ (left) and the entanglement entropy $S(A)$ (right). The latter one also computes the entanglement of purification ${\mathcal{E}}(A:B)$ as the part of the flux that goes through $\Sigma_{AB}^{(1)}$. The shaded regions correspond to different sections of the vector field that are glued together. The green region corresponds to a maximally packed flow and, as discussed in the text, it is constructed by exploiting the nesting property of entanglement entropy. The blue region inside $m(A)$ represents the continuation of the same thread bundle based on the geodesic flows constructed in section \ref{section2}.  The red region in the right figure corresponds to an extra bundle that is added, and is also based on a geodesic flow. The threads in this region necessarily end at the horizon and hence their flux has a purely thermal interpretation. The white regions correspond to patches where the vector field vanishes, i.e., $V=0$. In both figures we have chosen to symmetrize the vector field around $\Sigma_{AB}^{(1)}$ so the regions $x>\ell/2$ and $x<-\ell/2$ are obtained by a reflection around these surfaces.}
\end{figure}

We start by applying the same construction of section \ref{NPMPF} based on the nesting property of entanglement entropy. In order to do so, we take a family of geodesics that interpolate between $\Sigma_{AB}^{(1)}$ and $m(A)$ and are nested in the appropriate sense. Here, we recall that the geometry outside the bulk horizon has been conjectured to be dual to a pure state with extra degrees of freedom living at the (stretched) horizon \cite{Takayanagi:2017knl,Nguyen:2017yqw}. In this setup, geodesics that end at the horizon can indeed be interpreted as entanglement entropies in the purified state \cite{Espindola:2018ozt}, Moreover, the nesting property applies in the usual sense, but  one must in turn consider the horizon as part of the boundary of spacetime. With these points in mind, then, we can consider the family of geodesics with the right endpoint fixed at $x_R=\ell/2$ and either $(i)$ a second  endpoint at the horizon with $z_h\in[\ell/2,-\infty]$ or $(ii)$ a left endpoint at $x_L\in[-\infty,-\ell/2]$. In order to continuously parametrize this family of geodesics, we will use the notation of (\ref{fluxparaII}). We set $z_0=z_*$ so that we can vary $s_0\in[-\infty,0]$ continuously, and use the branch with $\sigma=1$. Furthermore, we set
\bea
x_0=\frac{\ell}{2}-z_h\log\left(\frac{z_h\sqrt{z_h^2-z_*^2}+z_*\sqrt{(1+s_0^2)z_h^2-z_*^2}}{z_h^2-s_0 z_h z_*-z_*^2}\right)\,,
\eea
to ensure that the right point is fixed at $x_R=\ell/2$. Given these conditions, the family of geodesics we look for is then fully specified by:
\bea\label{familybtzpuri}
x=\frac{\ell}{2}-z_h\log\(\frac{z_h+z_*\sqrt{1+\frac{s_0^2z_h^2}{z_h^2-z_*^2}}}{\sqrt{z_h^2-z^2}+\sqrt{z_*^2\(1+\frac{s_0^2z_h^2}{z_h^2-z_*^2}\)-z^2}}\)\,,\qquad s_0\in[-\infty,0]\,.
\eea
To find the vector field and corresponding integral curves in the region that interpolates between these two surfaces we have two options:
\begin{itemize}
  \item We find the outward-pointing unit normal vector $\hat{n}(s_0)$ for our family of geodesics. A brief calculation yields:
        \bea\label{nhatbtxpur}
        \hat{n}^a(s_0)=\frac{z}{z_*\sqrt{1+\frac{s_0^2z_h^2}{z_h^2-z_*^2}}}\(\sqrt{z_*^2\(1+\frac{s_0^2z_h^2}{z_h^2-z_*^2}\)-z^2},\frac{z\sqrt{z_h^2-z^2}}{z_h}\)\,.
        \eea
        Solving for $s_0(x,z)$ from equation (\ref{familybtzpuri}) and plugging it back into (\ref{nhatbtxpur}) we can then obtain $V^a(x,z)=\hat{n}^a(x,z)$.
        From (\ref{nhatbtxpur}) we can also obtain a first order ODE to obtain the integral curves:
        \bea
        \frac{dz}{dx}=\frac{z}{z_h}\sqrt{\frac{z_h^2-z^2}{z_*^2\(1+\frac{s_0^2z_h^2}{z_h^2-z_*^2}\)-z^2}}\,,
        \eea
        which can be solved after substituting $s_0(z,x)$. Since the equation is of first order, we only need to give one boundary condition, e.g., $z(\ell/2)=z_a$ for $z_a\leq z_h$.
  \item Alternatively, we can find the integral curves directly by matching the fluxes across a given cross-section of the bulk region between $\Sigma_{AB}^{(1)}$ and $m(A)$. More specifically, we consider the integral along $\Sigma_{AB}^{(1)}$ from $z=\epsilon$ to a point $z=z_a\leq z_h$ and match it with the integral along a geodesic of the form (\ref{familybtzpuri}) from $z=\epsilon$ to a given $z$. This is
 \bea\label{Areas}
\int_{\ee}^{z_a}\frac{ dz}{z \sqrt{z_h^2-z^2}}=  \int_{\ee}^{z}\frac{ z_h \tilde{z}_* d z}{z \sqrt{(z_h^2-z^2)(\tilde{z}_*^2-z^2)}}\,,
 \eea
 where
 \bea
 \tilde{z}_*=z_*\sqrt{1+\frac{s_0^2z_h^2}{z_h^2-z_*^2}}\,.
 \eea
 After integration the above equality yields
 \bea\label{Azazb}
 \log\(\frac{z_a}{z_h+\sqrt{z_h^2-z_a^2}}\)=\log\(\frac{\tilde{z}_* z}{\tilde{z}_* \sqrt{z_h^2-z^2}+z_h \sqrt{\tilde{z}_*^2-z^2}}\)\,.
 \eea
Notice that we have cancelled the divergent term from both sides of equation (\ref{Areas}) since in the $\ee\to0$ limit the two integrals have the same divergence structure.
Finally, solving for $z$ in (\ref{Azazb}) gives us $z(s_0,z_a)$, which can be plugged into (\ref{familybtzpuri}) to obtain $x(s_0,z_a)$. The collection of points $(x(s_0,z_a),z(s_0,z_a))$ for $s_0\in[-\infty,0]$ give us the corresponding integral curves. If desired one could also solve for $s_0(z,x)$ in (\ref{familybtzpuri}) a plug it into (\ref{Areas}) to obtain either $x(z;z_a)$ or $z(x;z_a)$.
\end{itemize}
In figure \ref{EoP-BTZ} (left) we plot the integral curves of our maximally packed flow based on the above construction. Notice that we have chosen a particular continuation for the thread configuration inside $m(A)$, which is based on the geodesic flows of section \ref{sec2btz} (region shaded in blue). We have also chosen to symmetrize the vector field around $\Sigma_{AB}^{(1)}$ so the region $x>\ell/2$ (and hence $x<-\ell/2$) is obtained by a reflection around this surface. However, these choices are highly non-unique. Finally, notice that the portion of the thread configuration that is maximally packed (shaded in green) covers only a portion of the minimal surface $m(A)$, as expected, since we are studying the case where $\ell>\ell_c$. The endpoint of the thread bundle is located at a point $(x_e,z_e)$ on the minimal surface, with
\bea
x_e=z_h\log\(\frac{z_h^3+2z_*^3-z_h^2z_*}{\sqrt{(z_h^2-z_*^2)(z_h^4+4z_*^2)}}\)\,,\qquad z_e=\frac{z_h}{\sqrt{1+\( \frac{z_h^2}{2 z_*^2} \)^2}}\,.
\eea
It is easy to check that $z_e<z_*$ for $z_*>z_*^c=z_h/\sqrt{2}$, which yields the critical length $\ell_c$. By construction, then, we have obtained a vector field that is orthogonal to $\Sigma_{AB}^{(1)}$ and has maximal norm there, i.e., $|V|=1$. Since this vector field configuration has only threads that start and end in regions $A$ and $B$, respectively, its flux through $A$ computes the entanglement of purification ${\mathcal{E}}(A:B)$. In figure \ref{EoP-BTZ} (right) we have also shown a configuration where we have added an extra thread bundle in the central region (shaded in red), based on the geodesic flow construction of section \ref{section2}. As advertised, this vector field configuration computes simultaneously $S(A)$ (as the flux through the boundary region $A$) and the entanglement of purification ${\mathcal{E}}(A:B)$ (as the part of the flux that goes through $\Sigma_{AB}^{(1)}$). The extra threads that we have added in the central bundle cross the minimal surface at points $x<x_e$ and eventually reach the horizon. Hence, these threads have a purely thermal interpretation.

Before finishing this example, let us offer more insights on the interpretation. Since we have obtained both $S(A)$ and ${\mathcal{E}}(A:B)$ from the same vector field configuration, then it is reasonable to subtract these two quantities. The difference corresponds to the smallest amount of flux that goes thorugh the horizon and therefore has a natural interpretation as the minimal part of $S(A)$ whose nature is undoubtedly thermal, this is
\bea
S^{\text{min}}_{\text{th}}(A)=S(A)-{\mathcal{E}}(A:B)=\frac{c}{3}\log\left[ \sinh\(\frac{\pi \ell}{\beta}\) \right ] \,.
\eea
Similarly, one can interpret ${\mathcal{E}}(A:B)$ itself as the maximum amount of quantum entanglement present in $S(A)$, distilled as Bell pairs,
\bea
S^{\text{max}}_{\text{qm}}(A)={\mathcal{E}}(A:B)=\frac{c}{3}\log\left(  \frac{\beta} {\pi\epsilon}\right )\,.
\eea
Both expressions are valid for $\ell>\ell_c$. As expected, for $\ell<\ell_c$ one obtains that $S^{\text{min}}_{\text{th}}(A)=0$ and hence $S^{\text{max}}_{\text{qm}}(A)=S(A)$. In this latter case one can easily show by similar constructions that there are vector field configurations where all threads completely avoid the horizon.

\subsubsection{Two intervals in pure AdS \label{twointervals}}
The second example illustrated in figure \ref{Fig:Purif}, consists of two strips in a pure AdS geometry. As for the BTZ case, we have a situation where there is a bottle neck in the entanglement wedge associated to $\rho_{AB}$, and therefore we can use the nesting property of entanglement entropy to construct a thread configuration that simultaneously computes entanglement entropy of one of the regions, say $S(B)$, and the entanglement of purification ${\mathcal{E}}(A:B)$. For the sake of simplicity we will focus on AdS$_3$, in which case the strips are just 1-dimensional intervals. A similar construction would also work for higher dimensions and we expect the results to be qualitatively the same.

We will begin by considering two disjoint and sufficiently close intervals, so that the corresponding entanglement wedge is connected. As we will see below, we will need a slight generalization of the method used in the BTZ example. Namely, the set of nested intervals needed to construct the maximally packed flow will need both endpoints to vary in a delicate synchronized fashion, as was considered at the end of section \ref{NPMPF} (see figure \ref{nesting2}). We will show, however, that the geodesic flows constructed in section \ref{section2} provide an appropriate family of nested boundary regions needed for this example. We will conclude by considering the case of two adjacent intervals, which can be easily obtained by a limiting case of the disconnected configuration.

\vspace{3mm}
\textbf{Disjoint intervals:}
For clarity of exposition we will discuss only the symmetric case, where the two intervals are of equal size $\ell_1=\ell_2=\ell$, and are separated a distance $\Delta$ from center to center.\footnote{We can recover the case of intervals of different sizes by applying the bulk coordinate transformation induced by a conformal transformation in the boundary.} When the separation is small enough, the entanglement wedge is a connected surface whose boundary consists of two concentric semi-circles of radii $R_1$, $R_2$ with $R_2 > R_1$, such that $\ell=R_2-R_1$ and $\Delta=R_2+R_1$. The condition to obtain a connected minimal surface in terms of these parameters is
\bea
z=\frac{(R_2-R_1)^2}{4R_1 R_2}>1 \,.
\eea
In this set up we have that both $I(A,B) \neq 0$ and ${\mathcal{E}}(A:B)\neq 0$. The minimal cross-section $\Sigma_{AB}$ that computes entanglement of purification is the vertical line at $x=0$ and $z\in[R_1,R_2]$. A brief calculation leads to
\bea
{\mathcal{E}}(A:B)=\frac{c}{6}\log\(1+2z+2\sqrt{z(z+1)}\)=\frac{c}{6}\log\(\frac{R_2}{R_1}\)\,.
\eea
In the following, we will show how to construct the vector field configuration whose flux across $\Sigma_{AB}$ computes the entanglement of purification for this system.

Since the problem has reflection symmetry around $x=0$, we will focus on the region $x>0$ for simplicity. The first step is to consider a family of nested intervals that interpolate between the minimal surface $m(B)$ and the vertical surface $\Sigma_{AB}$. There are infinitely many ways to pick a family, since the two endpoints have variable locations. In order to choose one, we consider an auxiliary minimal surface $m(\tA)$ for an interval $\tilde{A}$ centered at the origin. We impose that $m(\tilde{A})$ intersects normally the surface $m(B)$ (and by symmetry, the surface $m(A)$), which implies that $m(\tilde{A})$ is a semi-circle with radius
\bea
\tilde{R}=\sqrt{R_1 R_2}\,.
\eea

\begin{figure}[t!]
\centering
 \includegraphics[width=5in]{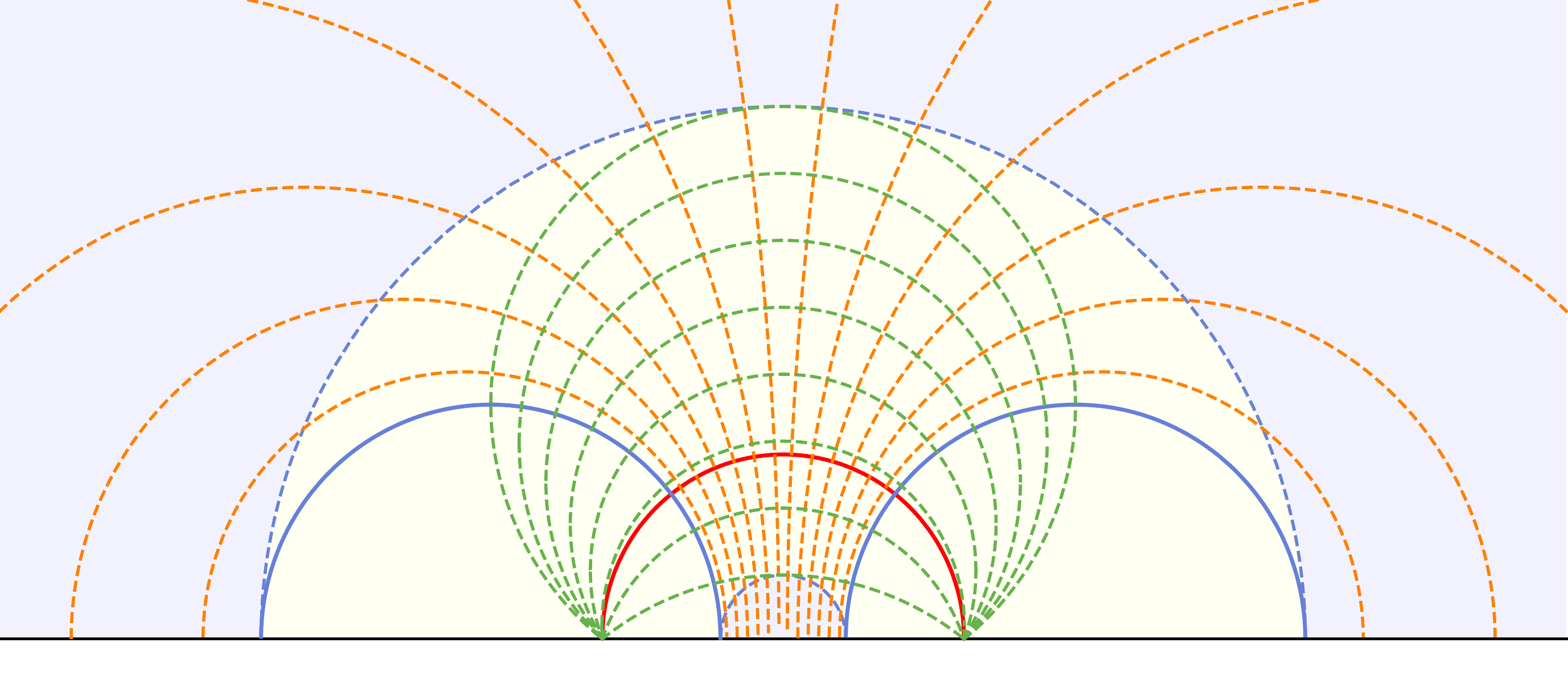}
 \put(-250, -5){{\small ${A}$}}
\put(-115, -5){{\small $B$}}
 \caption{\label{EoP-disjoint} A graphic illustration of the construction of a maximally packed flow for two disjoint intervals in AdS. The two intervals are of equal size
 and are close enough so that the entanglement wedge $r(AB)$ (shaded in yellow) is connected. The auxiliary minimal surface $m(\tilde{A})$ is shown in red, and is picked such that it intersects orthogonally both $m(A)$ and $m(B)$, represented in blue. The set integral curves of the auxiliary vector field $\tilde V$ are shown as orange dashed lines. These integral curves interpolate smoothly between $m(A)$ ($m(B)$) and $\Sigma_{AB}$ so they provide a natural family of nested intervals. The level set surfaces of $\tilde{V}$ are represented by green dashed lines. They correspond to the integral curves of the maximally packed flow $V$ that computes entanglement of purification ${\mathcal{E}}(A:B)$.
}
\end{figure}

Let us now take $\tilde{V}$ to be a geodesic flow (associated to region $\tilde{A}$) of the kind constructed in section \ref{section2}. The full set of integral curves of $\tilde{V}$ between $m(B)$ and $\Sigma_{AB}$ gives us a natural family of nested intervals, which are specified by their endpoints. Therefore, using the nesting property one can construct a maximally packed flow which simultaneously maximizes the flux through $B$ and all the intervals of the family. The associated integral curves of this maximally packed flow are simply given by the level set curves, with $|\tilde{V}|=$ constant. We illustrate this explicitly in figure \ref{EoP-disjoint}. More specifically, the level set curves are naturally parametrized by and angular parameter $\chi$ in terms of which $|\tilde{V}|=\sin\chi$, where
\bea
\chi=\arctan\({\frac{2\tilde{R}z}{\tilde{R}^2-x^2-z^2}} \)\in[0,\pi/2]\,.
\eea
This is a good parametrization for region inside $m(\tilde{A})$, namely, for $x^2+z^2<\tilde{R}^2$. Alternatively, one can chose to parametrize the level set curves with a parameter $\psi$ related to $\chi$ by
\bea
-\frac{1}{\psi}=\frac{2\tilde{R}z}{\tilde{R}^2-x^2-z^2}=\tan\chi
\eea
in terms of which the $|\tilde{V}|=$ constant surfaces adopt the simple form
\bea\label{acurves}
x^2+(z-\psi \tilde{R})^2=(1+\psi^2)\tilde{R}^2\,.
\eea
This is, they correspond to circles centered at $(r_c,z_c)=(0,\psi \tilde{R})$ and radius $R_c=\sqrt{1+\psi^2}\tilde{R}$. If one focus only on those integral curves lying entirely within the entanglement wedge associated to $AB$, $r(AB)$, then, their flux will compute the entanglement of purification ${\mathcal{E}}(A:B)$.

Two final comments are in order: $(i)$ Notice that all integral lines of this maximally packed flow end at a single point of region $A$ and $B$, respectively. This is perfectly valid. However, we can alternatively delete the portion of the threads inside $m(A)$ and $m(B)$ and put there instead a part of a geodesic flow. This is in complete analogy to the construction done in the BTZ case. And $(ii)$ If desired, one can add an extra thread bundles that connect $AB$ and $\overline{AB}$ and cross the regions of $m(A)$ and $m(B)$ that are not covered by the maximally packed flow. Such construction would give us a vector field that simultaneously computes $S(A)$ (or $S(B)$) and the entanglement of purification ${\mathcal{E}}(A:B)$.

\vspace{3mm}
\textbf{Adjacent intervals:}
When the two intervals share a boundary, the family of nested intervals required to construct our maximally packed flow is fully determined, since one of the endpoints is fixed. Therefore, one can easily obtain the corresponding flow by following the BTZ construction. However, since we already have a general construction for the disconnected case, one should be able to recover the case of adjacent intervals by taking an specific limit of the above.

\begin{figure}[t!]
\centering
 \includegraphics[width=5in]{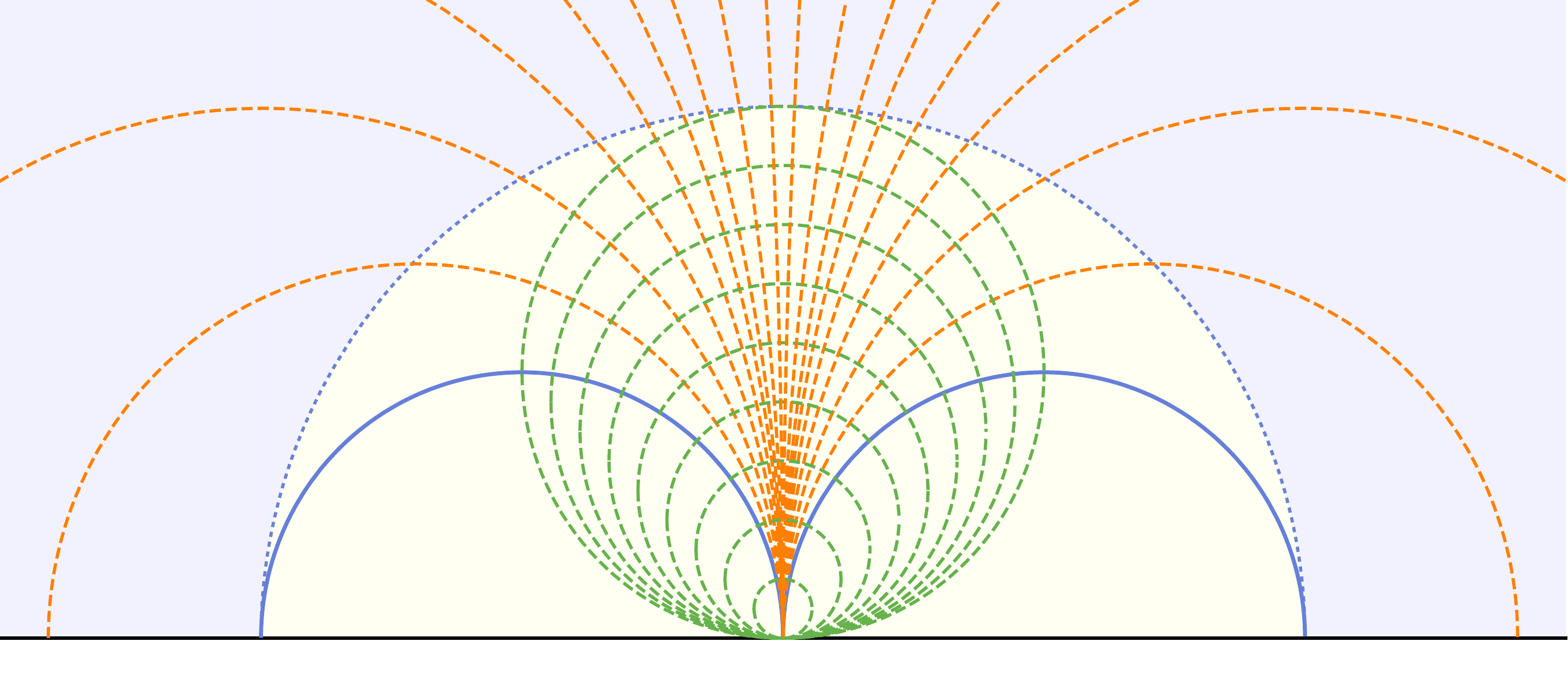}
  \put(-245, -5){{\small ${A}$}}
\put(-125, -5){{\small $B$}}
 \caption{\label{EoP-adjacent} A graphic illustration of the construction of a maximally packed flow for two adjacent intervals in AdS. The flow is obtained as a specific limit of the disjoint case, presented in figure \ref{EoP-disjoint}. All conventions of figure \ref{EoP-disjoint} apply to the present one as well.}
\end{figure}

Indeed, by considering the limit in which $\tilde{R}\to 0$ while keeping $\psi\tilde{R}$ finite and arbitrary, it is possible to show that we can recover the correct flow for adjacent intervals. In this limit equation (\ref{acurves}) becomes:
\bea\label{acurves2}
x^2+(z-\psi\tilde{R})^2=\psi^2\tilde{R}^2\,.
\eea
The integral lines in this case correspond to circles centered at $(r_c,z_c)=(0,\psi \tilde{R})$ and radius $R_c=\psi \tilde{R}$.
We illustrate this limit in figure \ref{EoP-adjacent}. We notice again that all integral lines of this maximally packed flow end at a single point, although, in this case at a coincident point $(x,z)=(0,0)$ that lies in the boundary of $A$ and $B$. This seems problematic at first glance, but all issues are alleviated by the presence of the UV cutoff $\epsilon$. Alternatively one can delete the portion of the threads inside $m(A)$ and $m(B)$ and continue them with part of a geodesic flow, as discussed in the previous example.

\subsection{Monogamy of mutual information}

Very recently, the authors of \cite{Cui:2018dyq} introduced the notion of \emph{multicomodity flows} or \emph{multiflows}: a set of flows that can coexist simultaneously in a given Riemannian geometry while satisfying some non-trivial properties that define them. Furthermore, they showed a set of theorems about multiflows that were used to prove the monogamy property of holographic mutual information. In this section we will give a quick review of the multiflow proposal and briefly summarize the connection with the monogamy property.  We refer the reader to \cite{Cui:2018dyq} for a complete presentation, a detailed proof of the theorems as well as some conjectures about the entanglement structure of holographic states suggested by the bit thread picture of holographic entanglement entropy.

Later in the same section, we will also show that our previous constructions can be used as individual components of such multiflows. As an application, we will show how the geodesic and maximally packed flows presented in the previous sections can be used to illustrate the monogamy property of holographic mutual information in concrete settings, complementing the alternative constructive proof presented in \cite{Hubeny:2018bri}. We do so by explicit construction in the case of two disjoint intervals in AdS$_3$. However, there is in principle no limitation in considering more regions, other geometries or strips in higher dimensions.

\subsubsection{Quick review of the multiflow proposal\label{rev:multiflow}}

Consider a Riemannian manifold $\mathcal{M}$ with boundary $\partial {\mathcal M}$ and take a set of non-overlapping boundary regions $\{A_1, \cdots A_n \}$ such that they cover the full boundary $\partial {\mathcal M}$ ($\cup_i A_i=\partial  {\mathcal M}$). A multiflow is a set of vector fields $V_{ij}$ on ${\mathcal M}$ that satisfy:
\bea
&&V_{ij}=-V_{ji}\,,\qquad \hat{n} \cdot V_{ij} =0 \quad {\rm on}\,\,  A_k \quad (\forall\, k\neq i,j)\,,\label{flow-mf0}
\\ &&\nabla\cdot V_{ij}=0\,, \qquad \sum_{i<j}^n |V_{ij}|\leq \frac{1}{4G_N}\label{flow-mf}\,.
\eea
Each $V_{ij}$ is by itself a flow which has non-zero flux only in the regions $A_i$ and $A_j$ and therefore, from (\ref{flow-mf0}), it follows that the flux through each of these two regions is equal in magnitude but opposite in sign,
\bea
\int_{A_i}V_{ij}=-\int_{A_j}V_{ij}\,.
\eea
Furthermore, the condition (\ref{flow-mf}) implies that the set of flows $V_i$ defined by
\bea\label{flow-mf2}
V_i\equiv\sum_{j=1}^{n}V_{ij}
\eea
are also good flows and therefore satisfy the bound
\bea
\int_{A_i} V_i\leq S(A_i)\,.
\eea
A non-trivial property of these multiflows, proved in \cite{Cui:2018dyq}, is the existence of what is called a \emph{max multiflow}.

\vspace{3mm}
\textbf{Max multiflow:}
is a multiflow $\{V_{ij}\}$ such that for each $i$, the flow $\{V_i\}$ given in (\ref{flow-mf2})
is a max flow for $A_i$, i.e.,
\bea
\int_{A_i}V_i=S(A_i)\,.
\eea
Based only on the existence of this max multiflow, the MMI follows almost trivially. In order to simplify the presentation we will use a slight generalization of the bit thread picture introduced in \cite{Cui:2018dyq}. According to them, the threads are non-oriented one dimensional objects with a density $\rho$ bounded by the Planck scale as\footnote{Recall that in the usual bit threads picture one consider each discrete thread as having a transverse area given by the Planck scale.}
\bea\label{density}
\rho\equiv\frac{\rm length \, of\, threads \,inside \,a\,small\, ball}{\rm volume\, of\, the\, ball }\leq \frac{1}{4G_N}\,.
\eea
The individual threads are allowed to intersect each other as long as the density bound is not violated. In turn, the correspondence with a smooth vector field is hence not one to one.

Denoting $N_{A:\bar{A}}$ as the number of threads connecting $A$ and its complement $\bar{A}$, then
\bea\label{maxthread}
S(A)={\rm max\,}N_{A:\bar{A}}\,.
\eea
This is, the entanglement entropy of $A$ is given by the maximum $N_{A:\bar{A}}$ subject to the density constrain (\ref{density}). The max multiflow (or multi-thread) theorem states that there exists a multiflow configuration $N_{A_{i}:A_{j}}$ obeying the density bound (\ref{density}) such that
\bea\label{maxthread2}
S(A_i)={\rm max\,} N_{A_i: \bar{A}_i}=\sum_{j\neq i} N_{A_i:A_j}\,,
\eea
where $N_{A_i:A_j}$ is the number of threads connecting the region $A_i$ with $A_j$, not necessarily the maximal one.

Lets consider a partition of the boundary in $4$ regions and label them as $A_1=A$, $A_2=B$, $A_3=C$ and $A_4=\overline{ABC}=D$. Then by considering the combined regions $AB$, $AC$ and $BC$, and using (\ref{maxthread}) one gets:
\bea\label{entropies}
S(AB)\geq N_{AB:CD}\,, \qquad S(AC)\geq N_{AC:BD}\,,\qquad S(BC)\geq N_{BC:AD}\,,
\eea
where the thread configuration we are considering corresponds to the one that satisfies (\ref{maxthread2}). Expanding each term in (\ref{entropies}), one gets:
\bea
S(AB)&\geq& N_{A:C}+N_{A:D}+N_{B:C}+N_{B:D}\,, \\
S(AC)&\geq& N_{A:B}+N_{A:D}+N_{C:B}+N_{C:D}\,, \\
S(BC)&\geq& N_{B:A}+N_{B:D}+N_{C:A}+N_{C:D}\,.
\eea
Adding all the above inequalities, using $N_{A_i:A_j}=N_{A_j:A_i}$,  the last equality in (\ref{maxthread2}), and $S(ABC)=S(D)$ (which comes from purity), one concludes that
\bea
S(AB)+S(AC)+S(BC)&\geq& S(A)+S(B)+S(C)+S(ABC)\,.
\eea
The above inequality can be written in terms of mutual informations as
\bea\label{MMI-5}
I(A:BC)\geq I(A:B) +I(A:C)\,,
\eea
which is known as the \emph{monogamy of mutual information}.

\subsubsection{Max multiflow and MMI for disjoint intervals}

In this section we will present an explicit realization of the max multiflow for two disjoint intervals in pure AdS starting from the constructions of section \ref{twointervals}.

Let us consider two disjoint intervals in AdS$_3$. Let $A$ be the interval on the left and $B$ the one on the right. Between $A$ and $B$ there is an interval, which we denote by $C$. Finally, we call $D$ to the complement of $ABC$, namely, $D=\overline{ABC}$. We are interested in the minimal surface and entanglement wedge associated to the disjoint region $AB$. As discussed earlier in section \ref{sec:puri}, depending on the relative separation of the intervals there are two qualitatively different scenarios or phases, which yield a connected or disconnected entanglement wedge. These two scenarios are related by purity: if $m(AB)$ is connected then $m(CD)$ is disconnected and vice-versa. This means that the thread configuration that describes the max flow of $m(AB)$ in the connected case would correspond to the thread configuration that describes the max flow of $m(CD)$ in their disconnected configuration. For concreteness, then, we will concentrate on the connected phase only. Additionally, we will consider the simplest case of two intervals of equal sizes since the most generic case with different sizes can be obtained from the former by a specific conformal transformation.

Consider the set of intervals $X=\{A, AB, ABC\}$. Given the nesting property of entanglement entropy, there should be a flow that simultaneously maximizes the flux through $A$, $AB$ and $ABC$. Such a flow is fixed (up to a sign) at the location of the minimal surfaces $m(A)$, $m(AB)$ and $m(ABC)$, this is $V|_{m(X_i)}=\pm\hat{n}|_{m(X_i)}$ where $\hat{n}|_{m(X_i)}$ is the outward-pointing unit vector normal to $m(X_i)$. The existence of such minimal surfaces $m(X_i)$ suggests a natural separation of the bulk for a given boundary region $X_i$ into the co-dimension one bulk region\footnote{Or co-dimension zero, if we are talking about a constant-$t$ slice of the bulk geometry.} $r(X_i)$ bounded by  $X_i\cup m(X)$ and its complement $\mathcal{M} \textbackslash r(X_i)$. The causal domain of dependence of $r(X_i)$ is the so called entanglement wedge and plays a crucial role in the subregion duality of the AdS/CFT correspondence. However, $r(X_i)$ is also informally called entanglement wedge so we do so throughout this paper. The idea now is to construct a flow maximizing the flux through $A$, $AB$  and $ ABC$ by considering the various patches of the bulk that are naturally separated, this is $r(A)$, $r(AB) \textbackslash r(A)$, $r(ABC)\textbackslash r(AB)$ and $\mathcal{M} \textbackslash r(ABC)$. We do so by imposing the appropriate boundary conditions at $m(A)$, $m(AB)$ and $m(ABC)$ and constructing flows for the individual patches.

First, let us consider one single interval $X_i$, as given in section \ref{strips} for $d=1$. In this case, the bulk is divided into two parts: ${\mathcal M}=r(X_i) \cup ({\mathcal M}\textbackslash  r(X_i) )$. We will call the part of $V$ inside $r(X_i)$ as the {\em interior flow} associated to $X_i$ and the part of $V$ inside ${\mathcal M}\textbackslash  r(X_i) $  as the {\em exterior flow}. Now, consider the system of two disjoint regions $A$ and $B$ discussed above. Given the previous arguments, we can then construct flows in the regions inside $r(A)$ and outside $r(ABC)$ by taking the inside and outside parts of a geodesic flow associated to a single interval, respectively. Next, since we are considering the case with connected $r(AB)$, i.e., $m(AB)=m(ABC)\cup m(C)$, it is easy to see that one must pick $V|_{m(C)}=-\hat{n}|_{m(C)}$. This implies that the flow inside of $r(C)$ can be chosen to be the negative of a geodesic flow of an interval of size $C$.\footnote{One way to see this is by noticing that the flow leaving $AB$ through $m(C)$ is entering $C$, minimizing the flux through $C$ instead of maximizing it.} In summary, using the geodesic flow constructions of section \ref{strips} we have found the part of the flow in the bulk regions $r(A)$, $r(ABC) \textbackslash r(AB)=r(C)$ and $\mathcal{M} \textbackslash r(ABC)$. The only missing piece now is to find the part of the flow inside region $r(AB) \textbackslash r(A)$, which we proceed to do next.

Before describing how to construct the flow inside region $r(AB) \textbackslash r(A)$ we can use simple symmetry arguments put some constraints. First, notice that since the two intervals are of equal size then there is a reflection symmetry around the origin.  Strictly speaking, this symmetry is broken by our choice of maximizing the flux through $A$ instead of $B$. However, if one ignores the orientation of the threads then the flow can indeed be taken to respect this symmetry. A number of consequences follow from it. For example, since the number of threads connecting the region $A$ and $\bar{A}$ is maximal the above symmetry implies that  the number of threads connecting $B$ and $\bar{B}$ is also maximal. This in turn implies that flow inside $m(B)$ can also be taken to be the interior flow of a single interval. 
Another consequence is that the threads connecting $D$ and $A$, and $D$ and $B$, through $m(ABC)$ are divided precisely by the symmetry preserving surface. The same also applies for the threads connecting $C$ and $A$, and $C$ and $B$, through $m(C)$. Indeed, the part of the symmetry preserving surface that lies inside $r(AB)$ is precisely equal to the cross-section whose area computes the entanglement of purification ${\mathcal{E}}(A:B)$. We therefore see that the entanglement of purification plays an interesting role in separating the multiflow components of a max multiflow. It would be very interesting to explore the extent to which this statement holds in more general situations, including general multiple regions, higher dimensions, and arbitrary geometries.

\begin{figure}[t!]
\centering
  \includegraphics[width=5in
 ]{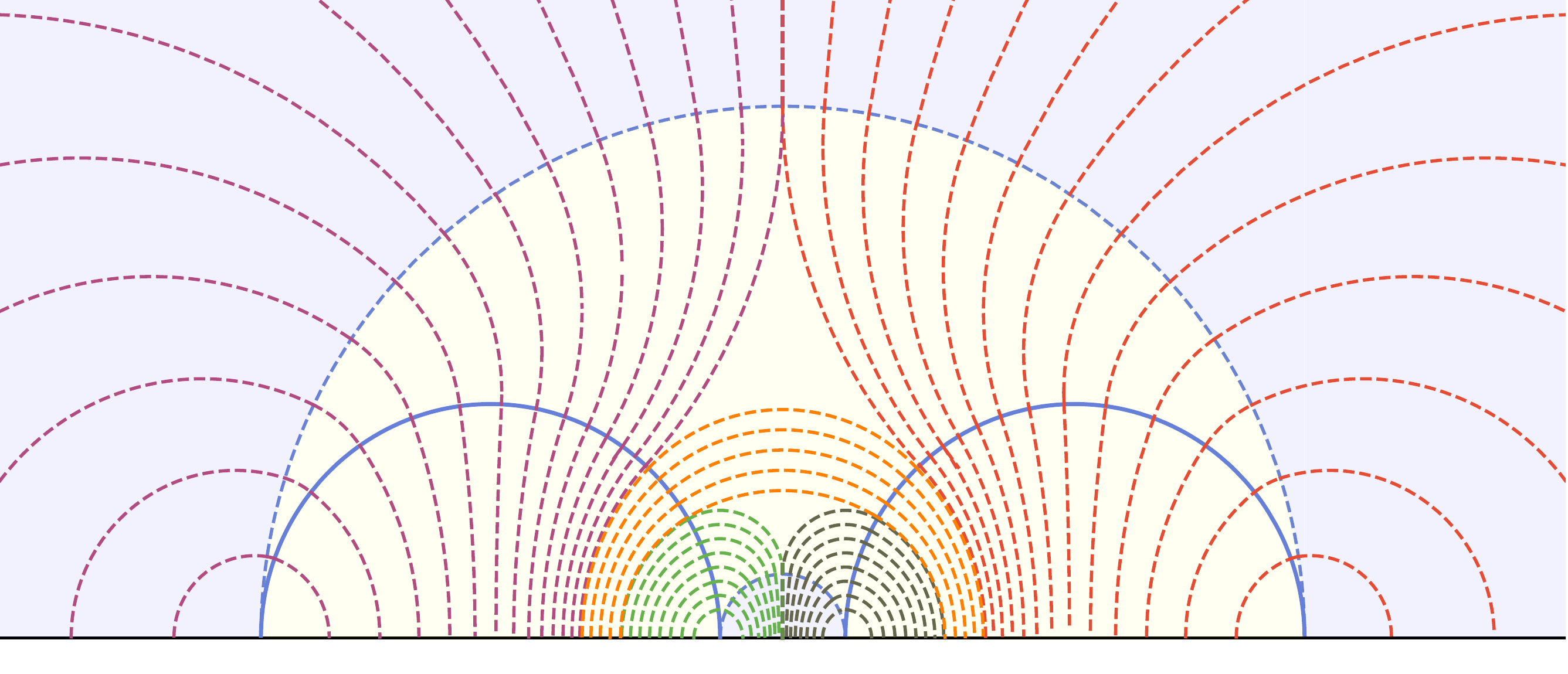}
 \put(-250, -5){{\small ${A}$}}
\put(-250, 80){{\small ${N_{AD}}$}}
\put(-120, -5){{\small $B$}}
\put(-130, 80){{\small ${N_{BD}}$}}
\put(-50, -5){{\small $D=\overline{ABC}$}}
\put(-358, -5){{\small $D=\overline{ABC}$}}
\put(-185, -5){{\small $C$}}
\put(-190, 47){{\small $N_{AB}$}}
\put(-202, 28){\scalebox{0.8}{ $N_{AC}$}}
\put(-182, 28){\scalebox{0.8}{ $N_{BC}$} }
  \caption{\label{MMI-fig} In this figure we represent in different colors the individual thread bundles of the max multiflow configuration connecting the various regions in in a system of two disjoint intervals. The threads connecting the regions $A$ and $B$ are labeled as $N_{AB}$, $A$ and $C$ as $N_{AC}$, $A$ and $D$ as $N_{AD}$, $B$ and $C$ as $N_{BC}$, and $B$ and $D$ as $N_{BD}$. The threads that connect $A$ and $B$ have the physical meaning of representing half the mutual information between $A$ and $B$, $I(A:B)$.}
\end{figure}
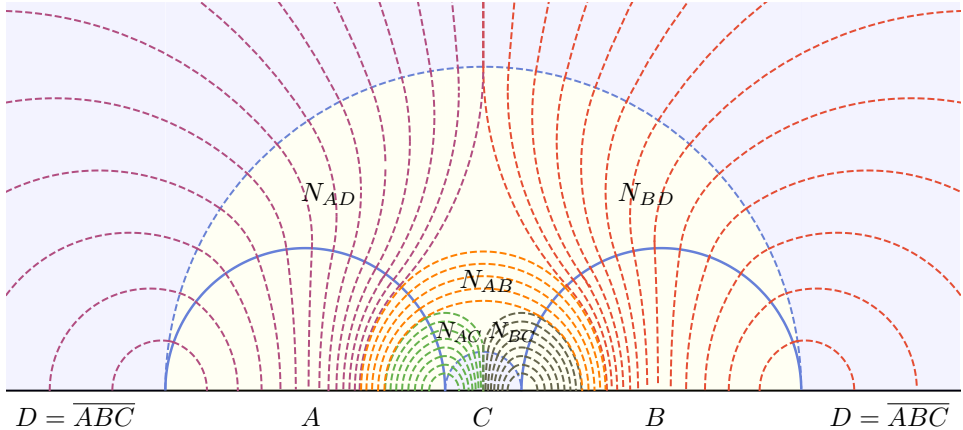
Finally, let us explicitly construct the advertised flow inside region $r(AB)/r(A)$, focussing on the $x<0$ part of the geometry.
Recall that in section \ref{twointervals} we constructed maximally packed flows connecting minimal surfaces of intervals that either share a common boundary or are separated by some distance ---see equations (\ref{acurves}) and (\ref{acurves2}). In fact, these are precisely the kind of flows that we need to connect the different minimal surfaces in $r(AB)/r(A)$. First, let us start from the threads connecting $m(A)$ to $m(ABC)$. These two surfaces share the left boundary, so it is easy to find a family of nested intervals whose minimal surfaces interpolate between the two. The threads that correspond to such maximally packed flow are called $N_{A:D}$. Similarly, $m(A)$ and $m(C)$ share a common boundary so we can repeat the same process to find a maximally packed flow that connects them. We refer to these threads as $N_{A:C}$.
Finally, we need to find threads connecting $m(A)$ and the symmetry preserving surface at $x=0$ (or, equivalently, threads connecting $m(A)$ and $m(B)$). These surfaces are separated by some distance, however, using the method explained in \ref{twointervals} we can propose a family of nested intervals whose minimal surfaces interpolate between the two, and hence construct the corresponding maximally packed flow. We use this construction for the portion of $m(A)$ that has not been covered so far. We call these new threads $N_{A:B}$. The parts of $r(AB)/r(A)$ that are not covered by one of these thread bundles are assumed to have a vanishing flow. In figure $\ref{MMI-fig}$ we represent graphically this construction.

\begin{figure}[t!]
\centering
 \includegraphics[width=5in
 ]{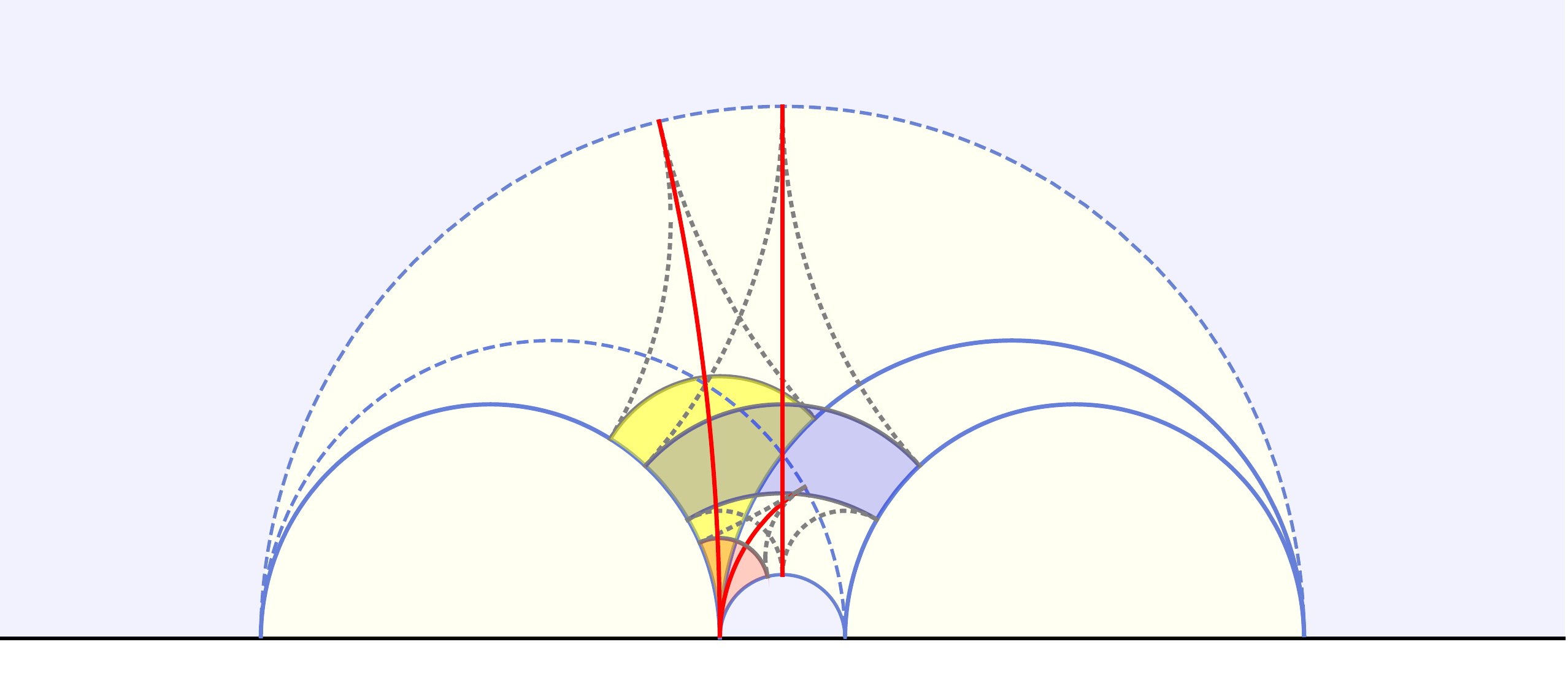}
  \put(-250, -5){{\small ${A}$}}
\put(-120, -5){{\small $B$}}
\put(-185, -5){{\small $C$}}
\put(-50, -5){{\small $D=\overline{ABC}$}}
\put(-358, -5){{\small $D=\overline{ABC}$}}
  \caption{\label{MMI-MI-fig} In this figure we plot the minimal surfaces of all the individual regions as well as the different pairwise combinations. In particular we are interested in $AB$,  $AC$, and $ABC$. We represent the different flows whose flux compute the mutual information between $A$ and $B$,  $A$ and $C$, and $A$ and $BC$ in blue, red and yellow, respectively.}
\end{figure}
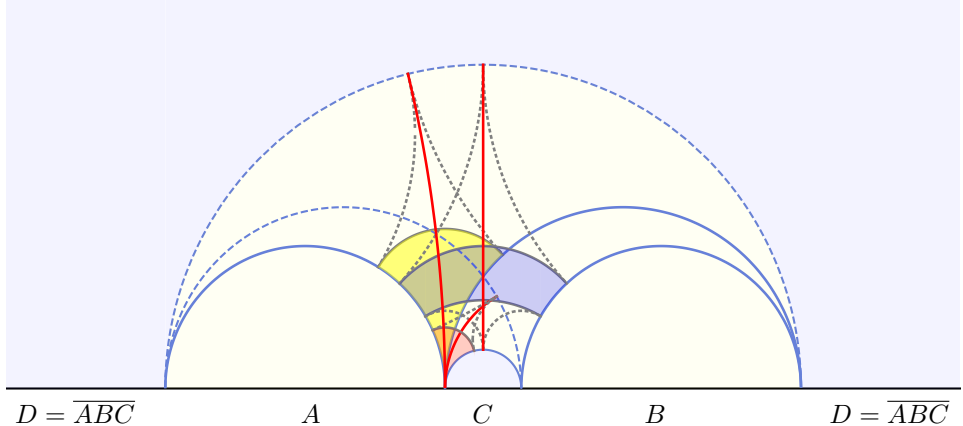

The flow constructed above is an explicit realization of the max multiflow theorem introduced in \cite{Cui:2018dyq}, for two disjoint intervals $A$ and $B$.
As discussed in section \ref{rev:multiflow}, the monogamy property of mutual information (\ref{MMI-5}) follows immediately from the mere existence of such a flow. Indeed,
one can illustrate the inequality by representing graphically the mutual information of the relevant pairs of intervals, as is done in figure \ref{MMI-MI-fig}. As it is shown in this figure, the flow that computes the mutual information between $A$ and $BC$, $I(A:BC)$, contains mutually disjoint flows that compute $I(A:B)$ and $I(A:C)$ separately. Therefore $I(A:BC)$ is manifestly larger or equal to $I(A:B)+I(A:C)$. This illustration has the flavor of the separation of flows that came from the commodity property in a slightly more detailed fashion.

\section{Summary and discussion\label{sec:con}}

In this paper, we have given explicit examples of vector field configurations that compute entanglement entropy in a variety of physical scenarios. As advertised in the introduction, our constructions can be categorized in three types: (i) symmetric or geodesic flows based on specific foliations of the bulk geometry, (ii) maximally packed flows, based on the nesting property of entanglement entropy, and (iii) mixed flows that are consistently composed of the two kinds of flows above.

The symmetric flows were used to explicitly construct flows for strips and spheres in empty AdS, in an arbitrary number of dimensions, and an interval in a BTZ background. In addition, we showed that a similar construction can be implemented in a more general background, provided that the metric satisfies certain curvature condition. For systems that can be dimensionally reduced to a $3$-dimensional gravitational setup, e.g., strips in a general translationally invariant background, we showed that the condition simplifies drastically and reduces to having negative spatial curvature in the effective lower dimensional metric. Intuitively, this result can be explained by the fact that minimal length curves in a space of negative curvature tend to deviate from each other, as is the case for hyperbolic space. This is in contrast to the more familiar case of positive curvature (e.g. a sphere), where the curves tend to focus.
For backgrounds supported by matter fields, we were also able to translate this statement in terms of the local energy density, for which we found a sharp upper bound.

We also presented two concrete applications of mixed flows to particular situations of interest: (i) a strip in a BTZ background and (ii) two disjoint intervals in empty AdS. The first case was already considered using geodesic flows; however, we showed that combining geodesic and maximally packed flows, it is possible to construct a configuration that simultaneously computes entanglement entropy and entanglement of purification of the system. Since the specific flow is found by maximizing the number of threads in the smallest cross-section of the entanglement wedge, we showed that the latter quantity can be naturally interpreted as the maximum number of Bell pairs that can be distilled from the mixed state. Finally, the case of two disjoint intervals was used to illustrate the monogamy property of mutual information. In order to do so, we provided an explicit example of the max multiflow theorem recently proposed in \cite{Cui:2018dyq}, and we showed that it can be graphically illustrated in an elegant way.

There are some open questions related to our work that are worth exploring:
\begin{itemize}
  \item \emph{Reduced symmetry.} It would be interesting to relax some of the symmetries that we have assumed in our constructions. For example, what can be said about situations in which the minimal surface is not known? This applies in particular to higher dimensional regions that cannot be dimensionally reduced.
      Are there special foliations that are useful in these cases? Perhaps by restricting the attention to families of lower dimensional cross-sections?
  \item \emph{Finite volume.} All of our explicit constructions were carried out in the Poincar\'e patch of AdS or in a black brane geometry, hence, the dual field theory is defined on $\mathbf{R}^{1,d+1}$. It would be interesting to study situations where the field theory lives on a compact space, such as $\mathbf{R}\times \mathbf{S}^{d+1}$. This would amount to study the problem in global AdS or in a black hole geometry. Both constructions, the one based on geodesics and the one based on the nesting property, are still valid in these scenarios. However, finite volume introduces new physical features such as the so-called entanglement plateaux \cite{Hubeny:2013gta}, entanglement shadows \cite{Freivogel:2014lja} and the existence of winding/entwinement geodesics \cite{Balasubramanian:2014sra}. It would be interesting to study the interplay between all these new phenomena, their role in the construction of bit threads configurations and their interpretation in this language.
  \item \emph{Covariant generalization.} In this paper we have focussed on static situations, where the entanglement entropy is computed by the standard RT prescription \cite{Ryu:2006bv,Ryu:2006ef}. In dynamical situations, however, one needs to upgrade the recipe and use the covariant HRRT formula instead \cite{Hubeny:2007xt}. A covariant proposal of bit threads is currently under development \cite{Headrick:toappear}. It would be very interesting to explicitly construct examples of thread configurations in relevant dynamical situations, e.g., a global quantum quench. The bit thread picture in this scenario would yield important insights regarding the \emph{entanglement tsunami} proposal of entanglement propagation \cite{Liu:2013iza,Liu:2013qca} and its breakdown \cite{Kundu:2016cgh}.
  \item \emph{Dynamics}: In perturbative excited states, entanglement entropy satisfies the so-called first law of entanglement, in which the change in entanglement entropy is given by the change in modular energy. In the bulk, this statement is mapped into the linearized Einstein equations around empty AdS \cite{Lashkari:2013koa,Faulkner:2013ica}. It would be very interesting to see how this translates into the bit thread picture. It would also be interested to go beyond linear level and investigate the imprints on quantities such as relative entropy and quantum Fisher information \cite{Blanco:2013joa,Lashkari:2015hha}. Some work in this direction is already underway \cite{Caceres:toappear}.
  \item \emph{Multipartite entanglement.} Our explicit construction for two disjoint intervals can be straightforwardly generalized to an arbitrary number of parties. It would be very interesting to explicitly do so, at least for 3 and 4 intervals, and find how the holographic inequalities that must be satisfied in each case \cite{Bao:2015bfa,Hubeny:2018trv} are reflected in terms of bit threads. In particular, it would be insightful to represent visually these inequalities in terms of thread bundles connecting the various regions, as we did for the monogamy of mutual information. We expect the construction of flows in these scenarios to be intrinsically related to the recent generalizations of entanglement of purification for multipartite systems \cite{Umemoto:2018jpc,Bao:2017nhh,Bao:2018gck}.
  \item \emph{Bulk reconstruction.} The program of hole-ography \cite{Balasubramanian:2013lsa,Myers:2014jia,Headrick:2014eia,Espindola:2017jil} aims to reconstruct arbitrary bulk surfaces by cleverly adding and subtracting the entanglement entropies of a family of intervals associated to the surface. Interestingly, after taking a continuum limit and shrinking the surface to a point it is possible to recover an intrinsic definition of a bulk point in terms of a CFT quantity dubbed as differential entropy \cite{Czech:2014ppa}. It would be interesting to explore similar questions in the language of bit threads and shed light on the issue of bulk reconstruction and bulk locality \cite{Lokhande:toappear}.
  \item \emph{Higher derivatives.} In higher curvature gravity, the holographic entanglement entropy is computed by the Dong-Camps prescription \cite{Dong:2013qoa,Camps:2013zua} which includes Wald's formula for black hole entropy, as well as corrections involving the extrinsic curvature.
      The bit thread proposal was recently generalized to arbitrary higher curvature gravity in \cite{Harper:2018sdd}. In would be interesting to construct explicit examples of flows using the generalized prescription to gain insight into the stringy or $\alpha'$ corrections to entanglement entropy.
  \item \emph{Quantum corrections.} The leading quantum or $1/N$ corrections to holographic entanglement entropy are given by bulk entanglement entropy across the RT surface \cite{Faulkner:2013ana}. One immediate consequence is that threads can now connect points in the bulk that are not necessarily at the boundary of AdS, and therefore one must relax the divergenceless condition. It would be interesting to investigate how different bulk fields couple to $V$ and how they source specific thread configurations. One specific example that might be of interest is the case of two disjoint intervals, in the case that the entanglement wedge is disconnected \cite{Agon:2015ftl}. For this system, the leading $1/N$ corrections to entanglement entropy gives the leading term in the mutual information $I(A,B)$.
  \item \emph{Tensor networks.} Finally, we remark that the bit thread picture of holographic entanglement entropy, as well as the proposal to compute entanglement of purification, were motivated in part by notions of cMERA and other holographic models tensor networks \cite{Swingle:2012wq,Miyaji:2015fia,Hayden:2016cfa,Miyaji:2016mxg}. An interesting avenue for further research would be to consider the explicit thread configurations constructed in this paper and extrapolate them back to the realm of tensor networks. For example, it would be natural to associate the magnitude $|V|$ to a density of tensors in a continuous network on AdS. The associated integral lines might in turn represent paths of optimal displacement along the RG scale. If so, do the geodesic and maximally packed flows constructed here have a specific interpretation?
\end{itemize}
We hope to come back to some of these points in the near future.

\section*{Acknowledgements}

We are especially indebted to Matthew Headrick for multiple discussions on multi-commodity flows and the monogamy of mutual information.  We would also like to thank Elena C\'aceres, Jonathan Harper, Sagar Lokhande, Martin Ro\v{c}ek, Bogdan Stoica and Julio Virrueta for useful discussions and comments on the manuscript.  CAA is supported by the National Science Foundation under CAREER award PHY16-20628. CAA also acknowledges
support from the $\Delta$-ITP visiting program and would like to thank the
Institute for Theoretical Physics at the University of Amsterdam for the warm hospitality during his
extended visit. JFP is supported by the Netherlands Organization for Scientific Research (NWO) under the VENI grant 680-47-456/1486.

\appendix

\section{Geodesic integral curves for strips? \label{A}}

A natural guess for the integral curves in the case of strip geometries are spacelike geodesics. Indeed, geodesics were shown to satisfy all the conditions for the case of spheres. In this appendix we will show that, contrary to the expectation, the integral curves obtained in this way fail to satisfy the nesting property (\ref{nested-1}), (\ref{nested-2}) for $d\geq3$.

The minimal surface associated to a strip in arbitrary dimensions was discussed in section \ref{strips}. We will use the same notation here; so we take the expressions for the minimal surface $x_m(z_m)$ and the outward-pointing unit normal vector $\hat{n}_m$ given in (\ref{minstrip}) and (\ref{nastrip}), respectively. As for the integral curves, we consider geodesics that lie in the $(x,z)$ plane. We require from these curves to intersect the minimal surface $m(A)$ at a point $(x_m,z_m)$ where its normal $\hat{n}_m$ coincides with the tangent to the curve $\hat{\tau}$.

The family of geodesics that lie in the $(x,z)$ plane is given by the two parameter family of circumferences
\bea\label{geodesic4}
(x-x_s)^2+z^2=R_s^2\,,
\eea
where $x_s$ is the center of the circle in the $x$-axis and $R_s$ is its radius. The tangent vector to the curve at an arbitrary point is given by
\bea
\hat{\tau}=\frac{z}{R_s}\(z\, , \, -(x-x_s)\)\,.
\eea
Enforcing that $\hat{\tau}=\hat{n}_m$ at a point $(x_m,z_m)$ on the minimal surface leads to
\bea\label{xsRs-strip}
R_s=\frac{z_*^d\zm}{\sqrt{z_{*}^{2d}-\zm^{2d}}}\qquad {\textrm{and}} \qquad x_s=x_m(\zm)+\frac{\zm^{d+1}}{\sqrt{z_{*}^{2d}-\zm^{2d}}}\,,
\eea
where $z_{*}$ is the parameter given in (\ref{zstar}). Plugging (\ref{xsRs-strip}) into (\ref{geodesic4}) gives us the one parameter family of integral curves label by $z_m$. The integral curves intersect $A$ at
\bea
x_a=x(\zm)+\frac{\zm^{d+1}-z_*^{d} \zm}{\sqrt{z_{*}^{2d}-\zm^{2d}}}\,,
\eea
and $\bar{A}$ at
\bea
x_{\bar{a}}=x(\zm)+\frac{\zm^{d+1}+z_*^{d} \zm}{\sqrt{z_{*}^{2d}-\zm^{2d}}}\,.
\eea
With these expressions we can check if the curves parametrized by $z_m$ are properly nested. Since the curves are geodesics, this is guaranteed provided that $dx_a/dz_m<0$ and $dx_{\bar{a}}/dz_m>0$. A quick calculation shows that
\bea
\frac{dx_0}{d\zm}=-\frac{z_*^d(z_*^d-(d-1)\zm^d)}{(z_*^d+\zm^d)\sqrt{z_*^{2d}-\zm^{2d}}}\qquad {\textrm{and }} \qquad \frac{d\bx}{d\zm}=\frac{z_*^d(z_*^d+(d-1)\zm^d)}{(z_*^d-\zm^d)\sqrt{z_*^{2d}-\zm^{2d}}}\,.
\eea
Unfortunately, the above expressions satisfy the nesting conditions only for $d=1$ and $d=2$. This means that for arbitrary $d$ one cannot construct $V$ for strips based on geodesics as their integral curves.

\subsection{A geodesic flow for strips in $d=2$}

As stated above, the vector field $V$ associated to strips can be constructed by taking geodesics as the integral curves only for $d=1$ and $d=2$. The $d=1$ case corresponds to an interval in a CFT$_2$, and was already considered in the spheres section \ref{spheres}. The $d=2$ corresponds to a strip in a CFT$_3$. While this case was already considered in section \ref{strips}, the construction there was based on minimal surfaces, instead of geodesics. The new integral curves based on geodesics will therefore provide an alternative vector field for a strip geometry. We will explicitly construct this vector field in the remaining part of this section.

Having the integral curves, the next step to determine the vector field $V$ is to compute its magnitude, using (\ref{magnitudeV}).
The process to obtain the orthogonal metric follows exactly as in section (\ref{spheres}), leading to
\bea
ds_\perp^2\equiv h_{ab}dx^a dx^b=\frac{1}{z^2}\frac{1}{R_s^2}\left[(x-x_s)dx+z dz \right]^2\,.
\eea
From implicit differentiation of (\ref{geodesic4}) together with (\ref{xsRs-strip}), the above expression can be written explicitly in terms of the $(z_m, x_m)$ coordinates as follows:
\bea
ds_\perp^2=\frac{L^2}{z^2}\left[\frac{z_*^4}{z_m^4} \(\frac{z_*^{4}+\zm^{4}-2z_*^2 \zm^2\sqrt{1-\frac{z^2}{\zm^2}\( \frac{z_*^{4}-\zm^{4}}{z_*^{4}}\)} }{z_*^4-\zm^4}\)^2dx_m^2+dy^2\right] \,.\nonumber \\
\eea
In the above equation $y$ represents the transverse coordinate, in which the strip extends infinitely.
Finally, using equation  (\ref{magnitudeV}) one obtains the norm of the vector field
\bea
|V|&=&\(\frac{z}{\zm}\)^{2} \(\frac{z_*^{4}-\zm^{4}}{z_*^{4}+\zm^{4}-2z_*^2 \zm^2\sqrt{1-\frac{z^2}{\zm^2}\( \frac{z_*^{4}-\zm^{4}}{z_*^{4}}\)}}\)\,,
\eea
which together with the unit tangent vector,
\bea
\htt=\frac{z}{L}\(\frac{z}{\zm}\frac{\sqrt{z_*^{4}-\zm^{4}}}{z_*^{2}}\,,\, \sqrt{1-\frac{z^2}{\zm^2}\( \frac{z_*^{4}-\zm^{4}}{z_*^{4}}\)}\)\,,
\eea
gives us the full vector field $V=|V|\hat{\tau}$. One can check that this flow respects the bound, $|V|\leq1$, everywhere away from the minimal surface.

\bibliographystyle{ucsd}
\bibliography{refs-flows}

\end{document}